\documentclass[aps,onecolumn,10pt,superscriptaddress]{revtex4-2}

\usepackage[english]{babel}
\usepackage[utf8]{inputenc}
\usepackage[colorinlistoftodos, color=green!40, prependcaption]{todonotes}
\usepackage{comment}
\usepackage{array}
\usepackage{multirow}
\usepackage{booktabs}
\usepackage{rotating}
\usepackage{amsthm}
\usepackage{physics}
\usepackage{amsmath}
\usepackage{amssymb}
\usepackage{etoolbox}
\usepackage{optidef}
\usepackage{adjustbox}
\usepackage{graphicx}
\usepackage{bm}
\usepackage{placeins}
\usepackage[colorlinks=true, linkcolor=blue,
            citecolor=blue, urlcolor=blue]{hyperref}

\allowdisplaybreaks

\newcommand{\ceil}[1]{\lceil #1 \rceil}

\begin{document}

\title{Efficient Block Encodings of Vibrational Hamiltonians}

\author{Shashank~G.~Mehendale}
\affiliation{Chemical Physics Theory Group, Department of Chemistry, University of Toronto, Toronto, Ontario M5S 3H6, Canada}
\affiliation{Department of Physical and Environmental Sciences, University of Toronto Scarborough, Toronto, Ontario M1C 1A4, Canada}

\author{Ignacio~Loaiza}
\affiliation{Xanadu, Toronto, Ontario M5G 2C8, Canada}

\author{Shreyas~Malpathak}
\affiliation{Chemical Physics Theory Group, Department of Chemistry, University of Toronto, Toronto, Ontario M5S 3H6, Canada}
\affiliation{Department of Physical and Environmental Sciences, University of Toronto Scarborough, Toronto, Ontario M1C 1A4, Canada}

\author{Stepan~Fomichev}
\affiliation{Xanadu, Toronto, Ontario M5G 2C8, Canada}

\author{Danial~Motlagh}
\affiliation{Xanadu, Toronto, Ontario M5G 2C8, Canada}

\author{Artur~F.~Izmaylov}
\affiliation{Chemical Physics Theory Group, Department of Chemistry, University of Toronto, Toronto, Ontario M5S 3H6, Canada}
\affiliation{Department of Physical and Environmental Sciences, University of Toronto Scarborough, Toronto, Ontario M1C 1A4, Canada}

\begin{abstract}
Simulation of molecular vibrations is emerging as a promising application of quantum computing in quantum chemistry.
In this work, we develop new strategies for block-encoding the second-quantized $n$-mode representation of the vibrational Hamiltonian, whose algebraic structure closely parallels that of the second-quantized electronic Hamiltonian.
We derive linear-combination-of-unitaries (LCU) representations of this Hamiltonian based on Pauli decomposition and tensor hypercontraction (THC) and construct efficient circuits for the corresponding block encodings.
We also extend the Block-Invariant Symmetry Shift (BLISS) technique to vibrational Hamiltonians, yielding a lower LCU 1-norm and, consequently, a lower overall circuit cost.
For the molecules considered here, BLISS-THC reduces the Toffoli cost of quantum phase estimation by more than two orders of magnitude relative to existing block-encoding approaches. 
In comparison with recent Trotter-based methods, it shows up to a ninefold reduction in the T-gate count, with the relative savings tending to increase with system size.
Asymptotically, and up to polylogarithmic factors, both the Pauli and THC LCUs trade additional ancillas for a quadratic improvement in scaling with the number of modes, while THC provides an additional quadratic improvement in scaling with the number of modals.
These results highlight the importance of Hamiltonian factorizations and Toffoli--qubit tradeoffs for reducing the computational cost of quantum simulations of molecular vibrations.
\end{abstract}

\maketitle

\section{Introduction}
Understanding molecular dynamics from first principles has many applications across chemistry, including molecular identification from vibrational spectra \cite{spectroscopy, loaiza2025NIR} and prediction of reaction pathways \cite{reaction_pathway, reaction_pathway_2}.
For example, the infrared spectrum not only functions as a molecular fingerprint for identifying unknown species, but it also dictates a molecule's ability to absorb thermal radiation, which can be used to evaluate its broader atmospheric and environmental impact \cite{global_warming, global_warming_2}.
Within the Born--Oppenheimer approximation, nuclear and electronic motion are separated, allowing the construction of a molecular vibrational Hamiltonian from the electronic potential energy surface (PES) evaluated at various nuclear geometries.
Accurate construction of electronic PESs has long been a central goal of \textit{ab initio} molecular simulations.
While achieving accurate PESs is essential, the subsequent vibrational structure problem remains a formidable computational challenge for many systems of interest.
Indeed, for floppy systems such as water clusters, the key bottleneck in obtaining accurate vibrational spectra often lies not in the quality of the PES but rather in solving the vibrational Schr\"{o}dinger equation \cite{sawaya2021near,Carrington2017}. 
This bottleneck motivates the need for further development of efficient algorithms for vibrational structure calculations.

Assuming a PES has been obtained to sufficient accuracy, various classical computational methods exist to describe the evolution of nuclei on the PES \cite{mctdh, mlmctdh} or to obtain the associated vibrational spectra \cite{Bowman20082008, VSCF, VSCF_VCC}.
However, as the number of vibrational modes and the strength of their coupling grow, these techniques become prohibitively expensive \cite{loaiza2025NIR}.
While quantum computing has been proposed and studied extensively for the electronic structure problem \cite{aspuruguzik, McArdle_RevModPhy}, its application to the vibrational problem remains comparatively underexplored.
Encoding the bosonic representation of the vibrational Hamiltonian natively on a bosonic quantum computer has been explored in Refs.~\cite{vib_on_bosonic_QC_3, vib_on_bosonic_QC_2, vib_on_bosonic_QC}.
While such an approach enables encoding the underlying Hilbert space without truncation, bosonic quantum hardware is still in the early stages of development.
In this work, we focus instead on simulating vibrational systems on qubit-based digital quantum devices.

Over the last few years, numerous proposals for utilizing near-term quantum computers to calculate vibrational spectra have emerged \cite{mcardle2019digital, ollitrault2020hardware, fermires_CO2, lotstedt2022evaluation, vADAPTVQE, lotstedt2026h2o}.
Most of these methods rely on the variational principle or subspace expansion techniques and are limited to small molecules. 
They inherit the well-known difficulties of near-term methods \cite{nisq} and struggle as the density of states and anharmonicity increase deeper in the spectrum.
More recently, sample-based quantum diagonalization methods have also been proposed for vibrational problems \cite{vib_SQD1, vib_SQD2}.
Interest in fault-tolerant proposals has been on the rise as well, with the majority of work \cite{loaiza2025NIR, lotstedt2026simulation, sawayaTrotter, trenev2025refining, malpathak2025trotter} focusing on utilizing the Trotter approximation for Hamiltonian time evolution within quantum phase estimation (QPE) \cite{QPE}.
Although the Trotter approximation is space-efficient, using it for energy estimation or dynamics with guaranteed error bounds \cite{commutator_Trotter} often yields resource estimates that are overly pessimistic by many orders of magnitude \cite{pert_trotter}.
The authors of Refs.~\cite{loaiza2025NIR, malpathak2025trotter} utilize a perturbative estimate of Trotter error \cite{chemical_basis} to evaluate the resources required for vibrational time evolution.
While such estimates are known to be tighter than commutator-based upper bounds, evaluating and validating them for large vibrational systems remains computationally demanding. Despite the recent appearance of highly optimized computational approaches for estimating such errors \cite{maxwell2026practicalestimationtrottererror}, a more fundamental question remains: are Trotter methods the most efficient route for vibrational simulations?
Qubitization \cite{qubitization}, on the other hand, encodes the energies of a Hamiltonian exactly, up to rotation synthesis errors that are common to any digital quantum simulation approach.
It works by transforming a block-encoded Hamiltonian expressed as a linear combination of unitaries (LCU) \cite{LCU} into a Szegedy walk operator \cite{szegedy2004quantum}, whose eigenphases encode the arccosine of the rescaled eigenvalues of the Hamiltonian.
This enables estimating the energies by directly extracting the phase of the walk operator, achieving Heisenberg-limited scaling \cite{linearT}.
Moreover, given access to an efficient block-encoding oracle, quantum signal processing \cite{QSP} achieves better asymptotic scaling than Trotterization for approximating time evolution.
Consequently, there is an increasing need to develop efficient techniques for block-encoding vibrational Hamiltonians. 

A recent study in Ref.~\cite{malpathak2025trotter} highlighted how different representations of the vibrational Hamiltonian enable distinct strategies for encoding vibrational problems on quantum computers.
One of these is the real-space representation that encodes a discretized real-space grid for the vibrational coordinates into the computational basis states of the qubits.
Resource estimates for block-encoding such real-space vibrational Hamiltonians have recently been analyzed in Refs.~\cite{kamakari2025BEIT, szczepanik2025simulatingBEIT}.
In this work, we focus on an alternative representation, the second-quantized $n$-mode representation of the vibrational Hamiltonian \cite{christiansen2004second}.
It exploits an $n$-mode expansion of the PES along with a finite number of basis functions, often called modals, to discretize each mode.
Its complexity scales with the number of modes that are allowed to be explicitly coupled to each other (discussed in detail in Section~\ref{sec: vib ham}), making it appealing for systems that allow for an accurate description at low mode-coupling levels. 
This makes the use of localized modes particularly beneficial, as they have been shown to significantly reduce the need to include higher-order mode couplings \cite{mode_loc_1,mode_loc_2,mode_loc_3,mode_loc_4}. 

Simulating experimental IR spectra is an interesting application of quantum simulation that can be realized by performing QPE on an initial state obtained by applying the dipole operator to a reference state \cite{sawaya2021near}. The near-IR regime is a particularly interesting target because of its dense manifold of excited states and the greater prevalence of combination bands, overtones, and anharmonic effects.
Since the dominant computational cost in this procedure arises from QPE, reducing the underlying Hamiltonian block-encoding cost is important for lowering the overall resource requirements.
Therefore, here we consider block encodings of second-quantized vibrational Hamiltonians and assess their QPE resource requirements using ground-state energy estimation as a representative benchmark.
In Ref.~\cite{majland2025fault}, such block-encoding-based QPE costs were evaluated using LCUs derived from the factorization of Hamiltonian tensors.
The approach proposed there expresses the Hamiltonian as a sum-of-products (SOP) of one-mode operators, yielding the block encoding of the full Hamiltonian as an SOP of one-mode block encodings.
While this construction is space-efficient and straightforward to generalize, it relies on repeating the block-encoding primitive for every one-mode operator.
Because each primitive loads a separate, relatively short data table, it cannot fully exploit quantum read-only memory (QROM) constructions that utilize additional ancillas \cite{sparse_LCU, low2024tradingSELECTSWAP, halving_qrom}, whose benefit grows with table size.
However, Ref.~\cite{majland2025fault} explored a tradeoff between qubits and T-\textit{depth} by implementing block-encoding primitives for different modes in parallel, enabled by the commutativity of disjoint mode operators.
While T-depth plays an important role in the total execution time of a quantum algorithm, T-gate parallelization requires multiple magic-state factories operating simultaneously on the quantum chip alongside storage and teleportation, and is therefore tied to the specific fault-tolerant architecture.
Consequently, reducing the total T-gate count serves as a direct measure of total magic state cost, while still preserving the option for subsequent parallelization if permissible under specific architectural constraints.

In this work, we propose more efficient block encodings for second-quantized vibrational Hamiltonians that require fewer Toffoli gates, and consequently fewer T gates, than prior approaches for QPE-based energy estimation.
We first introduce a vibrational Pauli LCU that does not involve tensor factorizations, similar to the approach in Ref.~\cite{sparse_LCU}.
By utilizing the swap networks proposed in Ref.~\cite{prebornoppenheimer}, we overcome the need for repeated application of one-mode block-encoding primitives and show that the Pauli LCU achieves a quadratic reduction in the asymptotic Toffoli cost (up to polylogarithmic factors) with respect to the number of modes compared to the best method presented in Ref.~\cite{majland2025fault}.
Moreover, by exploiting the permutational symmetries of the mode-coupling tensors, we extend the tensor hypercontraction (THC) \cite{thc, org_thc} formalism to the second-quantized form of the vibrational Hamiltonian.
We explicitly formulate this for two- and three-mode-coupling interactions, while noting that the underlying factorization is generalizable to an arbitrary $n$-mode expansion.
Under the assumption that the average THC rank scales linearly with the number of modals, and up to polylogarithmic factors, THC further reduces the asymptotic Toffoli count quadratically with respect to the number of modals.
Notably, we show that modal rotations (i.e., basis rotations in the modal space) can be used to optimize vibrational block-encoding resources despite their overhead in the earlier architecture of Ref.~\cite{majland2025fault}.
Finally, as the total algorithmic cost scales as $\mathcal{O}(\lambda_{\mathrm{LCU}})$, where $\lambda_{\mathrm{LCU}}$ is the LCU 1-norm, we extend the Block-Invariant Symmetry Shift (BLISS) technique \cite{bliss} to systematically minimize $\lambda_{\mathrm{LCU}}$.
By exploiting the symmetries of the target eigenstates, BLISS modifies the Hamiltonian strictly outside the relevant symmetry sector to minimize $\lambda_{\mathrm{LCU}}$ without altering the desired eigenvalues.

To evaluate the methods proposed here, we compare the resources required for QPE with those of the method proposed in Ref.~\cite{majland2025fault} for a series of polycyclic aromatic hydrocarbon (PAH) Hamiltonians ranging from 30 to 138 vibrational modes.
These systems employ a PES truncated at the two-mode-coupling level.
Additionally, we test our methods on three-mode-coupling Hamiltonians for two small molecules.
We show that vibrational THC combined with BLISS achieves a reduction of more than two orders of magnitude in Toffoli count for many systems compared to prior art.
For the systems studied in Ref.~\cite{malpathak2025trotter}, we also demonstrate the advantages of the methods developed here over Trotterized QPE.

The remainder of the paper is organized as follows.
Section~\ref{sec: theoretical background} establishes the necessary theoretical background.
Section~\ref{sec: LCUs} introduces the new vibrational Hamiltonian LCUs and details the BLISS technique.
Section~\ref{sec: resource estimates} derives the asymptotic resource estimates for the vibrational LCUs, followed by the presentation of our numerical results in Section~\ref{sec: results}.
Finally, Section~\ref{sec: conclusion} concludes the paper with a discussion of open challenges and future research directions in the quantum simulation of vibrational Hamiltonians.

\section{Theoretical Background} \label{sec: theoretical background}
In this section, we provide some theoretical background for the techniques developed in this work.
We describe the $n$-mode expansion of the PES and the second-quantized form of vibrational Hamiltonians.
We also provide a brief overview of costs associated with QPE and qubitization.

\subsection{The Vibrational Hamiltonian} \label{sec: vib ham}
Under the Born--Oppenheimer approximation, a vibrational Hamiltonian with $M$ modes is
\begin{align}
    \hat H_{\mathrm {real}} = -\dfrac{1}{2}\sum_{i=0}^{M-1} {\Omega}_{i}\dfrac{\partial^2}{\partial \mathcal Q_i^2} + V(\bm {\mathcal Q}), 
\end{align}
where the $\mathcal{Q}_i$ are the mass-weighted dimensionless normal modes, $\Omega_{i}$ are the corresponding harmonic frequencies, and $V(\bm {\mathcal Q})$ is the electronic PES.
$\bm{\mathcal{Q}} = \bm{0}$ is the equilibrium geometry of the molecule.
The second-quantized representation \cite{christiansen2004second} is then obtained by first expanding the PES in an $n$-mode expansion as
\begin{align}
    V(\bm{\mathcal{Q}}) &= V(\bm 0) + \sum_{i=0}^{M-1} V^{(1)}_i(\mathcal{Q}_i) + \sum_{i<j=0}^{M-1} V^{(2)}_{i,j}(\mathcal{Q}_i, \mathcal{Q}_j) + \sum_{i<j<k=0}^{M-1} V^{(3)}_{i,j,k}(\mathcal{Q}_i, \mathcal{Q}_j,\mathcal{Q}_k) +  \dots,
\end{align}
where
\begin{align*}
 V^{(1)}_i(\mathcal{Q}_i)  &= V(0, \dots, \mathcal Q_i, \dots, 0) - V(\bm 0), \\
    V^{(2)}_{i,j}(\mathcal{Q}_i, \mathcal{Q}_j)  &= V(0, \dots, \mathcal Q_i, \dots, \mathcal Q_j, \dots, 0) - V^{(1)}_i(\mathcal{Q}_i) - V^{(1)}_j(\mathcal{Q}_j) - V(\bm 0), 
\end{align*}
and so on.
This representation expresses the PES as contributions obtained by displacing selected subsets of modes from equilibrium.
The next step is to perform Galerkin discretization \cite{low-depth_quantum} of the vibrational Schr\"odinger equation.
While the exact vibrational state is a function of continuous variables $\mathcal Q_i$, an approximation is constructed by using a set of $N$ real orthonormal single-mode basis states $\ket* {\phi^{(i)}_{p_i}}$, often called modals, for each mode $i$.
A full-space basis state is then constructed by a direct product of modals as $\ket*{\phi^{(0)}_{p_0}\phi^{(1)}_{p_1}\dots \phi^{(M-1)}_{p_{M-1}}}$, and a general vibrational state is represented as a linear combination of these basis states.
This allows one to project the Hamiltonian onto the discretized basis as
\begin{align}
\begin{split}
    \hat H_{\mathrm{disc}} &= \sum_{i=0}^{M-1}\sum_{p,q=0}^{N-1} h_{i,p,q} \ketbra*{\phi^{(i)}_{p_i}}{\phi^{(i)}_{q_i}} + \sum_{i<j=0}^{M-1}\ \sum_{\substack{p,q,\\r,s}=0}^{N-1} \mathrm{g}_{i,p,q,j,r,s} \ketbra*{\phi^{(i)}_{p_i}}{\phi^{(i)}_{q_i}} \otimes\ketbra*{\phi^{(j)}_{r_j}}{\phi^{(j)}_{s_j}}  \\
    &\  + \sum_{i<j<k=0}^{M-1}\ \sum_{\substack{p,q,r,\\s,t,u}=0}^{N-1} \mathrm{v}_{i,p,q,j,r,s,k,t,u} \ketbra*{\phi^{(i)}_{p_i}}{\phi^{(i)}_{q_i}} \otimes \ketbra*{\phi^{(j)}_{r_j}}{\phi^{(j)}_{s_j}} \otimes \ketbra*{\phi^{(k)}_{t_k}}{\phi^{(k)}_{u_k}}, 
\end{split} \label{Eq: discretized vibrational Hamiltonian}
\end{align}
where
\begin{align}
    \begin{split}
    h_{i,p,q} &= \int_{-\infty}^{\infty}\phi^{(i)}_{p_i}(\mathcal{Q}_i) \left( -\dfrac{\Omega_{i}}{2}\dfrac{\partial^2}{\partial \mathcal Q_i^2} + V^{(1)}_i(\mathcal Q_i)\right)\phi^{(i)}_{q_i}(\mathcal{Q}_i) \ d\mathcal{Q}_i, \\ 
    \mathrm{g}_{i,p,q,j,r,s}&= \int_{-\infty}^\infty \int_{-\infty}^{\infty}\phi^{(i)}_{p_i} (\mathcal{Q}_i)\ \phi^{(j)}_{r_j} (\mathcal{Q}_j)\ V^{(2)}_{i,j}(\mathcal Q_i, \mathcal Q_j)\ \phi^{(i)}_{q_i} (\mathcal{Q}_i)\ \phi^{(j)}_{s_j} (\mathcal{Q}_j)\ d\mathcal{Q}_i d\mathcal{Q}_j, \\
    \mathrm{v}_{i,p,q,j,r,s,k,t,u}&= \int_{-\infty}^\infty \int_{-\infty}^\infty \int_{-\infty}^{\infty}\phi^{(i)}_{p_i} (\mathcal{Q}_i)\ \phi^{(j)}_{r_j} (\mathcal{Q}_j) \phi^{(k)}_{t_k} (\mathcal{Q}_k)\ V^{(3)}_{i,j,k}(\mathcal Q_i, \mathcal Q_j, \mathcal Q_k)\ \phi^{(i)}_{q_i} (\mathcal{Q}_i)\ \phi^{(j)}_{s_j} (\mathcal{Q}_j) \phi^{(k)}_{u_k} (\mathcal{Q}_k)\ d\mathcal{Q}_i d\mathcal{Q}_j d\mathcal{Q}_k.
    \end{split} 
    \label{Eq: org integrals}
\end{align} 
We extend these tensors by setting their elements to zero outside the index ranges specified in Eq.~\eqref{Eq: discretized vibrational Hamiltonian}.
Note that we have restricted ourselves to three-mode couplings for the sake of brevity; the construction extends formally to higher-order couplings as well.
Since each vibrational mode occupies exactly one of the $N$ modals in any given basis state, and there are $M$ modes in total, the Hilbert space of the wavefunction has dimension $N^M$.
A binary encoding of the modal space requires only $M\lceil \log_2(N) \rceil$ qubits, but maps the Hamiltonian operators $\ketbra*{\phi^{(i)}_{p_i}}{\phi^{(i)}_{q_i}}$ to a sum of $\mathcal{O}(N)$ Pauli strings, each of length up to $\lceil \log_2(N) \rceil$.
Alternatively, a unary encoding requires $N$ qubits per mode, restricting physical states to a subspace where exactly one of those $N$ qubits is in the $\ket{1}$ state (indicating modal occupation).
We refer to this as the subspace with a single excitation per mode.
While this expands the total computational Hilbert space to $2^{NM}$, it drastically simplifies the operators $\ketbra*{\phi^{(i)}_{p_i}}{\phi^{(i)}_{q_i}}$ to at most four Pauli strings acting on just two qubits.
This unary approach naturally motivates a second-quantized representation.
By defining (hard-core bosonic) creation ($\hat a_{p_i}^\dagger$) and annihilation ($\hat a_{q_i}$) operators for the modals, we replace the operators $\ketbra*{\phi^{(i)}_{p_i}}{\phi^{(i)}_{q_i}}$ with the excitation operators $\hat E^{p_i}_{q_i} := \hat a_{p_i}^\dagger \hat a_{q_i}$.
The resulting Hamiltonian bears significant structural similarities to a second-quantized electronic Hamiltonian, allowing us to leverage existing quantum simulation tools.
Traditionally, one uses bosons instead of hard-core bosons, but this distinction can be ignored since their action on the physically meaningful subspace of single excitation per mode is identical: no physical process can take a modal's occupation number above $1$.
The second-quantized form of the Hamiltonian in Eq.~\eqref{Eq: discretized vibrational Hamiltonian} follows as
\begin{align}
\begin{split}
    \hat H &= \sum_{i=0}^{M-1}\sum_{p,q=0}^{N-1} h_{i,p,q} \hat E^{p_i}_{q_i}  + \sum_{i<j=0}^{M-1}\ \sum_{\substack{p,q,\\r,s}=0}^{N-1} \mathrm{g}_{i,p,q,j,r,s} \hat E^{p_i}_{q_i}  \hat E^{r_j}_{s_j} + \sum_{i<j<k=0}^{M-1}\ \sum_{\substack{p,q,r,\\s,t,u}=0}^{N-1} \mathrm{v}_{i,p,q,j,r,s,k,t,u} \hat E^{p_i}_{q_i}  \hat E^{r_j}_{s_j} \hat E^{t_k}_{u_k}.
\end{split} \label{Eq: second-quantized vibrational Hamiltonian}
\end{align}
This procedure for obtaining a second-quantized Hamiltonian can easily become computationally expensive as the number of modes and the mode-coupling level increase.
The problem stems from having to evaluate multi-dimensional integrals that require sampling the electronic PES at many points.
Sampling the surface on a predefined grid requires a large number of grid points to obtain a sufficiently accurate Hamiltonian representation.
The adaptive density-guided approach (ADGA) \cite{adga} is a method that determines the grid points adaptively and provides a Hamiltonian that can be expressed as an SOP of one-mode operators.
With this approach, a three-mode-coupling Hamiltonian is obtained as
\begin{align}
\begin{split}
       \hat H &= \sum_{i} \sum_{o^{i}} c^{i}_{o^{i}}\ \hat h^{i, o^{i}} + \sum_{i < j}\ \sum_{o^{i}, o^{j}} c^{i j}_{o^{i}o^{j}}\ \hat h^{i, o^{i}} \hat h^{j, o^{j}} + \sum_{i < j < k}\  \sum_{o^{i}, o^{j}, o^{k}} c^{i j k}_{o^{i}o^{j}o^{k}}\ \hat h^{i, o^{i}} \hat h^{j, o^{j}} \hat h^{k, o^{k}},
\end{split} \label{Eq: SOP Hamiltonian}
\end{align}
where the coefficients $c$ correspond to tensors of different order.
The operators $\hat h^{i,o^i}$ are one-mode operators; the indices $i$ and $o^i$ label the specific operator among the set of one-mode operators of mode $i$ used to fit the Hamiltonian.
They can be expressed as $\hat h^{i, o^{i}} = \sum_{p,q}h_{p,q}^{i,o^i} \hat E^{p_i}_{q_i}$.
By expanding these one-mode operators and contracting the tensor indices, we can recast the Hamiltonian of Eq.~\eqref{Eq: SOP Hamiltonian} into the form of Eq.~\eqref{Eq: second-quantized vibrational Hamiltonian}.
This involves approximating the associated integrals as
\begin{align}
 \begin{split}
 h_{i,p,q} &\approx \sum_{o^{i}} c^{i}_{o^{i}} h^{i,o^{i}}_{p,q},\\
    \mathrm g_{i,p,q,j,r,s} &\approx  \sum_{o^{i},o^{j}} c^{ij}_{o^{i}o^{j}} h^{i,o^{i}}_{p,q}h^{j,o^{j}}_{r,s}, \\
    \mathrm v_{i,p,q,j,r,s,k,t,u} &\approx  \sum_{o^{i},o^{j},o^{k}} {c}^{ijk}_{o^{i},o^{j},o^{k}} h^{i,o^{i}}_{p,q}h^{j,o^{j}}_{r,s}h^{k,o^{k}}_{t,u}.
\end{split} \label{Eq: ADGA to integrals}
\end{align}

The block-encoding procedures developed in this work apply, in general, to the standard second-quantized Hamiltonian of Eq.~\eqref{Eq: second-quantized vibrational Hamiltonian} and are independent of the exact method used to approximate the tensors.
However, to benchmark against the method developed in Ref.~\cite{majland2025fault}, which requires an SOP representation of the Hamiltonian, we utilize the integrals approximated as in Eq.~\eqref{Eq: ADGA to integrals}.

Although the second-quantization framework was derived starting from the Hamiltonian represented in terms of normal modes, one can apply a unitary mode-localization transformation to maximize the spatial locality of the displacement vectors, yielding new coordinates $\tilde{\mathcal{Q}}_i$.
This localization procedure has been shown to enable significantly faster convergence of the $n$-mode expansion \cite{mode_loc_1,mode_loc_2,mode_loc_3,mode_loc_4}.
Furthermore, for molecules exhibiting large-amplitude torsional motion, curvilinear coordinates may be required, while complete rovibrational Hamiltonians additionally include rotational and Coriolis couplings. 
More general curvilinear and rovibrational Hamiltonians can also be represented in a finite modal basis, but their detailed treatment is beyond the scope of this work.

\subsection{Energy Estimation via Quantum Phase Estimation}
Most fault-tolerant applications of energy estimation on a quantum computer rely on the QPE algorithm \cite{QPE, textbook_QPE} or its iterative variants \cite{iterative_QPE, RPE, LinTong}.
Given a unitary $\hat U$ with eigenpairs $\hat U\ket{\phi_j} = e^{2\pi \mathrm{i} \phi_j}\ket{\phi_j}$ (where $\phi_j\in[0,1]$), QPE applied to an input state $\ket{\psi}=\sum_k c_k \ket{\phi_k}$ effectively allows us to sample from the probability distribution
\begin{equation}
    P(\omega) = \sum_k |c_k|^2 \mathcal{L}(\omega-\phi_k).
\end{equation}
Here, $\mathcal{L}(\omega)$ is a lineshape function that typically converges to a Dirac delta function $\delta(\omega)$ as the algorithmic precision is increased.
While standard QPE yields a squared sinc function lineshape, convergence can be accelerated by using windowing techniques such as Kaiser lineshapes \cite{gqpe,qpe_lineshapes,dpss_prep}.
This technique can then be used for ground-state energy estimation or, with an appropriate initial state, for vibrational spectroscopy.
Typically, the unitary $\hat U$ is chosen as the time-evolution operator $e^{-\mathrm{i}\hat H \Delta t}$.
In this case, the time step $\Delta t$ and the total number of unitary queries directly dictate the overall accuracy of the QPE procedure through standard Fourier grid relationships \cite{loaiza2025NIR}.
An alternative approach is offered by qubitization \cite{qubitization}, which utilizes the linear combination of unitaries (LCU) query model \cite{LCU} to encode the arccosine of the scaled eigenvalues of the Hamiltonian in the eigenphases of the unitary $\hat U$.

A Hamiltonian can be decomposed as an LCU, $\hat H = \sum_{l=0}^{L-1}a_l \hat U_l$, where $\hat U_l$ is a unitary and the coefficients $a_l$ are strictly positive reals (any phases are absorbed into $\hat U_l$).
The induced 1-norm is then defined as $\lambda_{\mathrm{LCU}} = \sum_{l=0}^{L-1}a_l$.
Note that such an LCU expansion is not unique for a given Hamiltonian.
Given the operation PREPARE (PREP), defined as
\begin{align*}
    \mathrm{PREP}\ket{0} = \ket{\mathrm{PREP}} =  \dfrac{1}{\sqrt{\lambda_{\mathrm{LCU}}}}\sum_{l=0}^{L-1}\sqrt{a_l}\ket{l},
\end{align*}
and the operation SELECT (SEL), defined as
\begin{align*}
    \mathrm{SEL} = \sum_{l=0}^{L-1} \ketbra{l}{l} \otimes\hat U_l,
\end{align*}
it follows that
\begin{align*}
    \bra{0}\mathrm{PREP}^\dagger \cdot \mathrm{SEL}\cdot \mathrm{PREP}\ket{0} = \dfrac{\hat H}{\lambda_{\mathrm{LCU}}}.
\end{align*}
In the matrix representation of the operator $\mathrm{PREP}^\dagger \cdot \mathrm{SEL}\cdot \mathrm{PREP}$, this corresponds to encoding the operator $\hat H/\lambda_{\mathrm{LCU}}$ in its top-left corner, which is known as a block encoding.
By construction, $\mathrm{PREP}$ only acts on ancilla qubits whereas $\mathrm{SEL}$ acts on both ancilla and system qubits.
This embedding of the Hamiltonian into a larger unitary is necessary because Hamiltonians are generally non-unitary and hence cannot be directly implemented as quantum gates.
An associated (Szegedy) walk operator can be constructed from the Hamiltonian block encoding as
\begin{align}
    \hat W = (2\ketbra{\mathrm{PREP}}{\mathrm{PREP}} \otimes\hat{\mathbb{I}} - \hat{\mathbb{I}}_{\mathrm{F}}) \cdot \mathrm{SEL}, \label{Eq: Walk operator}
\end{align}
where $\hat{\mathbb{I}}$ and $\hat{\mathbb{I}}_{\mathrm{F}}$ denote the identity operators on the system register and the full ancilla–system Hilbert space, respectively. 
When the unitaries $\hat U_l$ in the Hamiltonian LCU are also Hermitian, this walk operator admits a two-dimensional irreducible representation in the subspace spanned by $\ket{\mathrm{PREP}}\ket{E}$ and a state orthogonal to it \cite{linearT}, with eigenvalues $e^{\pm \mathrm{i}\arccos(E/\lambda_{\mathrm{LCU}})}$, where $\hat H\ket{E}=E\ket{E}$.
Thus, we can directly perform QPE on $\hat W$ in order to recover our target property, with the only difference being a need to post-process the measured phase $\phi$ to recover the associated energy as $E=\lambda_{\mathrm{LCU}}\cos(2\pi\phi)$.
The cost of constructing the walk operator equals the sum of the costs of performing $\mathrm{PREP}$, $\mathrm{SEL}$, and $\mathrm{PREP}^\dagger$, denoted as $C^{\mathrm{PREP}}_{\mathrm{LCU}},\ C^{\mathrm{SEL}}_{\mathrm{LCU}},$ and $C^{\mathrm{PREP}^\dagger}_{\mathrm{LCU}}$, and the cost of the reflection on the encoding qubits in Eq.~\eqref{Eq: Walk operator}.
To estimate the energy to an accuracy of $\epsilon$, the walk operator needs to be applied $\lceil \pi \lambda_{\mathrm{LCU}}/2\epsilon \rceil$ times \cite{thc}.
Hence, assuming access to the exact eigenstate, the total gate cost of estimating an eigenvalue to accuracy $\epsilon$ using qubitization is given by \cite{thc}
\begin{align}
    C_{\mathrm{LCU}} = \left\lceil \dfrac{\pi \lambda_{\mathrm{LCU}}}{2 \epsilon} \right\rceil \left(C^{\mathrm{BE}}_{\mathrm{LCU}} + \log(L) + \mathcal{O}(1)\right), \label{Eq: QPE general cost}
\end{align}
where $C^{\mathrm{BE}}_{\mathrm{LCU}} = C^{\mathrm{PREP}}_{\mathrm{LCU}} + C^{\mathrm{SEL}}_{\mathrm{LCU}} + C^{\mathrm{PREP}^\dagger}_{\mathrm{LCU}}$ is the cost of block-encoding the Hamiltonian and the subscript $\mathrm{LCU}$ denotes that these costs depend on the choice of the LCU decomposition utilized.
In this work, we leverage different LCU decompositions of the vibrational Hamiltonian to build efficient circuits for qubitization.
For fault-tolerant architectures based on the surface code and the universal Clifford+T gate set, executing T gates is the most resource-intensive operation.
Because the circuits developed in this work are heavily dominated by Toffoli gates, each of which can be realized using four T gates assuming access to an ancilla qubit and feedforward operations \cite{t_to_toff}, we report our non-Clifford costs in terms of Toffoli count.
For comparison with the Trotter-approximation results, we convert the Toffoli counts to the corresponding T-gate counts.

\section{LCU Decompositions of Vibrational Hamiltonians} \label{sec: LCUs}
Many methods for decomposing a Hamiltonian into an LCU rely on factorizing the constituent Hamiltonian tensors.
The Triangular LCU proposed in Ref.~\cite{majland2025fault} utilizes the Hamiltonian representation of Eq.~\eqref{Eq: SOP Hamiltonian} and factorizes the $c^{ij}_{o^io^j}$ and $c^{ijk}_{o^io^jo^k}$ tensors for every set of mode indices.
The Hamiltonian is then block-encoded by forming SOPs of one-mode operator block encodings.
We revisit this procedure in Appendix~\ref{App: Triangular LCU}, where we also discuss additional algorithmic modifications.
Specifically, we derive the Triangular LCU starting from the Hamiltonian in Eq.~\eqref{Eq: second-quantized vibrational Hamiltonian} rather than Eq.~\eqref{Eq: SOP Hamiltonian} because this was observed to yield slightly lower 1-norms, and we utilize alternative three-mode-coupling tensor factorization routines to simplify the optimization problem. 

In this section, we derive two new LCUs inspired by electronic Hamiltonian block-encoding techniques that result in smaller 1-norms and more efficient block-encoding circuits.
In Ref.~\cite{majland2025fault}, another LCU called the Diagonal LCU was also proposed, which relied on diagonalizing each one-mode operator in the Hamiltonian.
That work concluded that the Diagonal LCU required substantially more resources than the Triangular LCU owing to the need to apply modal rotations every time a one-mode operator is implemented.
While this observation is correct in that context, we show that by limiting the architecture to at most $n$ basis transformations for a Hamiltonian of mode-coupling level $n$, one-mode diagonalization can indeed be used to achieve more efficient block encodings.

We start by rewriting the Hamiltonian of Eq.~\eqref{Eq: second-quantized vibrational Hamiltonian} as
\begin{align}
    \hat H &= \sum_{i=0}^{M-1}\sum_{p,q=0}^{N-1} h_{i,p,q} \hat E^{p_i}_{q_i}  + \sum_{i,j=0}^{M-1}\ \sum_{\substack{p,q,\\r,s}=0}^{N-1} g_{i,p,q,j,r,s} \hat E^{p_i}_{q_i}  \hat E^{r_j}_{s_j} + \sum_{i,j,k=0}^{M-1}\ \sum_{\substack{p,q,r,\\s,t,u}=0}^{N-1} v_{i,p,q,j,r,s,k,t,u} \hat E^{p_i}_{q_i}  \hat E^{r_j}_{s_j} \hat E^{t_k}_{u_k},
    \label{Eq: second-quantized symmetrized vibrational Hamiltonian}
\end{align}
where we have relaxed the restriction on the sums over the mode indices and introduced new tensors $g$ and $v$, which correspond to the symmetrized versions of the $\mathrm g$ and $\mathrm v$ tensors, respectively.
The symmetrization is done as follows:
\begin{align*}
    g_{i,p,q,j,r,s} &= (\mathrm g_{i,p,q,j,r,s} + \mathrm g_{j,r,s,i,p,q})/2, \\
    v_{i,p,q,j,r,s,k,t,u} &= \big(\mathrm v_{i,p,q,j,r,s,k,t,u} + \mathrm v_{k,t,u,i,p,q,j,r,s} + \mathrm  v_{j,r,s,k,t,u,i,p,q} \\
    &\quad  +  \mathrm  v_{j,r,s,i,p,q,k,t,u} + \mathrm
    v_{k,t,u,j,r,s,i,p,q} + \mathrm  v_{i,p,q,k,t,u,j,r,s}\big)/6,
\end{align*}
and the tensors are zero whenever any two mode indices are identical.
This definition, along with the fact that the excitation operators on different modes commute, allows for rewriting the Hamiltonian of Eq.~\eqref{Eq: second-quantized vibrational Hamiltonian} as given in Eq.~\eqref{Eq: second-quantized symmetrized vibrational Hamiltonian}.
Notice that the $g$ and $v$ tensors are invariant under the exchange of modal indices within a mode, as well as under swapping a mode-modal composite index $(i,p,q)$ with another set $(j,r,s)$.
For example, $g$ satisfies the eightfold symmetry
\begin{align*}
    g_{i,p,q,j,r,s} = g_{i,q,p,j,r,s} = g_{i,p,q,j,s,r} = g_{i,q,p,j,s,r} = g_{j,r,s,i,p,q} = g_{j,r,s,i,q,p} = g_{j,s,r,i,p,q} = g_{j,s,r,i,q,p}.
\end{align*}
Similarly, $v$ satisfies a 48-fold symmetry.
These symmetries enable easy manipulation of the tensors.
After arriving at an LCU, we will return to restricting the sums over mode indices to remove the redundancy in the LCU terms.

\subsection{Pauli LCU}
A straightforward LCU can be obtained by expressing the excitation operators, $\hat E^{p_i}_{q_i} = \hat a_{p_i}^{\dagger} \hat a_{q_i}$, in terms of Pauli operators, analogous to the Pauli LCU for electronic Hamiltonians \cite{sparse_LCU}.
Since these creation and annihilation operators are equivalent to Pauli spin raising/lowering operators, we have
\begin{align*}
    \hat a_{p_i}^\dagger = \dfrac{\hat X_{p_i} - \mathrm i \hat Y_{p_i}}{2}, \ \ \ \ \ \hat a_{p_i} = \dfrac{\hat X_{p_i} + \mathrm i \hat Y_{p_i}}{2},
\end{align*}
where $\hat X_{p_i}$ and $\hat Y_{p_i}$ are Pauli $X$ and $Y$ operators acting on the qubit corresponding to modal $p$ of mode $i$, and acting as the identity everywhere else.
The symmetries of the Hamiltonian tensors allow us to rewrite the Hamiltonian in Eq.~\eqref{Eq: second-quantized symmetrized vibrational Hamiltonian} as an SOP of symmetric combinations, $\hat E^{p_i}_{q_i} + \hat E^{q_i}_{p_i}$.
We provide the exact expression and more details in Appendix~\ref{App: Deriving Pauli LCU}.
Expanding this symmetric term using Pauli operators yields
\begin{align}
    \hat E^{p_i}_{q_i} + \hat E^{q_i}_{p_i} = \hat{\mathbb{I}} \delta_{p, q} + \dfrac{(-1)^{\delta_{p,q}}}{2}\sum_{b_1=0}^1 \hat Q_{i,p,q}^{(b_1)}, \label{Eq: excitation pair as a Pauli LCU}
\end{align}
where we have defined the operators
\begin{align}
\begin{split}
    \hat Q_{i,p,q}^{(0)} = 
   \begin{cases}
   \hat X_{p_i}\hat X_{q_i} & p\neq q \\
   \hat Z_{p_i}  & p = q
   \end{cases},
   \quad \quad \quad 
    \hat Q_{i,p,q}^{(1)} = 
   \begin{cases}
   \hat Y_{p_i}\hat Y_{q_i} & p\neq q \\
   \hat Z_{p_i}  & p = q
   \end{cases}.  
\end{split}
   \label{eq: Pauli Q_ipq}
\end{align}
Because the operators $\hat Q_{i,p,q}^{(b_1)}$ are unitary, Eq.~\eqref{Eq: excitation pair as a Pauli LCU} constitutes an LCU decomposition of $\hat E^{p_i}_{q_i} + \hat E^{q_i}_{p_i}$.
Substituting this into the Hamiltonian, which is expanded as an SOP of these excitation operators, yields an LCU of the full Hamiltonian.
Note, however, that when forming products such as $(\hat E^{p_i}_{q_i} + \hat E^{q_i}_{p_i}) (\hat E^{r_j}_{s_j} + \hat E^{s_j}_{r_j})$, the identity term in Eq.~\eqref{Eq: excitation pair as a Pauli LCU} generates terms that contain fewer than two $\hat Q_{i,p,q}^{(b_1)}$ operators.
Consequently, the two-mode-coupling terms in the Hamiltonian generate new one-mode-coupling and constant terms in the LCU.
Similarly, the three-mode couplings lead to new two-mode, one-mode, and constant terms.
Because of this, the resulting coefficients of the unitaries no longer correspond to the original tensors in Eq.~\eqref{Eq: second-quantized symmetrized vibrational Hamiltonian}, but rather to new tensors formed by partial resummations.
The final Pauli LCU of the Hamiltonian takes the form
\begin{align}
    \begin{split}
    \hat H_{\mathrm{Pauli}} &= \bm {\mathrm C}_{\mathrm{Pauli}} + \dfrac{1}{2}\sum_{i=0}^{M-1} \sum_{p\leq q=0}^{N-1}\sum_{b_1=0}^1 \bm{\mathrm{h}}_{i,p,q} \hat Q_{i,p,q}^{(b_1)} + \dfrac{1}{4}\sum_{i<j=0}^{M-1}\sum_{\substack{p \leq q,\\r \leq s}=0}^{N-1} \sum_{b_1, b_2=0}^1 \bm{\mathrm{g}}_{i,p,q,j,r,s} \hat Q_{i,p,q}^{(b_1)} \hat Q_{j,r,s}^{(b_2)} \\ 
    &\quad+ \dfrac{1}{8}\sum_{i<j<k=0}^{M-1}\sum_{\substack{p\leq q,\\ r\leq s,\ t\leq u}=0}^{N-1} \sum_{b_1, b_2, b_3=0}^1 \bm{\mathrm{v}}_{i,p,q,j,r,s,k,t,u} \hat Q_{i,p,q}^{(b_1)} \hat Q_{j,r,s}^{(b_2)} \hat Q_{k,t,u}^{(b_3)},
    \end{split} \label{Eq: Pauli LCU}
\end{align}
where the new coefficients $\bm{\mathrm{C}}_{\mathrm{Pauli}}$, $\bm{\mathrm{h}}$, $\bm{\mathrm{g}}$, and $\bm{\mathrm{v}}$ are defined in Appendix~\ref{App: Deriving Pauli LCU}.
The mode index sums have now been restored to the canonical restricted domains $i<j$ and $i<j<k$ as in Eq.~\eqref{Eq: discretized vibrational Hamiltonian}.
Because $\bm {\mathrm C}_{\mathrm{Pauli}}$ is a known global energy shift, we block-encode $\hat H_{\mathrm{Pauli}} - \bm {\mathrm C}_{\mathrm{Pauli}}$ and add $\bm {\mathrm C}_{\mathrm{Pauli}}$ to the energy returned by phase estimation.
In practice, one can prune tensor elements below a specified cutoff, increasing the sparsity of the encoded Hamiltonian without a significant loss in accuracy.
We verify this in Section~\ref{sec: results} for small systems where the ground-state energy can be obtained classically.
The resulting LCU 1-norm is given by
\begin{align}
    \lambda_{\mathrm{Pauli}} = \sum_{i=0}^{M-1} \sum_{p\leq q=0}^{N-1} |\bm {\mathrm h}_{i,p,q}| + \sum_{i<j=0}^{M-1}\sum_{\substack{p \leq q, \\ r \leq s}=0}^{N-1} |\bm {\mathrm g}_{i,p,q,j,r,s}| + \sum_{i<j<k=0}^{M-1}\sum_{\substack{p \leq q, \\ r \leq s,\ t \leq u}=0}^{N-1} |{\bm{\mathrm v}}_{i,p,q,j,r,s,k,t,u} |. \label{Eq: Pauli LCU 1-norm}
\end{align}
The exclusion of the constant term leads to a noticeable reduction in the 1-norm.

\subsection{THC LCU}
Tensor hypercontraction (THC) is primarily used in electronic structure theory for compressing the electron repulsion integral tensor from $\mathcal{O}(N^4)$ terms to $\mathcal{O}(N^2)$ terms \cite{org_thc}.
Reference~\cite{thc} proposed a technique to exploit THC to block-encode the electronic Hamiltonian efficiently.
The vibrational Hamiltonian in Eq.~\eqref{Eq: second-quantized symmetrized vibrational Hamiltonian} closely resembles the second-quantized electronic structure Hamiltonian. 
Moreover, its excitation operators $\hat E^{p_i}_{q_i}$ have Lie algebraic properties similar to those of their electronic counterparts \cite{malpathak2025trotter}. 
Motivated by these similarities, we extend THC to vibrational Hamiltonians, including tensors beyond two-mode coupling.
In the following, all identities involving the excitation operators and modal rotations are understood in terms of their action on the physical subspace identified by having exactly one excitation per mode. 

THC relies on factorizing the two- and three-mode-coupling tensors in the vibrational Hamiltonian in a very specific fashion.
We begin with an approximate factorization of the three-mode-coupling tensor $v$ in Eq.~\eqref{Eq: second-quantized symmetrized vibrational Hamiltonian} as
\begin{align}
    v_{i,p,q,j,r,s,k,t,u}^{\mathrm{THC}} &= \sum_{\mu_i = 0}^{R_i-1} \sum_{\nu_j = 0}^{R_j-1} \sum_{\omega_k = 0}^{R_k-1}  \Gamma^{(i,j,k)}_{\mu_i,\nu_j,\omega_k}\ \chi^{(i)}_{p,\mu_i} \chi^{(i)}_{q,\mu_i} \ \chi^{(j)}_{r,\nu_j} \chi^{(j)}_{s,\nu_j} \ \chi^{(k)}_{t,\omega_k} \chi^{(k)}_{u,\omega_k}.\label{Eq: trbt THC factorization}
\end{align}
The factorization is done through a nonlinear optimization:
\begin{mini}
  {\Gamma, \chi}{{\norm{v - v^{\mathrm{THC}}(\Gamma,\chi)}_\mathrm{F}}}           % Optimization variable and objective
  {\label{eq: THC three mode optimization problem}}{}  % Label and optional result
  \addConstraint{\sum_{p=0}^{N-1}|\chi^{(i)}_{p,\mu_i}|^2}{= 1\ \ \forall\ i,\mu_i}
  \addConstraint{\Gamma^{(i,j,k)}_{\mu_i,\nu_j,\omega_k}}{= \Gamma^{(k,i,j)}_{\omega_k, \mu_i, \nu_j}}{= \Gamma^{(j,k,i)}_{\nu_j,\omega_k, \mu_i}}{= \Gamma^{(i,k,j)}_{\mu_i,\omega_k, \nu_j}}{= \Gamma^{(j,i,k)}_{\nu_j,\mu_i, \omega_k}}{= \Gamma^{(k,j,i)}_{\omega_k,\nu_j,\mu_i}}.
\end{mini}
Here $\norm{\cdot}_\mathrm{F}$ denotes the Frobenius norm and all variables are assumed to be real.
The first constraint is required so that each column of $\chi^{(i)}$ can be used as a modal rotation vector, as discussed below.
The second constraint is needed to ensure that the symmetries of the original tensor $v$ are respected by the THC-approximated tensor $v^{\mathrm{THC}}$.
$R_i$ stands for the THC rank for mode $i$.
By choosing a mode-dependent THC rank, the representation can be made more compact as some modes, which are weakly coupled to the rest, may require a smaller rank. 

Substituting the expression for $v^{\mathrm{THC}}$ into Eq.~\eqref{Eq: second-quantized symmetrized vibrational Hamiltonian}, we can write the three-mode-coupling term as an SOP of factors of the form $\sum_{p,q=0}^{N-1} \chi^{(i)}_{p,\mu_i} \chi^{(i)}_{q,\mu_i} \hat E^{p_i}_{q_i}$.
This is facilitated by the decoupling of modal indices in the THC factorization, which is one of its key benefits.
Since $\chi^{(i)}$ has normalized columns by construction, for every column $\mu_i$ of $\chi^{(i)}$, we can define a number-conserving modal rotation unitary $\hat V^{(i)}_{\mu_i}$ whose action in the single-excitation subspace maps modal $0_i$ onto the normalized modal vector specified by that column of $\chi^{(i)}$.
These modal rotations conserve the excitation number of each mode, leaving the physical subspace invariant \cite{malpathak2025trotter}. The result is the relation
\begin{align}
\begin{split}
\sum_{p,q=0}^{N-1} \chi^{(i)}_{p,\mu_i} \chi^{(i)}_{q,\mu_i} \hat a_{p_i}^\dagger \hat a_{q_i} &= \hat V^{(i)}_{\mu_i} \hat{ \tilde{n}}_{0_i}\hat V^{(i)^\dagger}_{\mu_i} \\
&= \dfrac{1}{2}\left(\hat{\mathbb{I}} - \hat V^{(i)}_{\mu_i} \hat Z_{0_i} \hat V^{(i)^\dagger}_{\mu_i} \right),
\end{split} \label{Eq: modal rotation of number operator}
\end{align}
where $\hat{\tilde{n}}_{0_i} = \hat{\tilde{a}}_{0_i}^\dagger \hat{\tilde{a}}_{0_i} = (\hat{\mathbb{I}} - \hat Z_{0_i})/2$ is the number operator on modal $0$ of mode $i$ in the rotated basis. 
A derivation constructing the unitary $\hat V^{(i)}_{\mu_i}$ and expressing it in terms of two-qubit rotations is presented in Appendix C of Ref.~\cite{majland2025fault}. 
Specifically, it is shown that each unitary $\hat V_{\mu_i}^{(i)}$ can be decomposed into $N-1$ pairs of two-qubit rotations of the form $\exp(\mathrm{i}\theta_{p_i,p'_i}^{\mu_i}\hat X_{p_i}\hat Y_{p'_i})\exp(-\mathrm{i}\theta_{p_i,p'_i}^{\mu_i}\hat Y_{p_i}\hat X_{p'_i})$, where $p_i$ and $p'_i$ label neighboring qubits of mode $i$, and the rotation angles $\theta_{p_i,p'_i}^{\mu_i}$ are determined by the corresponding column of $\chi^{(i)}$.

Using Eq.~\eqref{Eq: modal rotation of number operator} after substituting the THC factorization of Eq.~\eqref{Eq: trbt THC factorization} into the Hamiltonian, it follows that the three-mode-coupling term of the Hamiltonian can be written as an SOP of terms of the form $(\hat{\mathbb{I}}-\hat V^{(i)}_{\mu_i}\hat Z_{0_i}
\hat V^{(i)\dagger}_{\mu_i})$.
Due to the identity operator in each factor of the SOP, expanding the products yields terms of all mode-coupling levels up to three.
Thus, in obtaining an LCU for a three-mode-coupling term, corrections to lower mode-coupling terms need to be introduced, which are incorporated by modifying the corresponding mode-coupling tensors.
The explicit forms of the corrections are derived in Appendix~\ref{App: Deriving THC LCU}.
Following this, the new two-mode-coupling tensor is also approximated using a THC factorization.
The factorized tensor is given as
\begin{align}
    \tilde{g}_{i,p,q,j,r,s}^{\mathrm{THC}} &= \sum_{\mu_i = 0}^{R_i-1}\sum_{\nu_j = 0}^{R_j-1} \zeta^{(i,j)}_{\mu_i,\nu_j}\ \xi^{(i)}_{p,\mu_i} \xi^{(i)}_{q,\mu_i}\ \xi^{(j)}_{r,\nu_j} \xi^{(j)}_{s,\nu_j}, \label{Eq: tbt THC factorization}
\end{align}
where the modal rotation vectors $\xi^{(i)}_{\mu_i}$ satisfy the normalization condition $\sum_{p}\vert{}\xi^{(i)}_{p,\mu_i}\vert{}^2 = 1$, and we ensure $\zeta^{(i,j)}_{\mu_i,\nu_j}=\zeta^{(j,i)}_{\nu_j,\mu_i}$.
By the same reasoning as above, writing the two-mode-coupling term of the Hamiltonian as an LCU leads to further corrections to the one-mode tensor.
These corrections from higher-order terms are completely analogous to the one-electron correction arising from the two-electron term in LCUs of the electronic structure Hamiltonian \cite{thc,loaiza2025majorana}.
The new one-mode-coupling tensor can be diagonalized for each mode $i$, with the corresponding eigenvectors used to construct modal rotation unitaries and the eigenvalues serving as the coefficients of Pauli $Z$ operators.
Details of all these steps are explicitly provided in Appendix~\ref{App: Deriving THC LCU}.
Finally, the THC LCU can be written as
\begin{align}
\begin{split}
    \hat H_{\mathrm{THC}} &=  \bm{\mathrm{C}}_{\mathrm{THC}} + \dfrac{1}{2}\sum_{i=0}^{M-1}\sum_{\mu_i=0}^{N-1} \bm \eta^{(i)}_{\mu_i} \hat W^{(i)}_{\mu_i} \hat Z_{0_i} \hat W^{(i)^\dagger}_{\mu_i}  + \dfrac{1}{4}\sum_{i<j=0}^{M-1}\sum_{(\mu_i, \nu_j)= 0}^{(R_i-1, R_j-1)} \bm{\zeta}_{\mu_i,\nu_j}^{(i,j)}\ \hat U^{(i)}_{\mu_i} \hat Z_{0_i}\hat U^{(i)^\dagger}_{\mu_i} \hat U^{(j)}_{\nu_j} \hat Z_{0_j}\hat U^{(j)^\dagger}_{\nu_j}\\
    &~+ \dfrac{1}{8}\sum_{i<j<k=0}^{M-1} \sum_{(\mu_i, \nu_j , \omega_k)= 0}^{(R_i-1, R_j-1, R_k-1)}  \bm{\Gamma}_{\mu_i,\nu_j,\omega_k}^{(i,j, k)}\ \hat V^{(i)}_{\mu_i} \hat Z_{0_i}\hat V^{(i)^\dagger}_{\mu_i} \hat V^{(j)}_{\nu_j} \hat Z_{0_j}\hat V^{(j)^\dagger}_{\nu_j} \hat V^{(k)}_{\omega_k} \hat Z_{0_k} \hat V^{(k)^\dagger}_{\omega_k}, \label{Eq: THC LCU}
\end{split}
\end{align}
where $\mathrm{\bm{C}}_{\mathrm{THC}}$ is a constant resulting from summing the identities in all higher mode-coupling terms. 
The coefficient tensors $\bm{\eta},\ \bm{\zeta}, \text{ and } \bm{\Gamma}$ and the associated modal rotation unitaries $\hat{W},\ \hat{U}, \text{ and } \hat{V}$ arise from diagonalizing/factorizing one-, two-, and three-mode-coupling tensors, respectively. 
As in the case of the Pauli LCU, we block-encode $\hat H_{\mathrm{THC}}-\bm{\mathrm{C}}_{\mathrm{THC}}$ and add $\bm{\mathrm{C}}_{\mathrm{THC}}$ to the eigenvalues obtained from the QPE workflow.
The associated LCU 1-norm is given by
\begin{align}
    \lambda_{\mathrm{THC}} = \dfrac{1}{2}\sum_{i=0}^{M-1}\sum_{\mu_i=0}^{N-1} |\bm \eta^{(i)}_{\mu_i}|  + \dfrac{1}{4}\sum_{i<j=0}^{M-1}\sum_{(\mu_i, \nu_j)= 0}^{(R_i-1, R_j-1)} |\bm \zeta_{\mu_i,\nu_j}^{(i,j)}|
    + \dfrac{1}{8}\sum_{i<j<k=0}^{M-1} \sum_{(\mu_i, \nu_j , \omega_k)= 0}^{(R_i-1, R_j-1, R_k-1)} |\bm{ \Gamma}_{\mu_i,\nu_j,\omega_k}^{(i,j, k)}|. \label{Eq: THC 1-norm}
\end{align}
Compared to the Pauli LCU of Eq.~\eqref{Eq: Pauli LCU}, which has two modal indices per mode index, the LCU of Eq.~\eqref{Eq: THC LCU} requires only one index per mode.
We observed that the average THC rank was less than $2N$ for the systems benchmarked in this work, implying that THC provides a substantially more compact representation of the Hamiltonian tensors.
It has also been observed in electronic THC that the ranks are typically between two and five times the number of orbitals \cite{blissthc}.
As such, here we assume the average THC rank to scale as $\mathcal{O}(N)$.
The 1-norm of the THC LCU can be further reduced by using a 1-norm-weighted regularization in the tensor factorization \cite{blissthc}.
Since the number of THC parameters to be optimized grows rapidly with the number of modes, it is often convenient to start with an efficiently obtainable initial guess for the THC factorization.
For the two- and three-mode-coupling coefficient tensors, we first perform a third-order CANDECOMP/PARAFAC (CP3) decomposition \cite{CP3} to obtain an initial guess.
This method is highly efficient as it decouples the search for an initial guess in the mode degrees of freedom, allowing parameters for different modes to be obtained in parallel.
We provide details of the procedure for obtaining these initial guesses in Appendix~\ref{App: Initializing THC}.

Lastly, we note that we use different modal rotation vectors $\xi$ and $\chi$ for two- and three-mode-coupling terms in the Hamiltonian.
Using the same modal rotation vectors for both terms can potentially simplify the circuit implementation.
However, due to the computational difficulty of obtaining such a factorization, we factorize these terms separately.

\subsection{Block-Invariant Symmetry Shift (BLISS)}
Block-encoding a Hermitian operator within a unitary requires rescaling its eigenvalues, and the cost of algorithms utilizing such a block-encoding oracle scales linearly with the rescaling factor.
When using an LCU-based approach, this rescaling factor is dictated by the LCU 1-norm.
Reference~\cite{symmetry_shift} proposed a method based on Hamiltonian symmetries to reduce this 1-norm, which was later generalized to the Block-Invariant Symmetry Shift (BLISS) technique \cite{bliss, lpbliss}.
Although originally proposed for electronic Hamiltonians, the concept applies generally to any block encoding where only the action of the operator on a specific symmetry sector is relevant.
In the second-quantized vibrational framework, one such symmetry sector is naturally identified by noting that physically meaningful vibrational states occupy exactly one modal per mode.
Accordingly, $(\hat{\mathrm{N}}_i-1)\ket{\Psi} = 0$ for any physically meaningful vibrational wavefunction $\ket{\Psi}$, where $\hat {\mathrm N}_i = \sum_{p=0}^{N-1} \hat E^{p_i}_{p_i}$ is the number operator counting excitations in mode $i$.
The BLISS technique modifies the Hamiltonian as $ \hat H  \to \hat H - \hat K$ by introducing a parameterized ``killer'' operator $\hat K$.
This operator is constructed to annihilate states within the target symmetry subspace (i.e., $\hat K\ket{\Psi}=0$) while acting non-trivially on states in other symmetry subspaces.
One can then optimize the parameters within the killer operator to minimize the overall LCU 1-norm of the modified Hamiltonian.
Thus, for the second-quantized vibrational Hamiltonian, $\hat{\mathrm{N}}_i-1$ acts as a killer operator that systematically modifies the one-mode-coupling tensor.
A more general one-mode killer takes the form
\begin{align*}
\hat{K_1} &= \sum_{i=0}^{M-1} \mathcal{A}^{(i)} (\hat{\mathrm N}_i - 1),
\end{align*}
where the $\mathcal{A}^{(i)}$ are scalar parameters.
Similarly, higher-order killers can be constructed by forming weighted sums of operators such as $\hat{\mathrm{N}}_i\hat{\mathrm{N}}_j-1$ and $\hat{\mathrm{N}}_i\hat{\mathrm{N}}_j\hat{\mathrm{N}}_k-1$, optionally multiplied by excitation operators.
For example, a sum of terms like $\hat E^{t_k}_{u_k}(\hat{\mathrm{N}}_i\hat{\mathrm{N}}_j-1)$ forms a three-mode-coupling killer. 
Hermiticity of the resulting Hamiltonian imposes constraints on the coefficients multiplying the killer operators.
The explicit forms of the killer operators employed in this work and the corresponding constraints on their coefficients are provided in Appendix~\ref{app: BLISS details}. 

For the Pauli LCU, BLISS is incorporated by adding killers up to mode-coupling level $n$ for an $n$-mode-coupling Hamiltonian.
This modifies the underlying Hamiltonian tensors, producing a parameterized Pauli LCU whose 1-norm is subsequently minimized.
We refer to this optimized representation as the BLISS-Pauli LCU.
Care must be taken when adding killers: minimizing the 1-norm can sometimes inadvertently decrease the sparsity of the Hamiltonian tensors, potentially increasing the overall quantum resource requirements.
To avoid this, we restrict the killer operators considered in this work as detailed in Appendix~\ref{app: BLISS details}.

Applying BLISS to the THC framework (BLISS-THC) is complicated by the nonlinear optimization required to obtain the THC factors.
We observed that performing BLISS prior to THC factorization yields tensors that require substantially higher ranks to achieve the same accuracy.
Simultaneous optimization of the BLISS and THC parameters, an approach used for electronic Hamiltonians \cite{blissthc}, also suffered from the same issue.
Consequently, we restrict the BLISS-THC optimization to the simplest killer operator: the one-mode-coupling shift.
In this work, the BLISS-THC protocol consists of first deriving the standard THC LCU, followed by optimizing the one-mode killer coefficients to minimize $\lambda_{\mathrm{THC}}$.
Further details are provided in Appendix~\ref{app: BLISS details}. 

Because BLISS primarily alters the 1-norm without changing the algebraic structure of the Hamiltonian, the block-encoding circuit remains conceptually unchanged.
Provided the sparsity of the Hamiltonian is not significantly reduced, the required qubit count is also preserved.

\section{Resource Estimation} \label{sec: resource estimates}
Using qubitization for quantum phase estimation with an LCU query model leads to the total Toffoli cost given in Eq.~\eqref{Eq: QPE general cost}.
In this section, we describe the block-encoding procedures for the Triangular, Pauli, and THC LCUs at a high level and identify the dominant contributions to their computational cost.
The resulting asymptotic costs are summarized in Table~\ref{tab: asymptotic LCU resources}.
Throughout this section, when comparing the asymptotic scaling with respect to $M$ and $N$, the mode-coupling level $n$, LCU coefficient precision, and rotation-angle precision are held fixed. 
We use $\widetilde{\mathcal O}$ to suppress polylogarithmic factors in $M$ and $N$.
Detailed quantum circuits together with the corresponding Toffoli and logical qubit costs for energy estimation using these LCUs are provided in Appendices~\ref{App: Triangular LCU}, \ref{App: Pauli LCU}, and \ref{App: THC LCU}.
The circuits used to block-encode the Pauli and THC LCUs of the vibrational Hamiltonian closely follow the methods presented in Ref.~\cite{thc}.

In Ref.~\cite{majland2025fault}, the Triangular LCU was block-encoded by constructing an SOP of block encodings of individual one-mode operators.
Each one-mode operator was block-encoded by first loading up to $N(N+1)/2$ unique coefficients and then applying the corresponding Pauli operators.
This results in a Toffoli cost of $\mathcal{O}(N^2)$ per one-mode operator.
Since the Triangular LCU factorizes the tensors $c^{ij}_{o^io^j}$ and $c^{ijk}_{o^io^jo^k}$ for every choice of mode indices, the number of products of one-mode operators for a fixed set of mode indices depends on the average factorization rank $\bar{\kappa}$. 
The scaling of $\bar{\kappa}$ with $N$ is generally problem dependent, and we therefore retain it explicitly in the asymptotic resource estimates.
For a Hamiltonian with fixed mode-coupling level $n$, the number of possible mode index combinations scales as $\mathcal{O}(M^{n})$.
Therefore, the total number of one-mode operators scales as $\mathcal{O}\left(\bar{\kappa}M^{n}\right)$.
The logical qubit cost is dominated by the system qubits, since forming sums and products of block encodings requires only a small number of ancillas.
The resulting asymptotic block-encoding cost is
\begin{align}
\mathrm{T}_{\Delta}^{\mathrm{BE}}
    = \mathcal{O}\left(\bar{\kappa}M^{n}N^2\right), \quad \quad
\mathrm{L}_{\Delta}^{\mathrm{BE}}
    = \mathcal{O}(MN),
\label{Eq: Triangular LCU asymptotic BE cost}
\end{align}
where $\mathrm{T}$ and $\mathrm{L}$ denote the Toffoli and logical qubit costs, respectively.
A detailed derivation is given in Appendix~\ref{App: Triangular LCU}.

Unlike the Triangular LCU, the Pauli and THC block encodings do not require repeated applications of the PREP and SELECT primitives.
Instead, all coefficients are loaded once at the beginning, followed by a more expensive SELECT operation. 
Loading all coefficients at once produces a larger data table, which benefits more from advanced QROM techniques that reduce the leading dependence of the Toffoli cost on the table size from linear to square-root, up to polylogarithmic factors, in exchange for additional qubits. 
Such techniques include the Select-Swap method \cite{sparse_LCU, low2024tradingSELECTSWAP} and the more recently proposed Select-Copy method which improves on the prefactors \cite{halving_qrom}.
These extra qubits may be dirty ancillas that are already available elsewhere in the algorithm.
For the THC LCU, performing SELECT only once also reduces the number of basis transformations.
Only one transformation is required per mode-coupling level, rather than one for every one-mode operator appearing in the LCU.
Since practical problems generally scale with the number of modes while the number of modals remains relatively small, reducing the dependence on the number of modes has the greatest impact on the overall computational cost.

\begin{table}[!t]
\centering
\renewcommand{\arraystretch}{1.3}
\begin{tabular}{|c|c|c|c|}
\hline
\textbf{Resources} & \textbf{Triangular} & \textbf{Pauli} & \textbf{THC} \\ \hline \hline

Toffoli gates &
$\mathcal{O}\left(\lambda_{\Delta}\bar{\kappa}M^{n}N^2\epsilon^{-1}\right)$ &
$\widetilde{\mathcal O}\left(\lambda_{\mathrm{Pauli}}M^{n/2}N^{n}\epsilon^{-1}\right)$ &
$\widetilde{\mathcal O}\left(\lambda_{\mathrm{THC}}(MN)^{n/2}\epsilon^{-1}\right)$
\\ \hline

Logical qubits &
$\mathcal{O}(MN)$ &
$\widetilde{\mathcal O}\left(M^{n/2}N^{n}\right)$ &
$\widetilde{\mathcal O}\left((MN)^{n/2}\right)$
\\ \hline

\end{tabular}
\caption{Asymptotic quantum resources required for energy estimation of vibrational Hamiltonians using different LCUs. 
Here, $\lambda_\Delta$, $\lambda_\mathrm{Pauli}$, and $\lambda_\mathrm{THC}$ denote the LCU 1-norms of the Triangular, Pauli, and THC LCUs, respectively. 
The parameter $n \ge 2$ is the mode-coupling level of the Hamiltonian. 
The quantity $\bar{\kappa}$ is the average factorization rank of the Hamiltonian coupling tensor slices whose dependence on $N$ can be system dependent. 
The target energy precision is denoted by $\epsilon$, and $M$ and $N$ represent the numbers of modes and modals, respectively. 
The Pauli and THC scalings assume the asymptotically optimal QROM ancilla tradeoff, while the THC scaling additionally assumes an average THC rank of $\mathcal{O}(N)$. 
Here, $\widetilde{\mathcal O}$ suppresses polylogarithmic factors in $M$ and $N$, including those arising from QROM register sizes.
When comparing the asymptotic scaling with respect to $M$ and $N$, the mode-coupling level $n$, coefficient and rotation-angle precisions are held fixed.}
\label{tab: asymptotic LCU resources}
\end{table}

Physical Hamiltonians are typically sparse and often contain many terms with negligible coefficients.
Rather than encoding every possible index combination, the Pauli LCU can be truncated to include only those terms whose coefficients have magnitude above a predefined threshold.
However, in the worst case, for a Hamiltonian with mode-coupling level $n$, the number of unique coefficients scales as $\mathcal{O}\left((MN^2)^n\right)$.
Without additional ancillas, the PREP circuit is dominated by coefficient loading and has a Toffoli cost of $\mathcal{O}\left((MN^2)^{n}\right)$ \cite{linearT}.
If sufficiently many clean ancillas, initialized in the $\ket{0}$ state, or dirty ancillas, initialized in arbitrary unknown states, are available, this cost can be reduced to $\widetilde{\mathcal O}\!\left((MN^2)^{n/2}\right)$.
The corresponding logical qubit cost scales in the same way \cite{sparse_LCU}.
The SELECT circuit is considerably cheaper, since it only applies the appropriate Pauli operator conditioned on the index register.
This requires $\mathcal{O}(MN)$ Toffoli gates.
After accounting for the system qubits, the asymptotic block-encoding cost becomes
\begin{align}
\mathrm{T}_{\mathrm{Pauli}}^{\mathrm{BE}}
    = \widetilde{\mathcal O}\left((MN^2)^{n/2}\right), \quad \quad
\mathrm{L}_{\mathrm{Pauli}}^{\mathrm{BE}} = \widetilde{\mathcal O}\left((MN^2)^{n/2}\right).
\label{Eq: Pauli LCU asymptotic BE cost}
\end{align}
A complete derivation of these costs, together with the corresponding quantum circuits, is provided in Appendix~\ref{App: block encoding Pauli LCU}.

The THC LCU factorizes the modal space, which reduces the number of coefficients that must be loaded.
In this case, the number of coefficients scales as $\mathcal{O}((MN)^n)$.
The PREP circuit therefore has a Toffoli cost of $\mathcal{O}((MN)^n)$ without additional ancillas and $\widetilde{\mathcal O}((MN)^{n/2})$ when an optimal number of ancillas is available.
In the latter case, the logical qubit cost also scales as $\widetilde{\mathcal O}((MN)^{n/2})$.
The THC SELECT circuit is more involved than its Pauli counterpart because it must implement modal rotations.
Specifically, the THC LCU consists of products of up to $n$ Pauli $Z$ operators conjugated by modal rotations (see Eq.~\eqref{Eq: THC LCU}), so the SELECT circuit must implement $n$ modal rotation unitaries.
These rotations are synthesized using the phase gradient technique \cite{phase_grad_rot}.
The SELECT circuit requires QROM-based loading of the rotation angles together with controlled swaps of the system registers. 
For fixed mode-coupling level and rotation-angle precision, these operations have a leading cost of $\mathcal O(nMN)$, while the rotations themselves require only $\mathcal O(nN)$ Toffoli gates. 
Thus, the rotations do not alter the leading $\mathcal O(nMN)$ scaling of SELECT.
The ancillas allocated for the PREP circuit are reused during the rotations, so no additional qubits are required.
Combining the PREP and SELECT costs gives
\begin{align}
\mathrm{T}_{\mathrm{THC}}^{\mathrm{BE}}
    = \widetilde{\mathcal O}\left((MN)^{n/2}\right), \quad \quad
\mathrm{L}_{\mathrm{THC}}^{\mathrm{BE}}
    = \widetilde{\mathcal O}\left((MN)^{n/2}\right).
\label{Eq: THC LCU asymptotic BE cost}
\end{align}
The corresponding circuits and explicit expressions for $\mathrm{T}_{\mathrm{THC}}^{\mathrm{BE}}$ and $\mathrm{L}_{\mathrm{THC}}^{\mathrm{BE}}$ are derived in Appendix~\ref{App: block encoding THC LCU}.

Up to polylogarithmic factors, Table~\ref{tab: asymptotic LCU resources} shows that the Pauli and THC LCUs reduce the leading dependence of the Toffoli cost on the number of modes from $M^n$ to $M^{n/2}$. Relative to the Pauli LCU, THC further reduces the leading dependence on the number of modals from $N^n$ to $N^{n/2}$.
In contrast, the Triangular LCU requires fewer logical qubits.
Nevertheless, as we demonstrate with constant-factor cost estimates, the Pauli and THC LCUs achieve more than an order-of-magnitude reduction in Toffoli cost while requiring only a modest increase in logical qubits.

\section{Results} \label{sec: results}
Following the work of Ref.~\cite{majland2025fault}, we evaluate the LCUs developed in this study on a series of seven polycyclic aromatic hydrocarbons (PAHs) with 30–138 vibrational modes.
Specifically, these systems are linear chains of fused benzene rings known as polyacenes. 
They provide a systematic means to scale problem complexity by increasing the number of benzene rings.
Additional details of these molecules are provided in Appendix~\ref{App: data}.
We compare our approach with the Triangular LCU.
The procedure for constructing the Triangular LCU and determining the associated resource estimates is detailed in Appendix~\ref{App: Triangular LCU}.
We analyze the two-mode-coupling Hamiltonian costs for all seven systems using five modals per mode in the vibrational self-consistent field (VSCF) basis \cite{VSCF}.
For the three-mode-coupling Hamiltonians, we restrict our analysis to PAH1 and PAH2, employing three modals per mode.
All Hamiltonians are generated using MidasCpp \cite{midascpp}.
Tensor optimizations for THC and BLISS are performed using the Adam optimizer from Optax \cite{optax}.
The code used to obtain the LCUs and resource estimates reported herein is available on the associated GitHub repository \cite{github_link}.

\begin{figure}[!t]\centering    \includegraphics[width=1\linewidth]{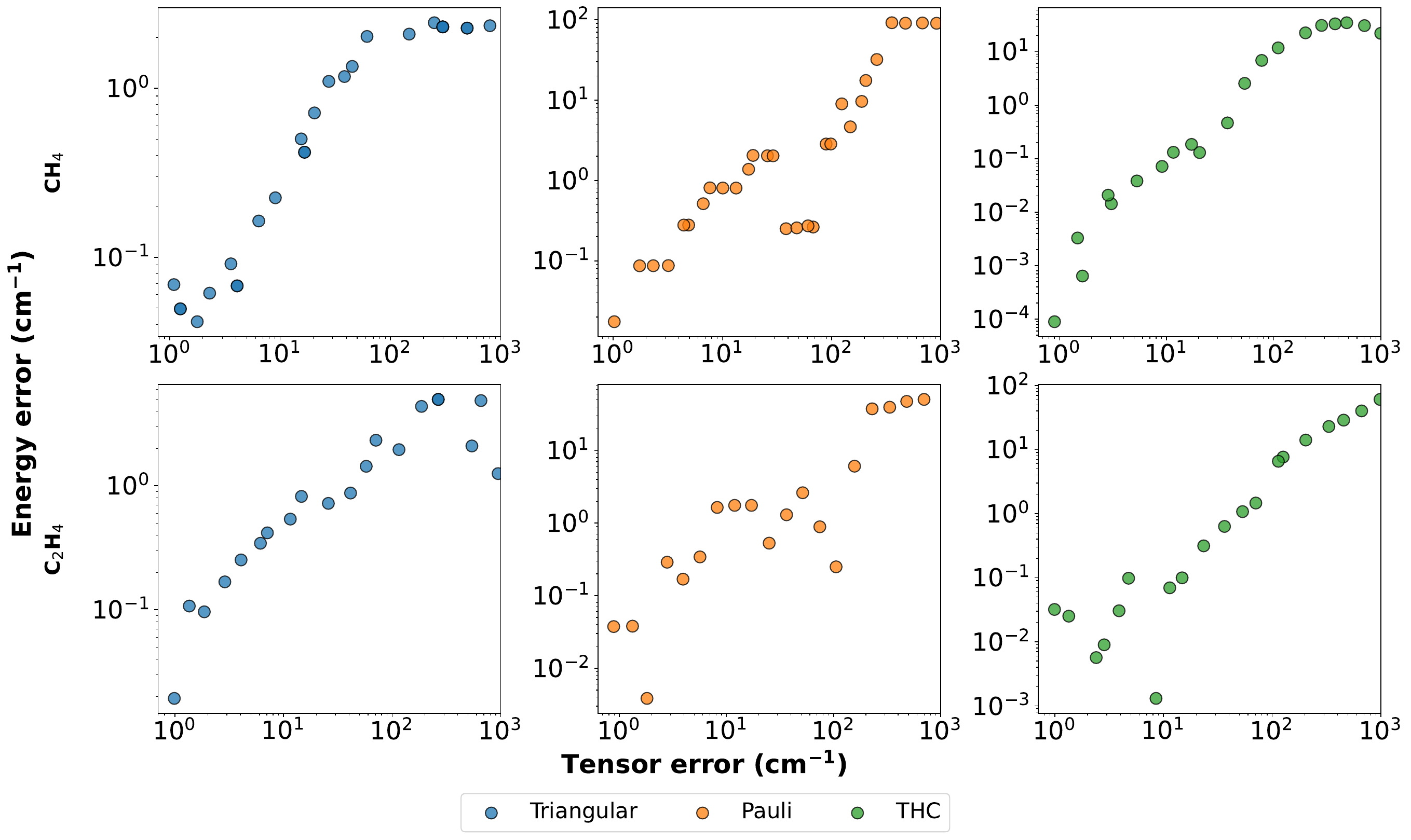}
\caption{Ground-state energy error versus tensor approximation error calculated using the Frobenius norm for the Triangular, Pauli, and THC LCUs, shown from left to right.
The top and bottom rows correspond to CH$_4$ (9 modes) and C$_2$H$_4$ (12 modes), respectively.
Both Hamiltonians include terms up to two-mode coupling and use 5 modals per mode.
}\label{Fig: energy vs tensor error}
\end{figure}

In all three LCUs, we utilize approximate tensors in place of their exact counterparts to enable efficient block encodings while maintaining the desired energy accuracy.
Because the approximation procedure differs between the LCUs, we quantify the tensor error using the Frobenius norm of the difference between the original and approximate tensors.
Figure~\ref{Fig: energy vs tensor error} shows the resulting ground-state energy error as a function of this tensor error for the two-mode-coupling $\text{CH}_4$ and $\text{C}_2\text{H}_4$ Hamiltonians.
The ground-state energies are computed using the Locally Optimal Block Preconditioned Conjugate Gradient (LOBPCG) algorithm as implemented in JAX \cite{jax}.
For the Triangular LCU, we generate the data by systematically increasing the truncation cutoff applied to the singular values of the tensor slices (see Appendix~\ref{App: Triangular LCU}).
For the Pauli LCU, we instead increase the cutoff used to discard small tensor elements.
For each resulting approximate tensor, we construct the Hamiltonian and compute the corresponding ground-state energy.
Nearly all THC data points are obtained by following the CP3-based initialization procedure described in Appendix~\ref{App: Initializing THC}.
The nonmonotonic behavior observed in some of the THC data potentially arises from the use of a 1-norm-regularized loss function rather than an objective based solely on tensor error.
We employ this regularized objective because it provides more reliable convergence at a fixed THC rank.

For the benchmark systems considered in Fig.~\ref{Fig: energy vs tensor error}, the resulting ground-state energy errors are typically one or more orders of magnitude smaller than the corresponding tensor errors.
Based on this empirical observation, we adopt a target tensor error of $1\,\mathrm{cm}^{-1}$ as a heuristic proxy for achieving an energy accuracy of $1\,\mathrm{cm}^{-1}$.
We apply the same tensor-error threshold to larger systems for which direct evaluation of the ground-state energy error is computationally impractical.
A rigorous procedure, however, would require a more direct estimator of the resulting energy error, which we leave for future work.

\begin{figure}[!t]\centering    \includegraphics[width=0.9\linewidth]{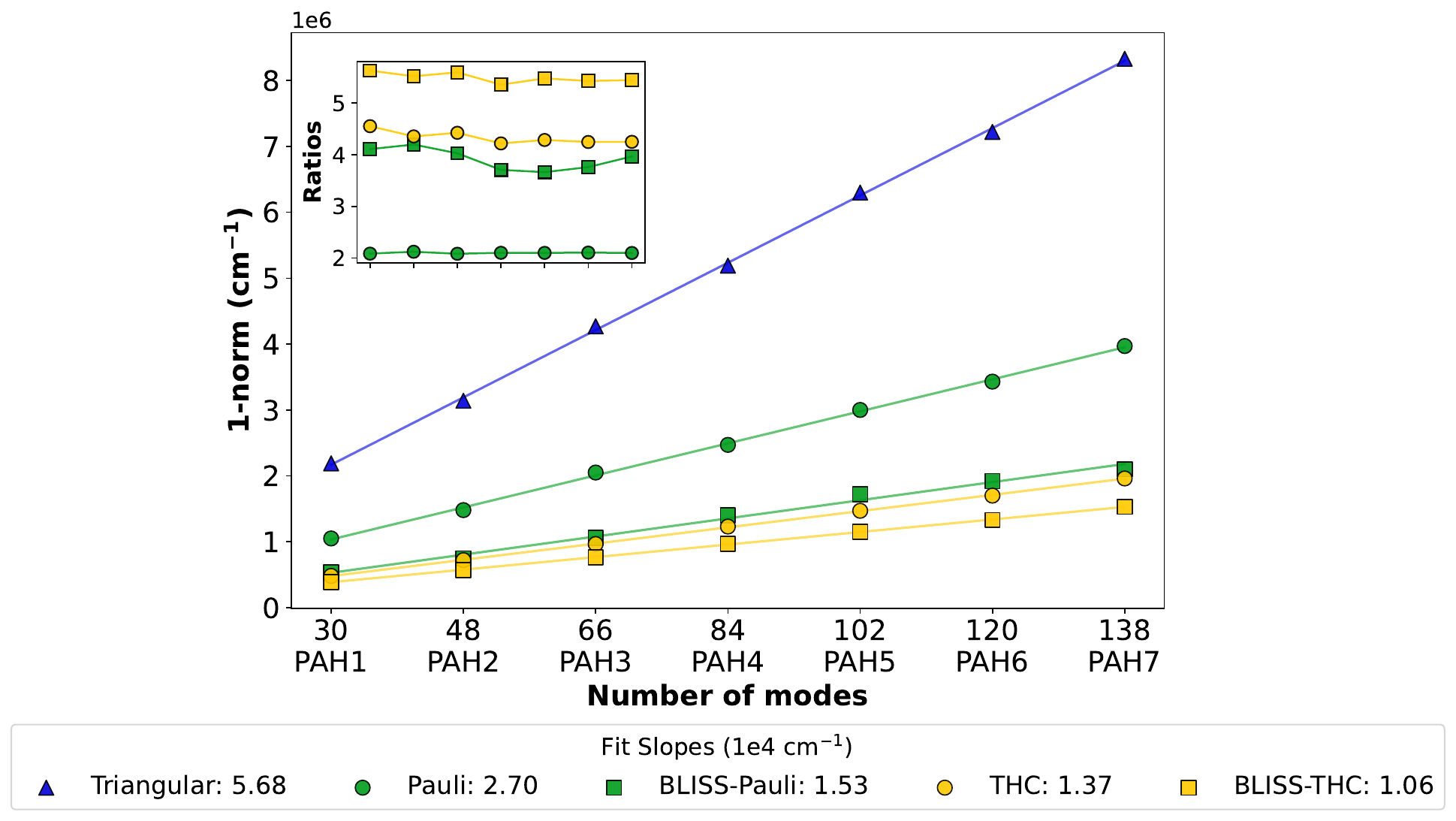}
\caption{Comparison of LCU 1-norms across seven two-mode-coupling PAH Hamiltonians. 
The main panel shows the 1-norms for the Triangular, Pauli, and THC LCUs, with the BLISS technique applied to further reduce the 1-norms for the Pauli and THC methods. 
Linear fits versus the number of modes are overlaid as guides to the observed scaling; the fitted slopes are reported in the legend. 
The inset shows the ratios of the Triangular LCU 1-norm to the Pauli, BLISS-Pauli, THC, and BLISS-THC 1-norms.
}\label{Fig: 1-norms}
\end{figure}

\begin{figure}[!t]\centering    \includegraphics[width=1\linewidth]{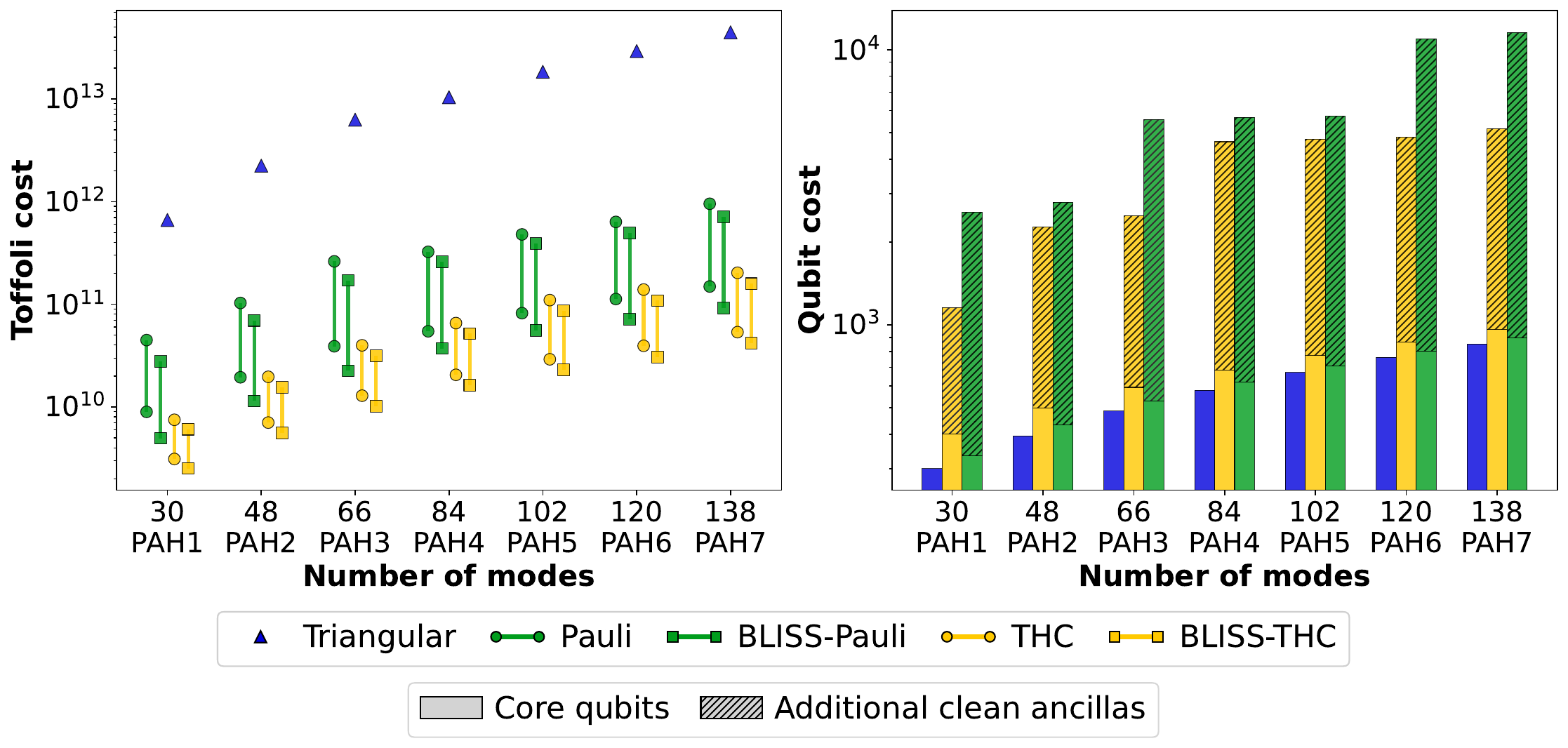}
\caption{Toffoli costs (left) and logical qubit costs (right) for QPE ground-state energy estimation of seven second-quantized, two-mode-coupling PAH Hamiltonians with a target QPE accuracy of $1\,\mathrm{cm}^{-1}$. 
Methods labeled ``BLISS'' use BLISS to reduce the LCU 1-norm. 
Because BLISS has a negligible effect on the logical qubit cost, separate BLISS qubit bars are omitted. A small x-axis offset between different methods was used for visual clarity.
The vertical ranges on the left show the Toffoli--qubit tradeoffs available through QROM for the Pauli and THC methods. 
The upper endpoint of each range corresponds to dirty-ancilla-assisted Select-Copy QROM \cite{halving_qrom}, while the lower endpoint corresponds to the largest clean-ancilla allocation considered \cite{sparse_LCU}. 
In the right panel, the solid portion of Pauli and THC bars shows the logical qubit cost of the dirty-ancilla configuration, and the hatched portion shows the additional clean ancillas required to reach the lower Toffoli endpoint. 
For the Triangular LCU, Ref.~\cite{majland2025fault} found no benefit from using either dirty or clean ancillas, so no Toffoli--qubit tradeoff is shown. 
The coefficient precision and, where applicable, the rotation-angle precision are both set to 20 bits.
}
\label{Fig: Toffoli and qubit costs for two-mode-coupling PAHs}
\end{figure}

As shown in Table~\ref{tab: asymptotic LCU resources}, the induced LCU 1-norm plays a crucial role in determining the overall Toffoli cost.
In Fig.~\ref{Fig: 1-norms}, we demonstrate that the Pauli and THC LCUs exhibit smaller 1-norms than the Triangular LCU for two-mode-coupling PAH Hamiltonians.
Furthermore, by applying the BLISS technique, the 1-norm of the Pauli LCU can be reduced by approximately 46\% on average, and that of the THC LCU by roughly 21\% on average.
Relative to the Triangular LCU, the inset in the figure shows that the BLISS-THC variant achieves more than a fivefold decrease in the 1-norm.
The total Toffoli and qubit costs for the same systems are presented in Fig.~\ref{Fig: Toffoli and qubit costs for two-mode-coupling PAHs}.
For these estimates, both the coefficient precision and the rotation-angle precision are set to 20 bits.
Across all evaluated systems with 5 modals per mode, the average THC rank lies between 6 and 7, substantially smaller than the ranks typically encountered for electronic structure Hamiltonians.
The Pauli and THC variants distinguish between calculations performed with and without BLISS.
Because the implementation of QROM admits a Toffoli--qubit tradeoff, the range of Toffoli costs obtained from different ancilla allocations is shown by vertical lines.
The upper endpoints correspond to Select-Copy-based QROM \cite{halving_qrom} using available dirty ancillas, with the associated qubit costs shown by the solid bars in the right panel.
The lower endpoints correspond to the costs obtained by using additional clean ancillas for minimizing the Select-Copy QROM's Toffoli counts, while the associated additional qubit overheads are shown by the hatched portions of the stacked bars.
Notably, even the Select-Copy-based Pauli LCU, which has the highest Toffoli cost among the Pauli and THC variants, achieves more than an order-of-magnitude reduction relative to the Triangular LCU.
The Select-Copy-based THC LCU reduces Toffoli costs by an additional factor of 4--7 relative to the corresponding Pauli variant.
Most importantly, both the Select-Copy-based Pauli and THC approaches achieve these Toffoli reductions with modest qubit overheads relative to the Triangular LCU, decreasing from approximately 11\% to 5\% for Pauli and from 33\% to 13\% for THC as the system size increases.
BLISS-THC achieves a 111- to 281-fold reduction in Toffoli cost relative to the Triangular LCU.
Further Toffoli reductions enabled by additional clean ancillas are accompanied by increased qubit overhead, with the relative increase in qubit requirements comparable to or larger than the corresponding Toffoli savings.
The numerical values used to plot Figs.~\ref{Fig: 1-norms} and \ref{Fig: Toffoli and qubit costs for two-mode-coupling PAHs} are presented in Appendix~\ref{App: data}.

Deriving the Triangular LCU for three-mode-coupling Hamiltonians involves factorizing order-three tensors.
Because this factorization is not unique, the resulting LCUs for the same Hamiltonian can yield varying resource estimates.
The authors of Ref.~\cite{majland2025fault} perform three-mode-coupling factorizations using the CANDECOMP/PARAFAC alternating least squares (CP-ALS) method, optionally initialized with an intermediate Tucker decomposition \cite{CP3}.
In this work, we also employ the CP-ALS method provided by the package TensorLy \cite{tensorly}; however, we use it on the decomposed THC core tensors instead as this is computationally more feasible.
The exact procedure is outlined in Appendix~\ref{App: Triangular LCU}. 
Thus, the three-mode-coupling Triangular LCU resource estimates reported here are based on the implementation described in Appendix~\ref{App: Triangular LCU} and do not constitute an exact reproduction of the procedure used in Ref.~\cite{majland2025fault}.
However, given the asymptotic improvements offered by the Pauli and THC LCUs (see Table~\ref{tab: asymptotic LCU resources}), we anticipate that differences in how the Triangular LCU is constructed will not produce a notable change to the trends presented in the analysis below. 
For all three LCUs, we ensure that the sum of the two- and three-mode-coupling tensor approximation errors is less than $1\,\mathrm{cm}^{-1}$.

\begin{table}[!t]
\centering
\begin{tabular}{|c|c|c|c|}
\hline
\textbf{Metric} & \textbf{Method} & \textbf{PAH1} & \textbf{PAH2} \\ \hline \hline

\multirow{5}{*}{\begin{tabular}[c]{@{}c@{}}1-norms \\ ($\times 10^{5}\,\mathrm{cm}^{-1}$)\end{tabular}}
& Triangular  & 9.69 & 16.1 \\ 
& Pauli       & 4.31 & 7.79 \\ 
& BLISS-Pauli & 3.35 & 6.41 \\ 
& THC         & 2.43 & 4.76 \\ 
& BLISS-THC   & \textbf{1.93} & \textbf{4.01} \\ \hline

\multirow{3}{*}{Qubit cost}
& Triangular & \textbf{244} & \textbf{302} \\ 
& Pauli      & 293 $\rightarrow$ 5711 & 360 $\rightarrow$ 12042 \\  
& THC        & 321 $\rightarrow$ 4938 & 385 $\rightarrow$ 10501 \\ \hline

\multirow{5}{*}{\begin{tabular}[c]{@{}c@{}}Toffoli cost \\ ($\times 10^{10}$)\end{tabular}}
& Triangular  & 1030 & 3810 \\ 
& Pauli       & 21.1 $\rightarrow$ 0.985 & 116 $\rightarrow$ 3.50 \\ 
& BLISS-Pauli & 14.9 $\rightarrow$ 0.728 & 86.4 $\rightarrow$ 2.74 \\ 
& THC         & 7.30 $\rightarrow$ 0.510 & 47.5 $\rightarrow$ 1.99 \\ 
& BLISS-THC   & \textbf{5.80 $\rightarrow$ 0.405} & \textbf{40.1 $\rightarrow$ 1.68} \\ \hline

\end{tabular}
\caption{LCU 1-norms and QPE resource estimates for ground-state energy estimation of the three-mode-coupling PAH1 and PAH2 Hamiltonians with a target QPE accuracy of $1\,\mathrm{cm}^{-1}$. 
The Hamiltonians are represented using three VSCF modals per mode. 
Arrows show the Toffoli--qubit tradeoff from Select-Copy QROM using available dirty ancillas on the left to the largest clean-ancilla allocation considered on the right. 
Providing additional clean ancillas reduces the Toffoli cost while increasing the logical qubit requirement \cite{low2024tradingSELECTSWAP,halving_qrom}. 
For the Triangular LCU, Ref.~\cite{majland2025fault} found no benefit from ancilla-assisted QROM, so no Toffoli--qubit tradeoff is reported. 
Because BLISS has a negligible effect on the logical qubit cost, separate BLISS qubit rows are omitted.
}
\label{Tab: PAH1 3 mode coupling resources}
\end{table}

The resulting resource estimates for the three-mode-coupling PAH1 and PAH2 Hamiltonians are provided in Table~\ref{Tab: PAH1 3 mode coupling resources}.
The arrows indicate the Toffoli--qubit tradeoff from the Select-Copy-based implementation using available dirty ancillas to the largest clean-ancilla allocation considered here.
For PAH1, Select-Copy-based BLISS-THC achieves an approximately 178-fold reduction in Toffoli cost relative to the Triangular LCU, with only a 1.3-fold increase in qubit requirements.
For PAH2, the corresponding reduction is approximately 95-fold, with a 1.27-fold increase in qubit requirements.
Further Toffoli reductions are possible by allocating additional clean ancillas, reaching reductions of approximately 2540-fold and 2270-fold for PAH1 and PAH2, respectively, albeit with substantially larger qubit overheads.
As in the two-mode-coupling results, THC provides lower Toffoli costs than the corresponding Pauli LCU, although the relative improvement is smaller for these three-mode-coupling Hamiltonians.
This can be attributed in part to the smaller modal space of three modals per mode used here, compared with five modals per mode for the two-mode-coupling Hamiltonians.
Since THC provides a near-quadratic asymptotic improvement in the Toffoli scaling with the number of modals relative to the Pauli LCU, its advantage is expected to grow as the modal space increases.
The reduction in the LCU 1-norm arising from tensor factorization may also become more pronounced with increasing modal-space size, further widening the gap between the Pauli and THC approaches.

\begin{figure}[!t]\centering    \includegraphics[width=0.9\linewidth]{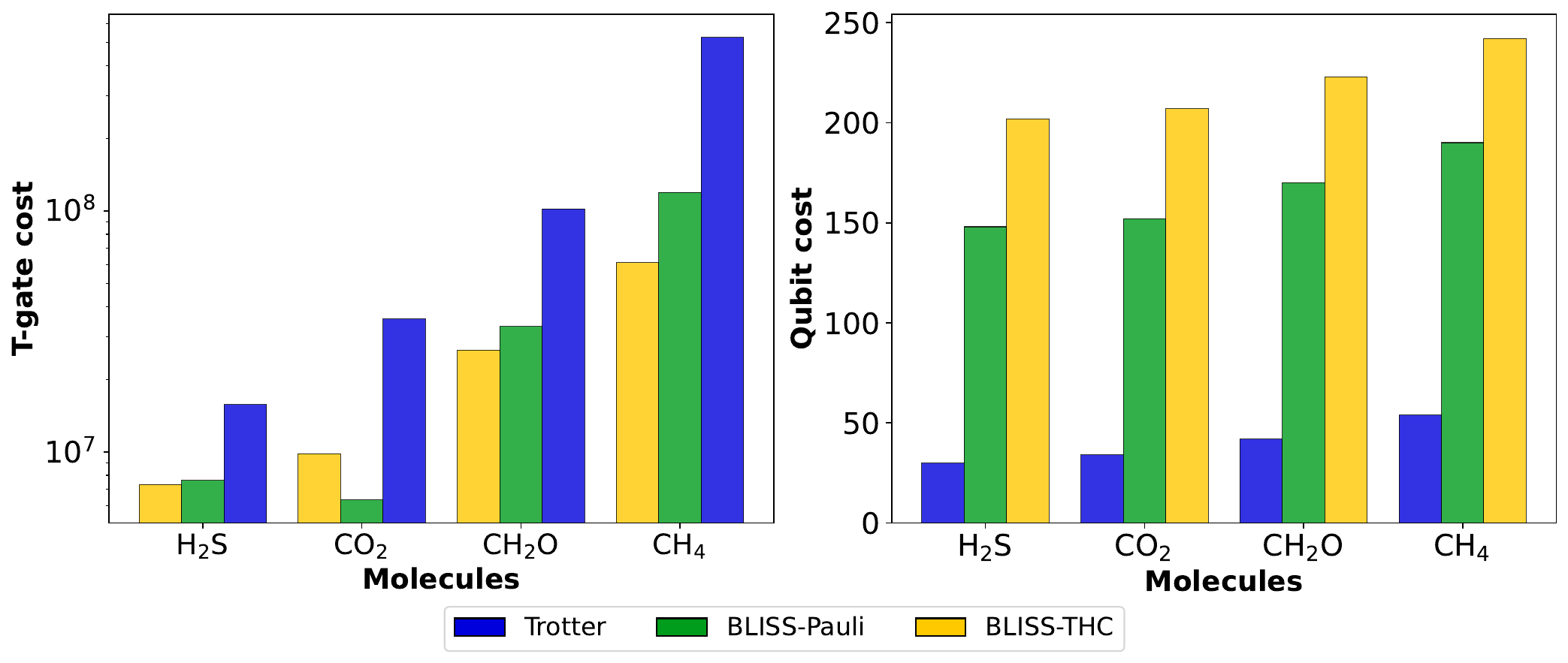}
\caption{Comparison of quantum resources required to estimate vibrational energies via QPE to an accuracy of $10\,\mathrm{cm}^{-1}$ for qubitization and Trotter-approximation-based approaches. 
For these small-system qubitization estimates, QROM without additional ancillas \cite{linearT} is used instead of its ancilla-assisted variants \cite{low2024tradingSELECTSWAP,halving_qrom}.
The Trotter estimates use the Christiansen Greedy Fragment (CGF) construction of Ref.~\cite{malpathak2025trotter}.
All Hamiltonians are represented in a local-mode harmonic-oscillator basis with four modals per mode.
}\label{Fig: qubitization_vs_trotter}
\end{figure}

To further contextualize the advantages of the methods developed here, we compare the resource requirements of qubitization-based quantum phase estimation with those of the Trotter-approximation approach reported in Table~3 of Ref.~\cite{malpathak2025trotter}.
This comparison considers four molecules: H$_2$S, CO$_2$, CH$_2$O, and CH$_4$, represented in a second-quantized local-mode harmonic-oscillator basis with four modals per mode.
The Trotter-approximation data correspond to the Christiansen Greedy Fragment (CGF) method of Ref.~\cite{malpathak2025trotter}.
As the non-Clifford cost reported in Ref.~\cite{malpathak2025trotter} is given in terms of T gates, we convert our Toffoli counts to T-gate counts and compare the resulting T-gate and qubit costs in Fig.~\ref{Fig: qubitization_vs_trotter}; each Toffoli gate was counted as 4 T gates.
Additional details of deriving resource estimates for Trotter-based QPE are provided in Appendix~\ref{App: Trotter}, and the numerical values used to plot Fig.~\ref{Fig: qubitization_vs_trotter} are presented in Appendix~\ref{App: data}.
As can be observed, Trotter-based QPE requires fewer qubits, whereas the qubitization-based approaches require fewer T gates.
Importantly, across the systems studied, the BLISS-THC T-gate savings increase from 2-fold to nearly 9-fold, while the ratio of qubit costs decreases from approximately 7-fold to 5-fold.
This trend suggests that the overall resource advantage of qubitization may increase with system size.
Beyond phase estimation, if the objective is to simulate quantum dynamics, the block-encoded Hamiltonian can be used to construct the time-evolution operator directly via quantum signal processing (QSP) \cite{QSP}.
Because QSP exhibits asymptotically superior scaling compared to standard Trotterization, we anticipate the computational advantage of our proposed methods to be even more pronounced for time evolution than in the comparisons shown in Fig.~\ref{Fig: qubitization_vs_trotter}.

Overall, the two-mode-coupling results, together with the three-mode-coupling benchmarks, demonstrate that the Pauli and THC LCUs consistently achieve substantial reductions in the number of non-Clifford operations across a range of problem sizes relative to both the Triangular LCU and Trotter-based approaches.

\section{Conclusion} \label{sec: conclusion}
In this work, we have developed improved block-encoding strategies for second-quantized vibrational Hamiltonians and benchmarked their resource requirements against those of Ref.~\cite{majland2025fault} for a series of polycyclic aromatic hydrocarbon Hamiltonians.
Exploiting similarities between the second-quantized electronic and vibrational Hamiltonians, we have generalized some of the block-encoding techniques developed for the electronic structure problem to vibrational quantum simulation.
We have derived Pauli and THC LCUs and extended the BLISS technique to reduce the LCU 1-norm.
Under the unary mapping, the vibrational creation and annihilation operators map directly to Pauli raising and lowering operators, making the construction of the corresponding qubit Hamiltonian straightforward.
Obtaining the THC representation, however, involves a nonlinear optimization and a basis transformation via modal rotation unitaries.
We have ensured efficient THC optimization by employing a CP3-based initialization that decouples the optimization problem for each mode.

For both the Pauli and THC LCUs, we have developed efficient block-encoding circuits that yield substantial reductions in Toffoli operations relative to the Triangular LCU proposed in Ref.~\cite{majland2025fault}. 
The Triangular LCU, however, uses the fewest logical qubits, with the Pauli LCU requiring only a small increase when no additional ancillas are used.
Up to polylogarithmic factors, allocating sufficient additional qubits allows both the Pauli and THC LCUs to achieve a quadratic improvement in the Toffoli scaling with the number of modes. 
Assuming that the average THC rank scales linearly with the number of modals, THC provides an additional quadratic improvement, up to polylogarithmic factors, in scaling with the number of modals.
In part, these improvements arise from the ability to better exploit Toffoli--qubit tradeoffs using ancilla-assisted QROM techniques \cite{sparse_LCU,low2024tradingSELECTSWAP,halving_qrom} compared to the Triangular LCU framework. 
We note that recently improved QROM implementations could reduce costs further \cite{halving_again}.
For a QPE-based workflow of two-mode-coupling Hamiltonians, our results demonstrate that BLISS-THC can achieve up to a 281-fold reduction in the total Toffoli count compared to the Triangular LCU for only a modest increase in the qubit count.
A similar analysis of small three-mode-coupling Hamiltonians indicates reductions by factors of 95--178, with the asymptotic scaling suggesting larger reductions as the number of modals increases.
In both cases, further reductions in Toffoli cost are possible, but only at the expense of substantially increased qubit requirements.
Furthermore, when benchmarked against recent Trotter-based approaches for a set of small systems with up to 9 modes, our qubitization methods achieve reductions of up to nearly ninefold in T-gate cost.
Taken together with our asymptotic resource analysis, these results identify BLISS-THC as the most resource-efficient block-encoding strategy for almost all benchmarked systems in this work.

Beyond quantum simulation, the vibrational THC factorization developed here may also be useful for classical methods for computing vibrational spectra.
Since its original development by Mart\'inez and coworkers \cite{org_thc}, THC has been used to accelerate several established electronic structure methods \cite{THC_CC2, THC_EOM_CC}.
Given the parallels between second-quantized electronic and vibrational structure, it would be interesting to investigate whether analogous THC-based improvements are possible for vibrational perturbation theory \cite{VMP2}, vibrational configuration interaction \cite{VCI}, and vibrational coupled-cluster theory \cite{vCC, VCC[2pt3]}.

Future work in quantum simulation could include end-to-end resource estimates for vibrational double factorization.
While straightforward to formulate for two-mode-coupling Hamiltonians, this approach will require nonlinear optimization for higher-order interactions.
Additionally, extending techniques such as sum-of-squares and spectral amplification, which were recently proposed for electronic Hamiltonian simulation \cite{SOSSA}, presents a promising avenue for further cost reductions.
Beyond optimizing our current framework, it is also valuable to establish numerically grounded comparisons of quantum resource requirements across different representations of the same vibrational problem at a fixed target accuracy.
While this study focuses on a second-quantized representation, an efficient block-encoding of a real-space representation of the rovibrational Hamiltonian was recently introduced in Ref.~\cite{szczepanik2025simulatingBEIT}.
A comparison across such representations for Trotter-based QPE was conducted in Ref.~\cite{malpathak2025trotter}, demonstrating that a second-quantized approach required fewer resources than a real-space formulation for spectral calculations considered in that work.
A similar comparison for block-encoding-based QPE will be important for identifying the most efficient pathway toward practical vibrational quantum simulation. 
Furthermore, given the rapid increase in research activity to develop fault-tolerant algorithms for molecular vibrations, there is a growing need to identify challenging target molecular systems against which algorithmic costs can be meaningfully benchmarked.
Such targets are likely to be large, strongly coupled systems with potential energy surfaces that do not admit compact SOP forms or $n$-mode expansions, making them challenging for leading tensor-network-based methods.

\begin{acknowledgments}
The authors would like to thank Ove Christiansen for providing the PAH Hamiltonians and assisting in configuring MidasCpp. S.M. appreciates financial support from the Research Excellence Postdoctoral Fellowship at the University of Toronto Scarborough.
We also acknowledge the Applied Quantum Computing Challenge program of the National Research Council of Canada (grant number AQC-103-2) and the NSERC Quantum Alliance program for financial support.
This research was enabled in part by computational resources provided by Compute Ontario (\url{https://computeontario.ca}) and the Digital Research Alliance of Canada (\url{https://alliancecan.ca}).
Some computations were performed on the Niagara and Trillium supercomputers at the SciNet High Performance Computing Consortium and on the NARVAL and RORQUAL supercomputers through Calcul Québec.
SciNet is funded by Innovation, Science, and Economic Development Canada, the Digital Research Alliance of Canada, the Ontario Research Fund: Research Excellence, and the University of Toronto.
\end{acknowledgments}

\appendix

\section{Details of Triangular LCU} \label{App: Triangular LCU}
The Triangular LCU relies on expressing the vibrational Hamiltonian as a sum-of-products (SOP) of one-mode operators.
The full Hamiltonian is then block-encoded by explicitly forming the SOP of one-mode block encodings.
Our implementation includes several modifications to the original approach; we first summarize the procedure described in Ref.~\cite{majland2025fault}.
Recall that the Hamiltonian obtained using the ADGA procedure is expressed as [see Eq.~\eqref{Eq: SOP Hamiltonian}]
\begin{align*}
       \hat H &= \sum_{i} \sum_{o^{i}} c^{i}_{o^{i}}\ \hat h^{i, o^{i}} + \sum_{i < j}\ \sum_{o^{i}, o^{j}} c^{i j}_{o^{i}o^{j}}\ \hat h^{i, o^{i}} \hat h^{j, o^{j}} + \sum_{i < j < k}\  \sum_{o^{i}, o^{j}, o^{k}} c^{i j k}_{o^{i}o^{j}o^{k}}\ \hat h^{i, o^{i}} \hat h^{j, o^{j}} \hat h^{k, o^{k}}.
\end{align*}
Block-encoding the Hamiltonian by encoding the individual operators $\hat h^{i,o^i}$ has a cost that scales with the number of terms in the coefficient tensors $c$.
To reduce this cost, the tensors can be factorized, resulting in fewer Hamiltonian terms at the expense of making the one-mode operators depend explicitly on the mode combinations.

The idea is to obtain the decompositions
\begin{align}
    c^{i j}_{o^{i} o^{j}} &= \sum_\alpha  \gamma^{\{i,j\}}_\alpha V^{i, \{i,j\}}_{o^{i},\alpha} V^{j, \{i,j\}}_{o^{j},\alpha}, \label{App Eq: cmomo factorization}\\
    c^{i j k}_{o^{i} o^{j} o^{k}} &= \sum_\alpha \gamma^{\{i,j,k\}}_\alpha V^{i, \{i,j,k\}}_{o^{i},\alpha} V^{j, \{i,j,k\}}_{o^{j},\alpha} V^{k, \{i,j,k\}}_{o^{k},\alpha}, \label{App Eq: cmomomo factorization}
\end{align}
for each mode pair $(i,j)$ and mode triplet $(i,j,k)$, respectively.
The factorization of the $c^{ij}$ tensor is a singular value decomposition (SVD) for each mode pair ($i$, $j$).
The factorization of $c^{ijk}$, however, is non-unique and can be challenging to obtain.
The authors of Ref.~\cite{majland2025fault} employ the CP-ALS method for each mode triplet $(i,j,k)$ to perform this factorization with optional initialization from a preliminary Tucker decomposition.
Substituting these factorizations into Eq.~\eqref{Eq: SOP Hamiltonian} modifies the Hamiltonian as
\begin{align}
    \hat H &= \sum_{i} \hat h^{i} + \sum_{i < j}\ \sum_{\alpha} \gamma^{\{i, j\}}_{\alpha}\ \hat h^{i, \{i,j\}}_\alpha \hat h^{j, \{i,j\}}_\alpha + \sum_{i < j < k}\ \sum_{\alpha} \gamma^{\{i, j, k\}}_{\alpha}\ \hat h^{i, \{i,j,k\}}_\alpha \hat h^{j, \{i,j,k\}}_\alpha \hat h^{k, \{i,j,k\}}_\alpha, \label{App Eq: new SOP Hamiltonian}
\end{align}
where the new one-mode operators are defined as
\begin{align*}
\hat h^{i} &= \sum_{o^{i}} c^{i}_{o^{i}}\ \hat h^{i, o^{i}}, \\
    \hat h^{i, \{i,j\}}_\alpha &= \sum_{o^{i}} V^{i,\{i,j\}}_{o^{i},\alpha} \hat h^{i, o^{i}}, \\
    \hat h^{i, \{i,j,k\}}_\alpha &= \sum_{o^{i}} V^{i,\{i,j,k\}}_{o^{i},\alpha} \hat h^{i, o^{i}}.
\end{align*}
The coefficients $\gamma^{\{i,j\}}$ and $\gamma^{\{i,j,k\}}$ can be absorbed into the one-mode operators.
The new SOP representation in Eq.~\eqref{App Eq: new SOP Hamiltonian} benefits from having fewer terms than the original representation of Eq.~\eqref{Eq: SOP Hamiltonian}.
The name Triangular LCU derives from the way each of the above one-mode operators is encoded.
Since these one-mode operators are Hermitian, their integrals accompanying the creation/annihilation operators constitute symmetric matrices.
As such, loading only the lower/upper triangle of these matrices is sufficient to construct the one-mode operators.

Our implementation of the Triangular LCU differs from the original approach in a few ways.
First, instead of starting from the Hamiltonian of Eq.~\eqref{Eq: SOP Hamiltonian} and factorizing $c^{ij}$ and $c^{ijk}$ tensors, we start from the Hamiltonian of Eq.~\eqref{Eq: second-quantized symmetrized vibrational Hamiltonian}, which is also in an SOP form where the one-mode operators are now excitation operators, and factorize the $g$ and $v$ tensors.
This choice was made since the resulting LCU 1-norm was often found to be slightly smaller.
For the two-mode-coupling tensor $g$, for each pair of $i$ and $j$ indices, we treat $(p,q)$ and $(r,s)$ as composite indices to obtain the SVD
\begin{align}
    g_{i,p,q,j,r,s} = \sum_{\alpha=0}^{\kappa^{\{i,j\}}-1}  \tilde \gamma^{\{i,j\}}_\alpha \tilde{V}^{i, \{i,j\}}_{(p,q),\alpha} \tilde{V}^{j, \{i,j\}}_{(r,s),\alpha}. \label{App Eq: g SVD}
\end{align}
 For the three-mode-coupling tensor $v$, we reuse the THC factorized form of Eq.~\eqref{Eq: trbt THC factorization}, which we rewrite below for convenience.
\begin{align*}
    v_{i,p,q,j,r,s,k,t,u} &= \sum_{\mu_i = 0}^{R_i-1} \sum_{\nu_j = 0}^{R_j-1} \sum_{\omega_k = 0}^{R_k-1}  \Gamma^{(i,j,k)}_{\mu_i,\nu_j,\omega_k}\ \chi^{(i)}_{p,\mu_i} \chi^{(i)}_{q,\mu_i} \ \chi^{(j)}_{r,\nu_j} \chi^{(j)}_{s,\nu_j} \ \chi^{(k)}_{t,\omega_k} \chi^{(k)}_{u,\omega_k}.
\end{align*}
Since the goal is to express this as a compact sum of products of one-mode integrals for each set of $i,j,$ and $k$ indices, we perform a CP-ALS decomposition of the $\Gamma^{(i,j,k)}$ tensor:
\begin{align}
    \Gamma^{(i,j,k)}_{\mu_i, \nu_j, \omega_k} = \sum_{\alpha=0}^{\kappa^{\{i,j,k\}}-1} \tilde \gamma^{\{i,j,k\}}_{\alpha} \tilde{\mathcal{V}}^{i,\{i,j,k\}}_{\mu_i,\alpha} \tilde{\mathcal{V}}^{j,\{i,j,k\}}_{\nu_j,\alpha} \tilde{\mathcal{V}}^{k,\{i,j,k\}}_{\omega_k,\alpha}.
\end{align}
By defining $\tilde{V}^{i,\{i,j,k\}}_{(p,q),\alpha} = \sum_{\mu_i} \tilde{\mathcal{V}}^{i,\{i,j,k\}}_{\mu_i,\alpha} \chi^{(i)}_{p,\mu_i} \chi^{(i)}_{q,\mu_i}$, we can rewrite the three-mode-coupling tensor as
\begin{align}
    v_{i,p,q,j,r,s,k,t,u} &= \sum_{\alpha=0}^{\kappa^{\{i,j,k\}}-1} \tilde{\gamma}^{\{i,j,k\}}_{\alpha} \tilde{V}^{i,\{i,j,k\}}_{(p,q),\alpha} \tilde{V}^{j,\{i,j,k\}}_{(r,s),\alpha}\tilde{V}^{k,\{i,j,k\}}_{(t,u),\alpha}. \label{App Eq: v CP-ALS}
\end{align}
Substituting Eqs.~\eqref{App Eq: g SVD} and \eqref{App Eq: v CP-ALS} back into the Hamiltonian of Eq.~\eqref{Eq: second-quantized symmetrized vibrational Hamiltonian} and restricting the sum over mode indices leads to
\begin{align}
    \hat H &= \sum_{i} \hat {h}^{i} + \sum_{i < j}\ \sum_{\alpha=0}^{\kappa^{\{i,j\}}-1} 2\tilde\gamma^{\{i, j\}}_{\alpha}\ \hat {\tilde{h}}^{i, \{i,j\}}_\alpha \hat {\tilde{h}}^{j, \{i,j\}}_\alpha + \sum_{i < j < k}\ \sum_{\alpha=0}^{\kappa^{\{i,j,k\}}-1} 6\tilde\gamma^{\{i, j, k\}}_{\alpha}\ \hat {\tilde{h}}^{i, \{i,j,k\}}_\alpha \hat {\tilde{h}}^{j, \{i,j,k\}}_\alpha \hat {\tilde{h}}^{k, \{i,j,k\}}_\alpha, \label{App Eq: this work SOP Hamiltonian}
\end{align}
where the new one-mode operators are defined as
\begin{align*}
    \hat {\tilde{h}}^{i, \{i,j\}}_\alpha &= \sum_{p,q} \tilde{V}^{i,\{i,j\}}_{(p,q),\alpha} \hat E^{p_i}_{q_i}, \\
    \hat {\tilde{h}}^{i, \{i,j,k\}}_\alpha &= \sum_{p,q} \tilde{V}^{i,\{i,j,k\}}_{(p,q),\alpha} \hat E^{p_i}_{q_i}. 
\end{align*}
Instead of the Hamiltonian of Eq.~\eqref{App Eq: new SOP Hamiltonian}, we use the Hamiltonian of Eq.~\eqref{App Eq: this work SOP Hamiltonian} to implement the Triangular LCU.
The resulting 1-norm is
\begin{align*}
    \lambda_\Delta &= \sum_{i} \norm{\hat {h}^{i}} + \sum_{i < j}\ \sum_{\alpha=0}^{\kappa^{\{i,j\}}-1} 2\left|\tilde\gamma^{\{i, j\}}_{\alpha}\right|\ \norm{\hat {\tilde{h}}^{i, \{i,j\}}_\alpha} \norm{\hat {\tilde{h}}^{j, \{i,j\}}_\alpha}\\ 
    &~+ \sum_{i < j < k}\ \sum_{\alpha=0}^{\kappa^{\{i,j,k\}}-1} 6\left|\tilde\gamma^{\{i, j, k\}}_{\alpha}\right|\ \norm{\hat {\tilde{h}}^{i, \{i,j,k\}}_\alpha} \norm{\hat {\tilde{h}}^{j, \{i,j,k\}}_\alpha} \norm{\hat {\tilde{h}}^{k, \{i,j,k\}}_\alpha}, 
\end{align*}
where $\norm{\cdot}$ denotes the induced 1-norm obtained by expanding the operator as a Pauli LCU.

To encode the full Hamiltonian, Ref.~\cite{majland2025fault} uses PREP-SEL primitives to explicitly encode one-mode operators, form their products, and finally their sums.
Note that a global PREP will be required when forming the final sum of operators to ensure correct rescaling of individual one-mode operators.
Though the details of the global PREP are not discussed in Ref.~\cite{majland2025fault}, its contribution to the total cost will be subdominant and hence will be ignored here.
In order to form the product of block encodings, the technique presented in Ref.~\cite{von2021quantum} is used.
Thus, the total number of Toffoli gates required to block-encode the Hamiltonian can simply be obtained as the cost of block-encoding a single one-mode operator times the number of one-mode operators in Eq.~\eqref{App Eq: this work SOP Hamiltonian} plus the cost of forming the products.

We now discuss the dominant contributors to the cost of block-encoding an arbitrary one-mode operator $\hat h$.
It can be expanded as $\hat h = \sum_{p,q}h_{p,q}\hat E^p_q$, where we have suppressed all except the modal indices for simplicity.
The qubit representation of the operator is given by
\begin{align*}
        \sum_{p,q=0}^{N-1} h_{p,q} \hat E^{p}_{q} &= \dfrac{1}{2} \sum_{p<q=0}^{N-1} h_{p,q} (\hat X_{p}\hat X_{q} + \hat Y_{p}\hat Y_{q}) + \dfrac{1}{2} \sum_{p=0}^{N-1} h_{p,p} (\hat {\mathbb{I}} - \hat Z_{p}).
\end{align*}

Since $h_{p,q} = h_{q,p}$, we only load $N(N+1)/2$ coefficients in the PREP, where $N$ is the number of modals.
In loading these values, it was reported in Ref.~\cite{majland2025fault} that ancilla-assisted QROM did not offer any benefit over the QROM of Ref.~\cite{linearT} for the PAH Hamiltonians.
Hence, the Toffoli cost of loading these values is equal to $N(N+1)/2-2$.
The cost of performing the comparator in coherent alias sampling is $\aleph$, where $\aleph$ is the number of bits used to approximate the $h_{p,q}$ values.
Note that this cost was assumed to be $2\aleph$ in Ref.~\cite{majland2025fault}.
Thus, the cost of PREP is dominated by
\[
C^{\text{PREP}(1)}_{\Delta} = \dfrac{N(N+1)}{2} + \aleph - 2,
\]
which is also the cost of PREP$^\dagger$, $C^{\text{PREP}^\dagger(1)}_{\Delta}$.
To perform SELECT, a controlled unary iteration over the $N(N+1)/2$ states was used.
Additionally, to distinguish between the application of $\hat X\hat X$ and $\hat Y \hat Y$, one needs to control on an ancilla along with the right branch of the unary iterator.
This does not appear to be accounted for in Ref.~\cite{majland2025fault}.
This additional control is only needed for $p<q$ values.
Hence, SELECT has a cost of 
\begin{align*}
    C^{\text{SEL}(1)}_{\Delta} &= \dfrac{N(N+1)}{2} + N(N-1) - 1 \\
    &= \dfrac{N(3N-1)}{2} - 1.
\end{align*}
Thus the total dominant cost of block-encoding a one-mode operator is
\begin{align*}
    C^{\text{BE(1)}} &= C^{\text{PREP}(1)}_{\Delta} + C^{\text{SEL}(1)}_{\Delta} + C^{\text{PREP}^\dagger(1)}_{\Delta} \\
    &= \dfrac{N(5N+1)}{2} + 2\aleph - 5.
\end{align*}
This cost ignores some of the logarithmic overheads.
While this cost was obtained by mainly following the procedure outlined in Ref.~\cite{majland2025fault}, in principle, it is possible to improve upon it by using the Pauli Sub-SELECT circuit of Fig.~\ref{Fig: Pauli Sub-SELECT} but without the outer mode-swap gates.

Another important contributor to the cost of block-encoding the whole Hamiltonian comes from controls used to form products of block encodings [see Fig. 1 of Ref.~\cite{majland2025fault}].
The qubits that need to be controlled are all the ancillas used when block-encoding a one-mode operator that remain entangled with the system qubits even after the application of PREP$^\dagger$.
This includes the $\aleph$ qubits used in coherent alias sampling.
Hence, the total Toffoli cost of block-encoding a two-mode-coupling Hamiltonian is dominated by
\begin{align*}
    C^{\text{BE}}_\mathrm{\Delta} &= \sum_{i=0}^{M-1}\left(\dfrac{N(5N+1)}{2} + 2\aleph - 5\right) + \sum_{i<j=0}^{M-1}\sum_{\alpha=0}^{\kappa^{\{i,j\}}-1}2\left(\dfrac{N(5N+1)}{2} + 3\aleph - 5\right) \\
    &= M\left(\dfrac{N(5N+1)}{2} + 2\aleph - 5\right) +M(M-1)\bar{\kappa}_2\left(\dfrac{N(5N+1)}{2} + 3\aleph - 5\right),
\end{align*}
where $\bar{\kappa}_2$ is the average of $\kappa^{\{i,j\}}$ for all $i<j$.
Treating $\aleph$ as a constant, the asymptotic Toffoli complexity of the Triangular LCU for a two-mode-coupling Hamiltonian is equal to $\mathcal{O}(\lambda_{\Delta}\bar{\kappa}_2 M^2N^2 /\epsilon)$.

Likewise, for a three-mode-coupling Hamiltonian, 
\begin{align*}
    C^{\text{BE}}_\mathrm{\Delta} &= M\left(\dfrac{N(5N+1)}{2} + 2\aleph - 5\right) + M(M-1)\bar{\kappa}_2\left(\dfrac{N(5N+1)}{2} + 3\aleph - 5\right)\\
    &\quad + \dfrac{M(M-1)(M-2)}{2}\bar{\kappa}_3\left(\dfrac{N(5N+1)}{2} + 3\aleph - 5\right),
\end{align*}
where $\bar{\kappa}_3$ is the average of $\kappa^{\{i,j,k\}}$ for all $i<j<k$.
Hence, for a Hamiltonian with a mode-coupling level equal to $n$, the asymptotic Toffoli complexity is given by $\mathcal{O}(\lambda_{\Delta}\bar{\kappa} M^{n}N^2/\epsilon)$, where $\bar\kappa = \max \{1,\bar\kappa_2, \dots, \bar\kappa_n\}$.
For both two- and three-mode-coupling Hamiltonians, the number of qubits required is dominated by the system qubits, and hence the asymptotic logical qubit cost scales as $\mathcal{O}(MN)$.

\section{Details of Pauli LCU} \label{App: Pauli LCU}
In this section, we provide details of the derivation of the Pauli LCU, along with the circuits used to block-encode it.
\subsection{Deriving Pauli LCU} \label{App: Deriving Pauli LCU}
The symmetries of the tensors in Eq.~\eqref{Eq: second-quantized symmetrized vibrational Hamiltonian} allow the Hamiltonian to be rewritten as
\begin{align*}
    \hat H & =\dfrac{1}{2} \sum_{i=0}^{M-1}\sum_{p,q=0}^{N-1} h_{i,p,q} (\hat E^{p_i}_{q_i} + \hat E^{q_i}_{p_i})  + \dfrac{1}{4}\sum_{i,j=0}^{M-1}\sum_{\substack{p,q,\\r,s}=0}^{N-1} g_{i,p,q,j,r,s} (\hat E^{p_i}_{q_i} + \hat E^{q_i}_{p_i})(\hat E^{r_j}_{s_j} + \hat E^{s_j}_{r_j}) \\ 
    &\quad+ \dfrac{1}{8}\sum_{i,j,k=0}^{M-1}\sum_{\substack{p,q,r,\\s,t,u}=0}^{N-1} v_{i,p,q,j,r,s,k,t,u} (\hat E^{p_i}_{q_i} + \hat E^{q_i}_{p_i})  (\hat E^{r_j}_{s_j} + \hat E^{s_j}_{r_j}) (\hat E^{t_k}_{u_k} + \hat E^{u_k}_{t_k}),
\end{align*}
where the Hermitian sum of excitation operators was decomposed into an LCU as
\begin{align*}
    \hat E^{p_i}_{q_i} + \hat E^{q_i}_{p_i} = \hat{\mathbb{I}} \delta_{p, q} + \dfrac{(-1)^{\delta_{p,q}}}{2}\sum_{b_1=0}^1 \hat Q_{i,p,q}^{(b_1)}.
\end{align*}
To illustrate the result of this substitution, consider the two-mode-coupling term of the Hamiltonian as an example.
\begin{align*}
    \sum_{i,j=0}^{M-1}\sum_{\substack{p,q,\\r,s}=0}^{N-1} & g_{i,p,q,j,r,s} (\hat E^{p_i}_{q_i} + \hat E^{q_i}_{p_i})(\hat E^{r_j}_{s_j} + \hat E^{s_j}_{r_j})  \\ 
    &= \sum_{i,j=0}^{M-1}\sum_{\substack{p,q,\\r,s}=0}^{N-1} g_{i,p,q,j,r,s} \left( \hat{\mathbb{I}} \delta_{p, q} + \dfrac{(-1)^{\delta_{p,q}}}{2}\sum_{b_1=0}^1 \hat Q_{i,p,q}^{(b_1)} \right) \left( \hat{\mathbb{I}} \delta_{r, s} + \dfrac{(-1)^{\delta_{r,s}}}{2}\sum_{b_2=0}^1 \hat Q_{j,r,s}^{(b_2)} \right) \\
    &= \sum_{i,j=0}^{M-1}\sum_{\substack{p,r}=0}^{N-1} g_{i,p,p,j,r,r} +  \sum_{i,j=0}^{M-1}\sum_{\substack{p,q,\\r,s}=0}^{N-1}  g_{i,p,q,j,r,s}\left( \sum_{b_1=0}^1 \dfrac{(-1)^{\delta_{p,q}}\delta_{r,s}}{2} \hat Q_{i,p,q}^{(b_1)} + \sum_{b_2=0}^1\dfrac{(-1)^{\delta_{r,s}}\delta_{p,q}}{2} \hat Q_{j,r,s}^{(b_2)}\right) \\
    &\quad + \sum_{i,j=0}^{M-1}\sum_{\substack{p,q,\\r,s}=0}^{N-1} \sum_{b_1,b_2=0}^1 \dfrac{(-1)^{\delta_{p,q} + \delta_{r,s}}g_{i,p,q,j,r,s}}{4}  \hat Q_{i,p,q}^{(b_1)} \hat Q_{j,r,s}^{(b_2)}.
\end{align*}
Most notably, the two terms that are linear in $\hat Q$ can be shown to be identical.
This follows from exchanging the dummy indices $\{i,p,q\}$ with $\{j,r,s\}$ and noticing that the $g$ tensor possesses the symmetry $g_{j,r,s,i,p,q} = g_{i,p,q,j,r,s}$.
Hence, the above relation reduces to
\begin{align*}
     \sum_{i,j=0}^{M-1} & \sum_{\substack{p,q,\\r,s}=0}^{N-1}  g_{i,p,q,j,r,s} (\hat E^{p_i}_{q_i} + \hat E^{q_i}_{p_i})(\hat E^{r_j}_{s_j} + \hat E^{s_j}_{r_j})  \\* 
     &= \sum_{i,j=0}^{M-1} \left( \sum_{\substack{p,r}=0}^{N-1} g_{i,p,p,j,r,r} +  \sum_{\substack{p,q,\\r}=0}^{N-1} \sum_{b_1=0}^1 (-1)^{\delta_{p,q}}g_{i,p,q,j,r,r} \hat Q_{i,p,q}^{(b_1)}  + \sum_{\substack{p,q,\\r,s}=0}^{N-1} \sum_{b_1,b_2=0}^1 \dfrac{(-1)^{\delta_{p,q}+\delta_{r,s}}g_{i,p,q,j,r,s}}{4}  \hat Q_{i,p,q}^{(b_1)} \hat Q_{j,r,s}^{(b_2)} \right).
\end{align*}
As shown, when the two-mode-coupling term is expanded in terms of Pauli unitaries, it generates new terms with a mode-coupling level of two or less.
The same principle applies to the one- and three-mode-coupling terms.
Grouping the $\hat Q$ unitaries based on their mode-coupling level thus yields new coefficient tensors and a global energy shift.
Because these new coefficients are constructed via partial summations over the original symmetric tensors, they inherit the same symmetries.
Specifically, they are symmetric under the exchange of composite mode-modal indices (e.g., swapping $\{i,p,q\}$ with $\{j,r,s\}$).
Since operators acting on different modes commute, this symmetry allows us to restrict the summation over the mode indices in the LCU.
Furthermore, because the tensors are invariant under the exchange of modal indices within a given mode, and since $\hat Q^{(b_1)}_{i,p,q} = \hat Q^{(b_1)}_{i,q,p}$, the summation over modal indices can be restricted to $p \leq q$.
These restrictions lead directly to the Hamiltonian in Eq.~\eqref{Eq: Pauli LCU}, with the new coefficients defined as
\begin{align*}
\begin{split}
   \bm {\mathrm C}_{\mathrm{Pauli}} &= \dfrac{1}{2}\sum_{i = 0}^{M-1}\sum_{p=0}^{N-1} h_{i,p,p} + \dfrac{1}{4}\sum_{i,j = 0}^{M-1} \sum_{p,r=0}^{N-1} g_{i,p,p,j,r,r} + \dfrac{1}{8}\sum_{i,j,k = 0}^{M-1} \sum_{p,r,t=0}^{N-1}  v_{i,p,p,j,r,r,k,t,t}, \\
    \bm{\mathrm{h}}_{i,p,q} &= \dfrac{1}{(-2)^{\delta_{p,q}}}\left(h_{i,p,q} + \sum_{j=0}^{M-1}\sum_{r=0}^{N-1} g_{i,p,q,j,r,r} + \dfrac{3}{4}\sum_{j,k = 0}^{M-1}\sum_{r,t=0}^{N-1}  v_{i,p,q,j,r,r,k,t,t} \right), \\
    \bm{\mathrm{g}}_{i,p,q,j,r,s} &= \dfrac{2}{(-2)^{\delta_{p,q}+\delta_{r,s}}}\left(g_{i,p,q,j,r,s} + \dfrac{3}{2}\sum_{k = 0}^{M-1}\sum_{t=0}^{N-1}  v_{i,p,q,j,r,s,k,t,t} \right), \\
    \bm{\mathrm{v}}_{i,p,q,j,r,s,k,t,u} &= \dfrac{6v_{i,p,q,j,r,s,k,t,u}}{(-2)^{\delta_{p,q}+\delta_{r,s}+\delta_{t,u}}}.
\end{split}
\end{align*}

\subsection{Complete cost of block-encoding Pauli LCU} 
\label{App: block encoding Pauli LCU}
For convenience, we rewrite the Pauli LCU of Eq.~\eqref{Eq: Pauli LCU} here.
\begin{align*}
    \begin{split}
    \hat H_{\mathrm{Pauli}} &= \bm {\mathrm C}_{\mathrm{Pauli}} + \dfrac{1}{2}\sum_{i=0}^{M-1} \sum_{p\leq q=0}^{N-1}\sum_{b_1=0}^1 \bm{\mathrm{h}}_{i,p,q} \hat Q_{i,p,q}^{(b_1)} + \dfrac{1}{4}\sum_{i<j=0}^{M-1}\sum_{\substack{p \leq q,\\r \leq s}=0}^{N-1} \sum_{b_1, b_2=0}^1 \bm{\mathrm{g}}_{i,p,q,j,r,s} \hat Q_{i,p,q}^{(b_1)} \hat Q_{j,r,s}^{(b_2)} \\ 
    &\quad+ \dfrac{1}{8}\sum_{i<j<k=0}^{M-1}\sum_{\substack{p\leq q,\\ r\leq s,\ t\leq u}=0}^{N-1} \sum_{b_1, b_2, b_3=0}^1 \bm{\mathrm{v}}_{i,p,q,j,r,s,k,t,u} \hat Q_{i,p,q}^{(b_1)} \hat Q_{j,r,s}^{(b_2)} \hat Q_{k,t,u}^{(b_3)}.
    \end{split}
\end{align*}
The constant term is ignored when block-encoding.
We use this three-mode-coupling Hamiltonian to draw PREP and SELECT circuits, but the costs are obtained for a general Hamiltonian of mode-coupling level $n$.
In the following, unless stated otherwise, QROM refers to the ancilla-assisted QROM of Ref.~\cite{sparse_LCU}.
The PREP circuit initializes a uniform index register of size $S$ \cite{thc}, where $S = \mathcal{O}\left((MN^2)^{n}\right)$ is the number of unique non-negligible coefficients, uses QROM to load the LCU coefficients, and then applies coherent alias sampling to extract the loaded values and realize the target LCU coefficients.
Assuming coefficients below a threshold are set to zero, the target PREP state for block-encoding a three-mode-coupling Hamiltonian is given by
\begin{align}
\begin{split}
\mathrm{PREP}\ket{0} &= \sum_{i=0}^{M-1}\ \sum_{p\leq q=0}^{N-1} \sqrt{\dfrac{|\bm {\mathrm h}_{i, p, q}|}{\lambda_{\mathrm{Pauli}}}}\ket{00}\ket{ipq\alpha}\ket{0000}\ket{0000}\ket{\theta}\ket{+}\ket{+}\ket{+}\ket{\mathrm{garb}} \\
    &\quad+ \sum_{i<j=0}^{M-1}\ \sum_{\substack{p \leq q, \\ r \leq s}=0}^{N-1}\sqrt{\dfrac{|\bm {\mathrm g}_{i, p, q, j, r, s}|}{\lambda_{\mathrm{Pauli}}}}\ket{01}\ket{ipq\alpha}\ket{jrs\beta}\ket{0000}\ket{\theta}\ket{+}\ket{+}\ket{+}\ket{\mathrm{garb}} \\
    &\quad+ \sum_{i<j<k=0}^{M-1}\ \sum_{\substack{p \leq q, \\ r \leq s,\ t \leq u}=0}^{N-1}\ \sqrt{\dfrac{|\bm {\mathrm v}_{i, p, q, j, r, s, k, t, u}|}{\lambda_{\mathrm{Pauli}}}}\ket{10}\ket{ipq\alpha}\ket{jrs\beta}\ket{ktu\gamma}\ket{\theta}\ket{+}\ket{+}\ket{+}\ket{\mathrm{garb}}.
\end{split}    \label{Eq: Pauli PREP state intermediate}
\end{align}
By expanding the qubits in the $\ket{+}$ state as $\ket{+} = (\ket{0} + \ket{1})/\sqrt{2}$, the PREP state can be rewritten as
\begin{align}
\begin{split}
    \mathrm{PREP}\ket{0} &= \sum_{i=0}^{M-1}\ \sum_{p\leq q=0}^{N-1}\sum_{b_1=0}^1\sqrt{\dfrac{|\bm {\mathrm h}_{i, p, q}|}{2\lambda_{\mathrm{Pauli}}}}\ket{00}\ket{ipq\alpha}\ket{0000}\ket{0000}\ket{\theta}\ket{b_1}\ket{+}\ket{+}\ket{\mathrm{garb}} \\
    &\quad+ \sum_{i<j=0}^{M-1}\ \sum_{\substack{p \leq q, \\ r \leq s}=0}^{N-1}\sum_{b_1,b_2=0}^1\sqrt{\dfrac{|\bm {\mathrm g}_{i, p, q, j, r, s}|}{4\lambda_{\mathrm{Pauli}}}}\ket{01}\ket{ipq\alpha}\ket{jrs\beta}\ket{0000}\ket{\theta}\ket{b_1}\ket{b_2}\ket{+}\ket{\mathrm{garb}} \\
    &\quad+ \sum_{i<j<k=0}^{M-1}\ \sum_{\substack{p \leq q, \\ r \leq s,\ t \leq u}=0}^{N-1} \sum_{b_1,b_2,b_3=0}^1\ \sqrt{\dfrac{|\bm {\mathrm v}_{i, p, q, j, r, s, k, t, u}|}{8\lambda_{\mathrm{Pauli}}}}\ket{10}\ket{ipq\alpha}\ket{jrs\beta}\ket{ktu\gamma}\ket{\theta}\ket{b_1}\ket{b_2}\ket{b_3}\ket{\mathrm{garb}}.
\end{split} \label{Eq: Pauli PREP state}
\end{align}

By comparing this with Eq.~\eqref{Eq: Pauli LCU}, we can see that the PREP state has the same coefficients as the Pauli LCU, up to the rescaling factor and the square roots which will be canceled on the application of PREP$^\dagger$.
The registers in Eq.~\eqref{Eq: Pauli PREP state} serve distinct functions.
The first two qubits enumerate the mode-coupling level, where the states $\ket{00}$, $\ket{01}$, and $\ket{10}$ correspond to one-, two-, and three-mode-coupling terms, respectively.
The next three multi-part registers are designed to distinguish terms based on their coupling level.
Consider the first of these registers, $\ket{ipq\alpha}$: the first sub-register encodes the mode index $i$, the next two encode the modal indices $p$ and $q$, and the final qubit $\alpha$ acts as a flag indicating whether the modal indices are identical (i.e., $\alpha = 1 \iff p = q$).
This $\alpha$ flag is later utilized by the SELECT circuit.
The qubit labeled $\theta$ encodes the sign of the coefficients, which depends on the mode and modal indices (this dependency is omitted in Eq.~\eqref{Eq: Pauli PREP state} for brevity).
The three subsequent qubits are used to distinguish the two Pauli operators defined in Eq.~\eqref{eq: Pauli Q_ipq}.
The value encoded by these qubits is used as a flag in the SELECT circuit to implement appropriate unitaries.
Note that the $b_1, b_2,$ and $b_3$ values are not loaded during QROM but rather realized by expanding the $\ket{+}$ states, which is evident from comparing Eq.~\eqref{Eq: Pauli PREP state} to Eq.~\eqref{Eq: Pauli PREP state intermediate}.
Thus, to realize the state in Eq.~\eqref{Eq: Pauli PREP state}, the loaded coefficients during QROM are those of Eq.~\eqref{Eq: Pauli PREP state intermediate} and not of Eq.~\eqref{Eq: Pauli PREP state}.
Finally, the garbage register, denoted as $\ket{\mathrm{garb}}$, contains all auxiliary qubits generated during state preparation that are not acted upon by the SELECT operation.
Due to this, it does not interfere with the block-encoded Hamiltonian.
To block-encode a Hamiltonian with different levels of mode coupling, we accordingly allocate or omit the extra registers required to encode the corresponding mode and modal indices.

\begin{figure}[!t]\centering    \includegraphics[width=0.7\linewidth]{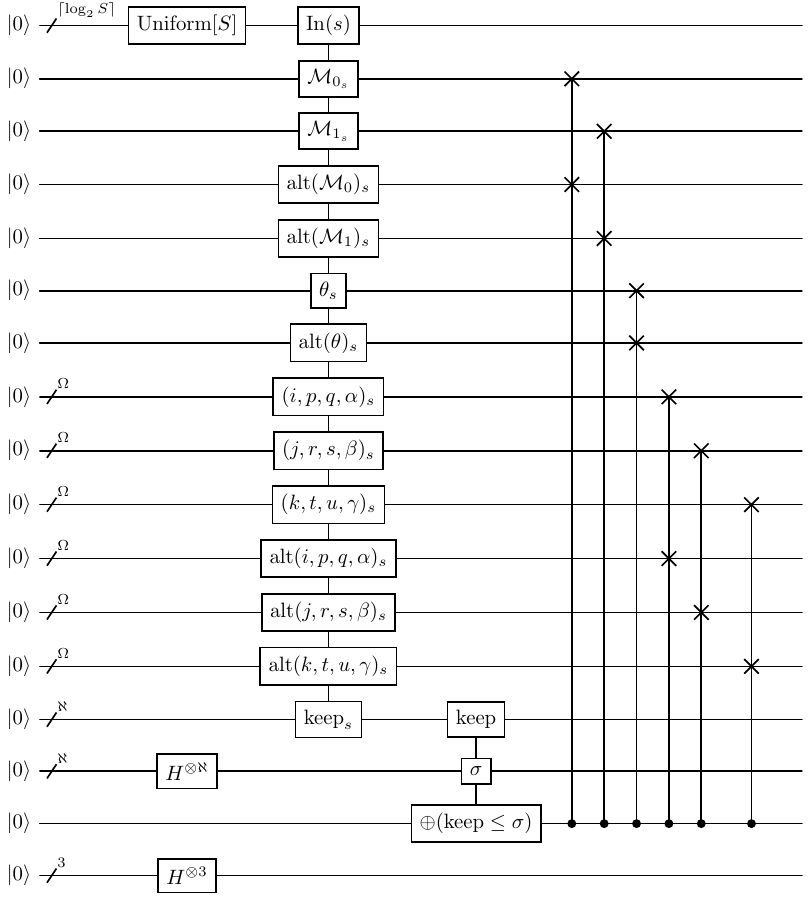}
\caption{Quantum circuit detailing the $\mathrm{PREP}$ operation for block-encoding a second-quantized vibrational Hamiltonian expressed as a Pauli LCU. 
The first register from the top is used to create a master index that acts as an address to load all the necessary indices. 
The next four qubits are used to load mode-coupling level flags and their alternate values. 
The $\theta$ and $\mathrm{alt}(\theta)$ qubits store the primary and alternate coefficient-sign bits used in alias sampling.
The six registers of size $\Omega$ are used to load tensor indices and their alternate values. 
$\Omega = \lceil\log_2 M\rceil + 2\lceil\log_2 N\rceil + 1$ accommodates a single mode index, two modal indices, and a Boolean flag denoting modal index equality, which is needed in the SELECT implementation. 
The parameter $\aleph$ represents the bit precision employed for the binary approximation of the coefficients. 
The three bottom qubits are prepared in $\ket{+}$ and subsequently provide the $b_1,b_2,$ and $b_3$ selector bits that choose between the Pauli operators of Eq.~\eqref{eq: Pauli Q_ipq} in SELECT.}\label{Fig: Pauli PREP}
\end{figure}

A circuit realizing the state in Eq.~\eqref{Eq: Pauli PREP state} is presented in Fig.~\ref{Fig: Pauli PREP}.
Below, we derive the Toffoli count associated with implementing the PREP circuit.
Although the circuit presented here corresponds to $n=3$, we will derive the cost for an arbitrary $n$.
\begin{itemize}
\item The initial step involves preparing a uniform superposition over the $S$ basis states.
Using the procedure of Ref.~\cite{thc}, this operation incurs a cost of $3\ceil{\log_2 S} - 3\eta + 2b_r - 9$, where $\eta$ represents the largest integer such that $2^{\eta}$ divides $S$, and $b_r$ denotes the number of precision bits used to specify the required rotation angles for the state preparation.
\item The subsequent step utilizes QROM to load the classical data.
The Toffoli cost of data loading depends on the availability of clean or dirty ancilla qubits.
Clean ancillas are strictly initialized in the $\ket{0}$ state, whereas dirty ancillas reside in an arbitrary, unknown state.
The cost of dirty-ancilla-assisted QROM is analyzed using the Select-Copy procedure of Ref.~\cite{halving_qrom}, whereas the cost of clean-ancilla-assisted QROM is analyzed using the theorems stated in Ref.~\cite{sparse_LCU}.
To determine these costs, we first define the QROM output size, $m$, which depends on the mode-coupling level $n$.
The circuit requires $2n$ copies of registers of size $\ceil{\log_2 M} + 2\ceil{\log_2 N} + 1$ each, two sets of $\lceil \log_2 n \rceil$ qubits to encode the mode-coupling level of the terms and their alternate values, two qubits to encode the sign and its alternate value, and an $\aleph$-bit register to store the $\mathrm{keep}$ probabilities.
Consequently, the total output size is
\begin{align}
m = 2n(\ceil{\log_2 M} + 2\ceil{\log_2 N} + 1) + 2\lceil \log_2 n \rceil + 2 + \aleph.\label{Eq: Pauli QROM volume}
\end{align} 
The Toffoli cost of QROM using clean ancillas is given by
\begin{align}
\mathrm{QRM}^{\mathrm T}_{\mathrm c}(S,m) = \left\lceil{\dfrac{S}{k_1}} \right\rceil + m(k_1 - 1), \label{Eq: QROM clean Toffoli cost}
\end{align}
 where $k_1$ is constrained to be a power of 2 such that $1 < k_1 < S$ and is chosen to minimize the Toffoli count subject to the availability of 
\begin{align}
    \mathrm{QRM}^{\mathrm N}_{\mathrm c}(S,m) = \left\lceil{\log_2\left(\dfrac{S}{k_1}\right)} \right\rceil + m(k_1 - 1) \label{Eq: QROM clean qubit cost}
\end{align}
clean ancillas.
The superscripts $\mathrm{T}$ and $\mathrm{N}$ in Eqs.~\eqref{Eq: QROM clean Toffoli cost} and \eqref{Eq: QROM clean qubit cost} denote the Toffoli and clean ancilla qubit counts, respectively. 
The Toffoli count is minimized when $k_1 \approx \sqrt{S/m}$.
Conversely, if dirty ancillas are utilized, the Toffoli cost according to the Select-Copy architecture is given by (see Theorem~1 of Ref.~\cite{halving_qrom})
\begin{align}
\mathrm{QRM}^{\mathrm T}_{\mathrm d}(S,m) = \left( \left\lceil \dfrac{m}{\rho} \right\rceil + 1 \right) \left( \left\lceil \dfrac{S}{k_1} \right\rceil + k_1 - 3 \right) + (k_1 - 1)\left( \rho \left(\left\lfloor  \dfrac{m}{\rho}\right\rfloor + 1\right) + m \bmod \rho  \right), 
\label{Eq: QROM dirty Toffoli cost}
\end{align}
where $k_1$ has the same constraints as above and $\rho$ is an integer satisfying $1\leq \rho \leq m$. $k_1$ and $\rho$ are chosen to minimize the Toffoli cost subject to the availability of $\rho(k_1 - 1)$ dirty qubits and
\begin{align}
    \mathrm{QRM}^{\mathrm N}_{\mathrm d}(S) = \max\left\{\left\lceil{\log_2\left(\dfrac{S}{k_1}\right)} \right\rceil,\ \log_2(k_1) \right\} \label{Eq: QROM dirty qubit cost}
\end{align}
clean ancillas.
\item The exact superposition over states with their target coefficients is realized via coherent alias sampling \cite{linearT}.
This necessitates an inequality test between registers of size $\aleph$, requiring exactly $\aleph$ Toffoli gates.
\item The Boolean output of this inequality test controls the swaps between the primary index registers and their alternate values.
This step requires $\lceil \log_2 n \rceil$ controlled swaps for mode-coupling qubits, $n$ sets of controlled swaps acting on registers of size $\ceil{\log_2 M} + 2\ceil{\log_2 N} + 1$, and one controlled swap between $\theta$ and $\mathrm{alt}(\theta)$.
Because each controlled swap on a single qubit incurs a cost of one Toffoli gate, the cumulative Toffoli cost for this step is $n(\ceil{\log_2 M} + 2\ceil{\log_2 N} + 1) + \lceil \log_2 n \rceil + 1$.
\end{itemize}
Combining these contributions, the total Toffoli cost of the PREP circuit, parameterized by the choice of clean or dirty (c/d) ancillas, can be expressed as
\begin{align}
\begin{split}
C^{\text{PREP}}_{\text{Pauli}} &= 3\ceil{\log_2 S} - 3\eta + 2b_r - 9 + \mathrm{QRM}^{\mathrm T}_{\mathrm c/\mathrm d}(S,m) + \aleph\\ 
&\quad+ n(\ceil{\log_2 M} + 2\ceil{\log_2 N} + 1) + \lceil \log_2 n \rceil + 1.   
\end{split}\label{Eq: Pauli PREP Toffoli cost}  
\end{align}

In principle, the $\mathrm{PREP}^\dagger$ operation can be implemented by strictly reversing the $\mathrm{PREP}$ circuit.
However, one can utilize the measurement-based uncomputation technique of Ref.~\cite{sparse_LCU} to uncompute the QROM lookup more efficiently.
By employing this approach, the Toffoli cost of clean-ancilla-assisted inverse QROM is reduced to
\begin{align}
\mathrm{QRM}^{\dagger \mathrm T}_{\mathrm c}(S) = \left\lceil{\dfrac{S}{k_2}} \right\rceil + k_2, \label{Eq: Inv QROM clean Toffoli cost}
\end{align}
where $k_2$ is a power of two such that $1<k_2<S$ and is chosen to minimize the Toffoli cost subject to the availability of 
\begin{align}
\mathrm{QRM}^{\dagger \mathrm N}_{\mathrm c}(S) = \left\lceil \log_2\left(\dfrac{S}{k_2}\right) \right\rceil + k_2 \label{Eq: Inv QROM clean qubit cost}
\end{align}
clean ancillas.  
Alternatively, using dirty ancillas for the inverse QROM gives a Toffoli cost of
\begin{align}
\mathrm{QRM}^{\dagger \mathrm T}_{\mathrm d}(S) = 2\left\lceil{\dfrac{S}{k_2}} \right\rceil + 4k_2, \label{Eq: Inv QROM dirty Toffoli cost}
\end{align}
where $k_2$ satisfies the same constraints as above and is chosen to minimize the Toffoli cost subject to the availability of  $k_2 - 1$ dirty ancillas and
\begin{align}
\mathrm{QRM}^{\dagger \mathrm N}_{\mathrm d}(S) = \left\lceil \log_2\left(\dfrac{S}{k_2}\right) \right\rceil + 1 \label{Eq: Inv QROM dirty qubit cost}
\end{align}
clean ancillas.
Here $k_2$ can be different from $k_1$ used by the QROM of the PREP circuit.
Apart from this substitution, the remaining Toffoli cost for $\mathrm{PREP}^\dagger$, denoted as $C^{\text{PREP}^\dagger}_{\text{Pauli}}$, is identical to that of the forward $\mathrm{PREP}$ circuit and is given as
\begin{align}
\begin{split}
C^{\text{PREP}^\dagger}_{\text{Pauli}} &= 3\ceil{\log_2 S} - 3\eta + 2b_r - 9 + \mathrm{QRM}^{\dagger \mathrm T}_{\mathrm c/\mathrm d}(S) + \aleph\\ 
&\quad+ n(\ceil{\log_2 M} + 2\ceil{\log_2 N} + 1) + \lceil \log_2 n \rceil + 1.   
\end{split}  
\label{Eq: Pauli PREP dagger Toffoli cost}  
\end{align}

The Toffoli complexities of the $\mathrm{PREP}$ and $\mathrm{PREP}^\dagger$ operations are asymptotically dominated by the QROM and inverse QROM subroutines, respectively.
In the absence of clean or dirty ancillas, these data-loading costs scale linearly with $S$, resulting in an overall complexity of $\mathcal{O}\left((MN^2)^{n}\right)$.
Conversely, with sufficient auxiliary qubits, the forward-QROM cost scales as $\mathcal{O}\left((MN^2)^{n/2}\sqrt{m}\right)$, whereas the inverse-QROM cost scales as $\mathcal{O}\left((MN^2)^{n/2}\right)$.
For fixed $n$ and precision parameters, suppressing the polylogarithmic factor arising from $\sqrt{m}$ allows both costs to be summarized as $\widetilde{\mathcal{O}}\left((MN^2)^{n/2}\right)$.

\begin{figure}[!t]\centering    \includegraphics[width=0.8\linewidth]{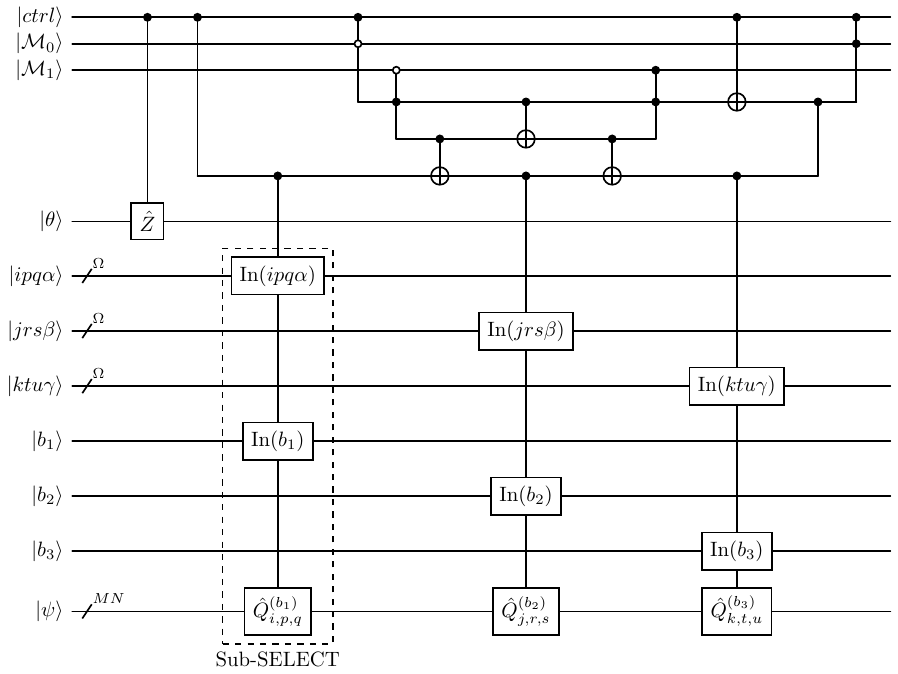}
\caption{Quantum circuit of the $\mathrm{SELECT}$ operation for block-encoding a second-quantized vibrational Hamiltonian expressed as a Pauli LCU. 
The controlled-$Z$ gate conditioned on the external control applies the phase associated with the sign of the selected LCU coefficient. 
A ranged operation involving mode-coupling level qubits $\ket{\mathcal{M}_0\mathcal{M}_1}$ governs the sequential activation of the Sub-SELECT primitives. 
Each activated primitive applies the corresponding indexed Pauli unitary $\hat{Q}$ defined in Eq.~\eqref{eq: Pauli Q_ipq} to the $MN$-qubit system register $\ket{\psi}$. 
The selector qubits $b_1,b_2,$ and $b_3$ correspond to the bottom three qubits of the PREP circuit (see Fig.~\ref{Fig: Pauli PREP}). 
The Sub-SELECT primitive is expanded in Fig.~\ref{Fig: Pauli Sub-SELECT}.}\label{Fig: Pauli SELECT}
\end{figure}

\begin{figure}[!t]\centering    \includegraphics[width=1\linewidth]{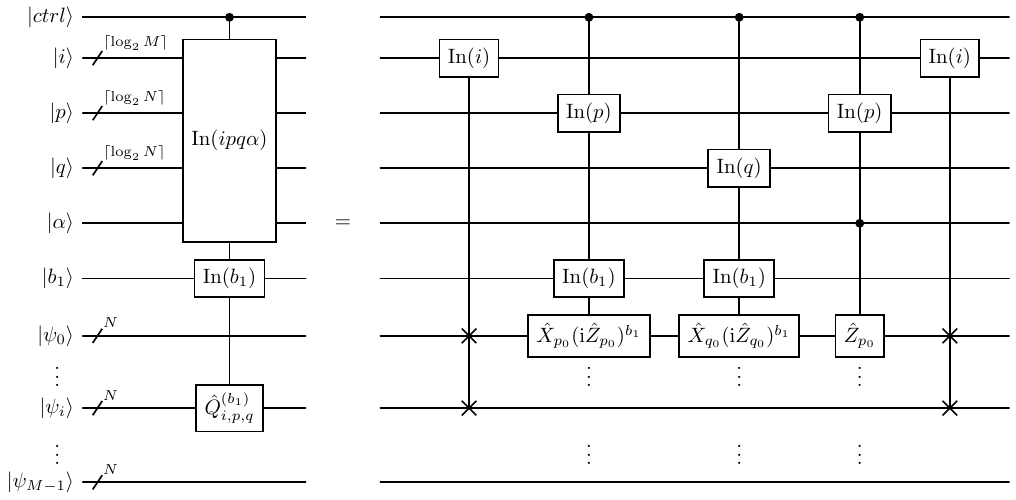}
\caption{Quantum circuit for the Pauli Sub-SELECT primitive used in the Pauli SELECT circuit of Fig.~\ref{Fig: Pauli SELECT}. 
The system qubits for mode $i$ are swapped with those for mode 0 at the beginning of the circuit, and the swap is undone at the end.
$\hat Q_{i,p,q}^{(b_1)}$ is implemented using separate unary iterators over $p$ and $q$, leading to a cost linear in the number of modals.
The qubit $\ket{b_1}$ is used to selectively implement $\hat X_{p_0} \hat X_{q_0}$ and $\hat Y_{p_0} \hat Y_{q_0}$. 
The equality flag $\alpha=1\iff p=q$ selects $Z_{p0}$ for the diagonal case.}\label{Fig: Pauli Sub-SELECT}
\end{figure}

We now detail the implementation of the $\mathrm{SELECT}$ oracle.
For a Pauli LCU, the construction of $\mathrm{SELECT}$ is conceptually straightforward: the objective is to apply the Pauli operators of the form $\hat Q_{i,p,q}^{(b_1)}$, defined in Eq.~\eqref{eq: Pauli Q_ipq}, conditioned on the indices $\{i,p,q,b_1\}$.
The corresponding quantum circuit for the $\mathrm{SELECT}$ operation is illustrated in Fig.~\ref{Fig: Pauli SELECT}.
To recover the correct sign in front of each term of the LCU, a controlled-$Z$ gate is applied in the SELECT circuit.
A primitive circuit, denoted as $\mathrm{Sub}$-$\mathrm{SELECT}$, is utilized to implement the controlled indexed unitaries $\hat Q$ of Eq.~\eqref{eq: Pauli Q_ipq}.
This primitive is repeated up to $n$ times, reflecting the maximum mode-coupling level of the Hamiltonian.
Because an individual LCU term may couple fewer than $n$ modes, we incorporate a ranged operation, as introduced in Ref.~\cite{linearT}, to ensure the sub-circuit is executed only as many times as required by the active term of the LCU.
This ranged operation introduces an additional Toffoli cost of $n - 1$.
Consequently, the total Toffoli cost of the complete $\mathrm{SELECT}$ circuit is expressed as
\[ 
C^{\text{SEL}}_{\text{Pauli}} = n C^{\text{Sub-SEL}}_{\text{Pauli}} + n - 1,
\]
where $C^{\text{Sub-SEL}}_{\text{Pauli}}$ represents the Toffoli cost of a single $\mathrm{Sub}$-$\mathrm{SELECT}$ primitive, which is expanded in Fig.~\ref{Fig: Pauli Sub-SELECT}.

The Sub-SELECT circuit implements the operator $\hat{Q}_{i,p,q}^{(b_1)}$ for an arbitrary mode $i$ by conjugating the operation $\hat{Q}_{0,p,q}^{(b_1)}$ with a block-swap network that exchanges the system qubits of the $i$-th mode with those of the zeroth mode.
To realize $\hat{Q}_{0,p,q}^{(b_1)}$, two independent unary iterations over the indices $p$ and $q$ are used to apply the two-qubit operators $\hat{X}_{p_0}\hat{X}_{q_0}$ or $\hat{Y}_{p_0}\hat{Y}_{q_0}$ conditioned on the value of $\ket{b_1}$.
Note that if $p=q$, this results in an identity operation.
To correct this, an additional unary iteration over $p$ indices controlled on $\ket{\alpha}$ is used to implement the $\hat Z_{p_0}$ operation.
The Toffoli complexity of the Sub-SELECT primitive can be obtained as follows.

\begin{itemize}
    \item Unary iteration over a mode index without an external control costs $M-2$ Toffolis.
    For each index of the iteration, we have controlled swaps over registers of size $N$ which cost $N$ Toffolis per index.
    There is no swap when the mode index is zero.
    Thus, the total cost is $(M-2) + (M-1)N = (M-1)(N+1)-1$.
    \item Controlled unary iteration over indices $p$ and $b_1$ costs $2N-1$ Toffolis.
    The controlled unary iteration over $q$ and $b_1$ also costs $2N-1$ Toffolis.
    \item Controlled unary iteration over $p$ with an additional control over $\alpha$ costs $N$ Toffolis.
    \item Finally, the controlled swap is identical to the first operation above and costs $(M-2) + (M-1)N = (M-1)(N+1)-1$ Toffolis.
\end{itemize}
Hence, the total cost of Sub-SELECT is
\begin{align*}
    C^{\text{Sub-SEL}}_{\text{Pauli}} &= 2((M-1)(N+1)-1) + 5N-2\\
    &= 2MN + 2M +3N - 6.
\end{align*}
This gives the total Toffoli count of SELECT as
\begin{align}
    C^{\text{SEL}}_{\text{Pauli}} = n(2MN + 2M +3N - 6) + n - 1. \label{Eq: Pauli SELECT Toffoli cost}
\end{align}

Thus, the cost of SELECT scales as $\mathcal{O}(nMN)$ and is asymptotically subleading to PREP for $n\geq2$.
The total cost of block-encoding the Hamiltonian is
\begin{align}
C^{\mathrm{BE}}_{\mathrm{Pauli}} = C^{\text{PREP}}_{\text{Pauli}} + C^{\text{SEL}}_{\text{Pauli}} + C^{\text{PREP}^\dagger}_{\text{Pauli}}. \label{Eq: Pauli BE Toffoli cost}
\end{align}
To build the walk operator after applying $\mathrm{PREP}^\dagger \cdot \mathrm{SELECT} \cdot \mathrm{PREP}$, one needs to reflect on all the qubits that are still entangled with the system qubits.
These qubits are as follows: $\ceil{\log_2 S}$ qubits storing the uniform superposition index, one qubit that stores the success of uniform superposition preparation, $\aleph$ qubits used in alias sampling, and $n$ qubits used to store $b_1, b_2, \dots$ values.
A reflection on these qubits has a cost of $\ceil{\log_2 S} + \aleph + n -1$.
We need one Toffoli gate to control SELECT on the successful preparation of the uniform superposition state and one Toffoli gate to control the reflection on the appropriate branch of the QPE unary iterator.
Therefore, the total Toffoli cost of performing quantum phase estimation using the Pauli LCU for a second-quantized vibrational Hamiltonian is given by
\begin{align}
    C_{\text{Pauli}} = \left\lceil\dfrac{\pi \lambda_{\mathrm{Pauli}}}{2\epsilon} \right \rceil \left(C^{\mathrm{BE}}_{\mathrm{Pauli}} + \ceil{\log_2 S} + \aleph + n + 1 \right),
    \label{Eq: Pauli LCU total Toffoli cost}
\end{align}
where $\epsilon$ is the required accuracy in the target energy and $\lambda_{\mathrm{Pauli}}$ is the 1-norm of the Pauli LCU, as given in Eq.~\eqref{Eq: Pauli LCU 1-norm}.
Hence, for optimal data-loading ancillas and fixed $n\ge 2$ and precision parameters, the asymptotic Toffoli cost of estimating the energy of a Hamiltonian expressed as a Pauli LCU scales as $\widetilde{\mathcal{O}}\left(\lambda_{\mathrm{Pauli}}(MN^2)^{n/2}/\epsilon\right)$.

We now evaluate the total qubit count for phase estimation using the Pauli LCU.
\begin{enumerate}
    \item The phase estimation requires a register of size $\lceil\log_2(\mathcal{I} + 1)\rceil$ where $\mathcal{I} = \left\lceil \dfrac{\pi\lambda_{\text{Pauli}}}{2\epsilon} \right \rceil$.
    \item A unary iterator on the phase estimation register requires $\lceil\log_2(\mathcal{I} + 1)\rceil - 1$ ancillas.
    \item Preparing the uniform superposition state requires $\ceil{\log_2(S)}$ qubits to store the uniform index, $1$ qubit for storing the success, $b_r$ qubits for the phase gradient state, and $1$ ancilla to perform the rotation on (see Appendix A of Ref.~\cite{thc}).
    Here $S$ is the number of terms in the LCU with unique coefficients.
    \item A register of size $m$ (defined in Eq.~\eqref{Eq: Pauli QROM volume}) is required to store the bit strings loaded by QROM.
    The QROM in PREP uses $\mathrm{QRM}^{\mathrm N}_{\mathrm c/ \mathrm d}(S,m)$ additional clean ancillas, as defined in Eqs.~\eqref{Eq: QROM clean qubit cost} and \eqref{Eq: QROM dirty qubit cost}, depending on whether clean- or dirty-ancilla-assisted QROM is employed.
    For notational convenience, we define $\mathrm{QRM}^{\mathrm N}_{\mathrm d}(S,m)\equiv \mathrm{QRM}^{\mathrm N}_{\mathrm d}(S)$, since the clean-ancilla requirement of the dirty-ancilla-assisted QROM is independent of the output size $m$.
    The inverse QROM in $\mathrm{PREP}^{\dagger}$ uses $\mathrm{QRM}^{\dagger \mathrm N}_{\mathrm c/\mathrm d}(S)$ additional clean ancillas, as defined in Eqs.~\eqref{Eq: Inv QROM clean qubit cost} and \eqref{Eq: Inv QROM dirty qubit cost}, depending on whether clean- or dirty-ancilla-assisted inverse QROM is employed.
    Because the clean-ancilla requirement of the inverse QROM is independent of the output size $m$, these ancillas can come from the QROM workspace or other clean qubits that are idle during uncomputation, without increasing the maximal logical qubit count.
    \item $\aleph$ qubits are required for storing the uniform superposition state $\sigma$ used in coherent alias sampling.
    Another qubit is needed to store the result of the inequality test in coherent alias sampling.
    \item $n$ qubits are needed to ensure Pauli $\hat X$ and $\hat Y$ are implemented in the SELECT circuit.
    \item To perform the ranged operation over the mode-coupling qubits in SELECT,
    QROM ancillas can be reused.
    \item Finally, the system wavefunction is stored in a register of size $MN$.
\end{enumerate}
In total, the logical qubit cost depending on whether clean or dirty (c/d) ancillas are utilized is given by
\begin{align}
\begin{split}
    N_{\text{Pauli}} &= 2\lceil\log_2(\mathcal{I} + 1)\rceil + \ceil{\log_2 S}+ b_r + m + \aleph + 2 + n + MN + \mathrm{QRM}^{\mathrm N}_{\mathrm c/\mathrm d}(S,m).
   \label{Eq: Pauli LCU total qubit cost}
\end{split}
\end{align}

For fixed $n\ge 2$ and precision parameters, if an optimal number of clean or dirty ancillas is available to minimize the Toffoli count, then the qubit cost scales as $\widetilde{\mathcal{O}}\left((MN^2)^{n/2}\right)$.
Equations \eqref{Eq: Pauli LCU total Toffoli cost} and \eqref{Eq: Pauli LCU total qubit cost} together determine the quantum resources required to estimate an eigenvalue of a vibrational Hamiltonian expressed as a Pauli LCU using QPE.
If the BLISS technique is employed, the same functional forms of the Toffoli and qubit counts can be used to estimate the resources but with a modified 1-norm and a potentially different $S$.

\section{Details of THC LCU} \label{App: THC LCU}
This section presents the derivation of the THC LCU together with the associated block-encoding circuits.
\subsection{Deriving THC LCU} \label{App: Deriving THC LCU}
As in the main text, all operator identities in this derivation are understood in terms of their action on the physical subspace with exactly one excitation per mode.
Recall that a three-mode-coupling THC factorized tensor has the form
\begin{align*}
    v_{i,p,q,j,r,s,k,t,u}^{\mathrm{THC}} &= \sum_{\mu_i = 0}^{R_i-1} \sum_{\nu_j = 0}^{R_j-1} \sum_{\omega_k = 0}^{R_k-1}  \Gamma^{(i,j,k)}_{\mu_i,\nu_j,\omega_k}\ \chi^{(i)}_{p,\mu_i} \chi^{(i)}_{q,\mu_i} \ \chi^{(j)}_{r,\nu_j} \chi^{(j)}_{s,\nu_j} \ \chi^{(k)}_{t,\omega_k} \chi^{(k)}_{u,\omega_k}.
\end{align*}
We look at the three-mode-coupling term of Eq.~\eqref{Eq: second-quantized symmetrized vibrational Hamiltonian} for a given set of modes $(i,j,k)$ when the tensor $v$ is approximated as above:
\begin{align*}
    \sum_{\substack{p,q,r,\\s,t,u}=0}^{N-1} v^{\mathrm{THC}}_{i,p,q,j,r,s,k,t,u} \hat E^{p_i}_{q_i}  \hat E^{r_j}_{s_j} \hat E^{t_k}_{u_k} &= \sum_{\substack{p,q,r,\\s,t,u}=0}^{N-1}\sum_{(\mu_i, \nu_j , \omega_k)= 0}^{(R_i-1, R_j-1, R_k-1)} \Gamma_{\mu_i,\nu_j,\omega_k}^{(i,j, k)}\ \chi_{p, \mu_i}^{(i)}\chi_{q, \mu_i}^{(i)}\chi_{r, \nu_j}^{(j)}\chi_{s, \nu_j}^{(j)} \chi_{t, \omega_k}^{(k)} \chi_{u, \omega_k}^{(k)}\  \hat E^{p_i}_{q_i} \hat E^{r_j}_{s_j} \hat E^{t_k}_{u_k}. 
\end{align*}
Using the fact that summation over modal indices has been mutually decoupled and the relation
\[\sum_{p,q=0}^{N-1} \chi^{(i)}_{p,\mu_i} \chi^{(i)}_{q,\mu_i} \hat a_{p_i}^\dagger \hat a_{q_i} = \dfrac{1}{2}\left(\hat{\mathbb{I}} - \hat V^{(i)}_{\mu_i} \hat Z_{0_i} \hat V^{(i)^\dagger}_{\mu_i} \right),\] 
which has been detailed in the main text, we can rewrite the three-mode-coupling term as
\begin{align*}
    \sum_{\substack{p,q,r,\\s,t,u}=0}^{N-1} v^{\mathrm{THC}}_{i,p,q,j,r,s,k,t,u} \hat E^{p_i}_{q_i}  \hat E^{r_j}_{s_j} \hat E^{t_k}_{u_k} & \\ = \dfrac{1}{8} \sum_{(\mu_i, \nu_j , \omega_k)= 0}^{(R_i-1, R_j-1, R_k-1)} &\Gamma_{\mu_i,\nu_j,\omega_k}^{(i,j, k)}\  \left(\hat{\mathbb{I}} - \hat V^{(i)}_{\mu_i}\hat Z_{0_i} \hat V^{(i)^\dagger}_{\mu_i} \right)  \left(\hat{\mathbb{I}} - \hat V^{(j)}_{\nu_j} \hat Z_{0_j}\hat V^{(j)^\dagger}_{\nu_j} \right)  \left(\hat{\mathbb{I}} - \hat V^{(k)}_{\omega_k} \hat Z_{0_k}\hat V^{(k)^\dagger}_{\omega_k} \right).
\end{align*}
We multiply the terms across the parentheses and substitute in place of the three-mode-coupling term in the Hamiltonian of Eq.~\eqref{Eq: second-quantized symmetrized vibrational Hamiltonian} to get
\begin{align}
    \begin{split}
    \hat H &\approx \tilde C + \sum_{i=0}^{M-1}\sum_{p,q=0}^{N-1} \tilde h_{i,p,q}\hat E^{p_i}_{q_i}  + \sum_{i,j=0}^{M-1}\sum_{\substack{p,q,\\r,s}=0}^{N-1} \tilde g_{i,p,q,j,r,s} \hat E^{p_i}_{q_i}  \hat E^{r_j}_{s_j}\\
    &~+ \dfrac{1}{8}\sum_{i<j<k=0}^{M-1} \sum_{(\mu_i, \nu_j , \omega_k)= 0}^{(R_i-1, R_j-1, R_k-1)}  \bm{\Gamma}_{\mu_i,\nu_j,\omega_k}^{(i,j, k)}\ \hat V^{(i)}_{\mu_i} \hat Z_{0_i}\hat V^{(i)^\dagger}_{\mu_i} \hat V^{(j)}_{\nu_j} \hat Z_{0_j}\hat V^{(j)^\dagger}_{\nu_j} \hat V^{(k)}_{\omega_k} \hat Z_{0_k} \hat V^{(k)^\dagger}_{\omega_k}, \label{Eq: THC LCU intermediate 1}
    \end{split}
\end{align}
where the tensors are redefined as 
\begin{align}
\begin{split}
\tilde C &= \dfrac{1}{8}\sum_{i,j,k=0}^{M-1} \sum_{p,r,t=0}^{N-1}  v^{\mathrm{THC}}_{i,p,p,j,r,r,k,t,t}, \\
    \tilde h_{i,p,q} &= h_{i,p,q} - \dfrac{3}{4} \sum_{j,k=0}^{M-1} \sum_{r,t=0}^{N-1} v^{\mathrm{THC}}_{i,p,q,j,r,r,k,t,t}, \\
    \tilde g_{i,p,q,j,r,s} &= g_{i,p,q,j,r,s} + \dfrac{3}{2} \sum_{k=0}^{M-1} \sum_{t=0}^{N-1} v^{\mathrm{THC}}_{i,p,q,j,r,s,k,t,t}, \\
    \bm{\Gamma}_{\mu_i,\nu_j,\omega_k}^{(i,j, k)} &= -6\Gamma_{\mu_i,\nu_j,\omega_k}^{(i,j, k)}.
\end{split} \label{Eq: THC tensor resum 1}
\end{align}
The multiplier of six in the definition of $\bm{\Gamma}$ comes from restricting the sum over mode indices in Eq.~\eqref{Eq: THC LCU intermediate 1}.
The next step is to obtain a THC factorization of the two-mode-coupling tensor $\tilde g$.
In analogy with Eq.~\eqref{eq: THC three mode optimization problem}, we solve
\begin{mini}
  {\zeta, \xi}{{\norm{\tilde g - \sum_{\mu_i = 0}^{R_i-1} \sum_{\nu_j = 0}^{R_j-1}  \zeta^{(i,j)}_{\mu_i,\nu_j}\ \xi^{(i)}_{p,\mu_i} \xi^{(i)}_{q,\mu_i} \ \xi^{(j)}_{r,\nu_j} \xi^{(j)}_{s,\nu_j}}_\mathrm{F}}}           % Optimization variable and objective
  {\label{eq: THC two mode optimization problem}}{}  % Label and optional result
  \addConstraint{\sum_{p=0}^{N-1}|\xi^{(i)}_{p,\mu_i}|^2}{= 1\ \ \forall\ i,\mu_i}
  \addConstraint{\zeta^{(i,j)}_{\mu_i,\nu_j}}{=\zeta^{(j,i)}_{\nu_j,\mu_i}},
\end{mini}
and the THC-approximated tensor is defined as
\begin{align*}
    \tilde{g}_{i,p,q,j,r,s}^{\mathrm{THC}} &= \sum_{\mu_i = 0}^{R_i-1}\sum_{\nu_j = 0}^{R_j-1} \zeta^{(i,j)}_{\mu_i,\nu_j}\ \xi^{(i)}_{p,\mu_i} \xi^{(i)}_{q,\mu_i}\ \xi^{(j)}_{r,\nu_j} \xi^{(j)}_{s,\nu_j}.
\end{align*}
Following the workflow for three-mode-coupling THC,  we define number-conserving unitaries $\hat U^{(i)}_{\mu_i}$ such that 
\begin{align*}
\sum_{p,q=0}^{N-1}\xi^{(i)}_{p,\mu_i}\xi^{(i)}_{q,\mu_i}\hat a_{p_i}^\dagger \hat a_{q_i} = \hat U^{(i)}_{\mu_i} \hat{\tilde{n}}_{0_i} \hat U^{(i)^\dagger}_{\mu_i},
\end{align*}
leading to the Hamiltonian
\begin{align}
    \begin{split}
    \hat H &\approx \tilde{\tilde C} + \sum_{i=0}^{M-1}\sum_{p,q=0}^{N-1} \tilde{\tilde h}_{i,p,q}\hat E^{p_i}_{q_i}  + \dfrac{1}{4}\sum_{i<j=0}^{M-1}\sum_{(\mu_i, \nu_j)= 0}^{(R_i-1, R_j-1)} \bm{\zeta}_{\mu_i,\nu_j}^{(i,j)}\ \hat U^{(i)}_{\mu_i} \hat Z_{0_i}\hat U^{(i)^\dagger}_{\mu_i} \hat U^{(j)}_{\nu_j} \hat Z_{0_j}\hat U^{(j)^\dagger}_{\nu_j}\\
    &~+ \dfrac{1}{8}\sum_{i<j<k=0}^{M-1} \sum_{(\mu_i, \nu_j , \omega_k)= 0}^{(R_i-1, R_j-1, R_k-1)} \bm{\Gamma}_{\mu_i,\nu_j,\omega_k}^{(i,j, k)}\ \hat V^{(i)}_{\mu_i} \hat Z_{0_i}\hat V^{(i)^\dagger}_{\mu_i} \hat V^{(j)}_{\nu_j} \hat Z_{0_j}\hat V^{(j)^\dagger}_{\nu_j} \hat V^{(k)}_{\omega_k} \hat Z_{0_k} \hat V^{(k)^\dagger}_{\omega_k}, \label{Eq: THC LCU intermediate 2}
    \end{split}
\end{align}
where the new coefficients are given by
\begin{align}
\begin{split}
    \tilde{\tilde C} &= \tilde C - \dfrac{1}{4}   \sum_{i,j=0}^{M-1}\sum_{p,r=0}^{N-1} \tilde{g}^{\mathrm{THC}}_{i,p,p,j,r,r}, \\
    \tilde{\tilde h}_{i,p,q} &= \tilde h_{i,p,q} + \sum_{j=0}^{M-1}\sum_{r=0}^{N-1} \tilde{g}^{\mathrm{THC}}_{i,p,q,j,r,r}, \\
    \bm{\zeta}_{\mu_i,\nu_j}^{(i,j)} &= 2 \zeta_{\mu_i,\nu_j}^{(i,j)}.
\end{split} \label{Eq: THC tensor resum 2}
\end{align}
Finally, the one-mode-coupling operator must also be expressed in diagonal form.
We write the eigendecomposition as \(\tilde{\tilde h}_{i,p,q} = \sum_{\mu_i=0}^{N-1} \eta^{(i)}_{\mu_i}\mathcal W^{(i)}_{p,\mu_i}\mathcal W^{(i)}_{q,\mu_i}\) and define the modal rotation unitary \(\hat W^{(i)}_{\mu_i}\) associated with column \(\mu_i\) of \(\mathcal W^{(i)}\). 
The resulting Hamiltonian is
\begin{align*}
\begin{split}
    \hat H_{\mathrm{THC}} &=  \bm{\mathrm{C}}_\mathrm{THC} + \dfrac{1}{2}\sum_{i=0}^{M-1}\sum_{\mu_i=0}^{N-1} \bm \eta^{(i)}_{\mu_i} \hat W^{(i)}_{\mu_i} \hat Z_{0_i} \hat W^{(i)^\dagger}_{\mu_i}  + \dfrac{1}{4}\sum_{i<j=0}^{M-1}\sum_{(\mu_i, \nu_j)= 0}^{(R_i-1, R_j-1)} \bm{\zeta}_{\mu_i,\nu_j}^{(i,j)}\ \hat U^{(i)}_{\mu_i} \hat Z_{0_i}\hat U^{(i)^\dagger}_{\mu_i} \hat U^{(j)}_{\nu_j} \hat Z_{0_j}\hat U^{(j)^\dagger}_{\nu_j}\\
    &~+ \dfrac{1}{8}\sum_{i<j<k=0}^{M-1} \sum_{(\mu_i, \nu_j , \omega_k)= 0}^{(R_i-1, R_j-1, R_k-1)}  \bm{\Gamma}_{\mu_i,\nu_j,\omega_k}^{(i,j, k)}\ \hat V^{(i)}_{\mu_i} \hat Z_{0_i}\hat V^{(i)^\dagger}_{\mu_i} \hat V^{(j)}_{\nu_j} \hat Z_{0_j}\hat V^{(j)^\dagger}_{\nu_j} \hat V^{(k)}_{\omega_k} \hat Z_{0_k} \hat V^{(k)^\dagger}_{\omega_k}, 
\end{split}
\end{align*}
where $\bm{\mathrm{C}}_\mathrm{THC} = \tilde {\tilde C} + \sum_{i=0}^{M-1}\sum_{\mu_i=0}^{N-1} \eta^{(i)}_{\mu_i}/2$ and $\bm \eta^{(i)}_{\mu_i} = - \eta^{(i)}_{\mu_i}$.
This recovers the THC LCU of Eq.~\eqref{Eq: THC LCU}, whose 1-norm is given in Eq.~\eqref{Eq: THC 1-norm}.
To prevent THC fitting errors from entering the lower-order tensor redefinitions directly, the contractions in Eqs.~\eqref{Eq: THC tensor resum 1} and \eqref{Eq: THC tensor resum 2} may instead be evaluated using the unfactorized tensors $v$ and $\tilde g$, respectively, rather than their THC approximations.
We adopt this prescription here, following the analogous treatment of the one-body correction in Ref.~\cite{thc}.

\subsection{Initializing THC factorization} \label{App: Initializing THC}
Here we derive the initializations used to obtain the THC LCUs for the PAH Hamiltonians.
To obtain a warm start for a three-mode-coupling Hamiltonian, we start by reshaping $v_{i,p,q,j,r,s,k,t,u}$ to a symmetric order-three tensor $V_{(i,p,q),(j,r,s),(k,t,u)}$ and perform a CP3 factorization as 
\begin{align}
    V_{(i,p,q),(j,r,s),(k,t,u)} \approx \sum_\alpha \mathcal V_{(i,p,q),\alpha}\mathcal V_{(j,r,s),\alpha}\mathcal V_{(k,t,u),\alpha}, \label{Eq: trbt symmetric CP3}
\end{align}
where we treat each of $(i,p,q)$, $(j,r,s)$ and $(k,t,u)$ as a single composite index.
This decouples the mode degrees of freedom, which allows us to perform a second CP3 factorization on individual modes as
\begin{align}
     \mathcal V_{(i,p,q),\alpha} \approx \sum_{\mu_i = 0}^{R_i-1}\chi^{(i)}_{p,\mu_i}\chi^{(i)}_{q,\mu_i}D^{(i)}_{\alpha, \mu_i}, \label{Eq: 2nd trbt CP3}
\end{align}
where $\chi$ is subjected to the same constraint as in Eq.~\eqref{eq: THC three mode optimization problem}.
Substituting this into Eq.~\eqref{Eq: trbt symmetric CP3} leads to the desired THC initialization:
\begin{align*}
        v_{i,p,q,j,r,s,k,t,u}^{\mathrm{THC}} = \sum_{\mu_i = 0}^{R_i-1} \sum_{\nu_j = 0}^{R_j-1} \sum_{\omega_k = 0}^{R_k-1}  
        \Big( \sum_\alpha D^{(i)}_{\alpha, \mu_i} D^{(j)}_{\alpha, \nu_j} D^{(k)}_{\alpha, \omega_k} \Big)\ \chi^{(i)}_{p,\mu_i} \chi^{(i)}_{q,\mu_i} \ \chi^{(j)}_{r,\nu_j} \chi^{(j)}_{s,\nu_j} \ \chi^{(k)}_{t,\omega_k} \chi^{(k)}_{u,\omega_k},
\end{align*}
where we identify $\Gamma^{(i,j,k)}_{\mu_i,\nu_j,\omega_k} = \sum_\alpha D^{(i)}_{\alpha, \mu_i} D^{(j)}_{\alpha, \nu_j} D^{(k)}_{\alpha, \omega_k}$.\\

The initialization for a two-mode-coupling tensor is more straightforward.
We start with the tensor $g_{i,p,q,j,r,s}$ and reshape it to a symmetric matrix $G_{(i,p,q),(j,r,s)}$.
Since $G$ is symmetric, it admits an eigendecomposition 
\[
G_{(i,p,q),(j,r,s)} = \sum_\alpha \sigma_\alpha \mathcal{G}_{(i,p,q),\alpha} \mathcal{G}_{(j,r,s),\alpha}.
\]
This decouples the mode indices, allowing us to perform a CP3 factorization for each mode as
\begin{align}
    \mathcal G_{(i,p,q),\alpha} \approx \sum_{\mu_i = 0}^{R_i-1}\xi^{(i)}_{p,\mu_i}\xi^{(i)}_{q,\mu_i}\bar D^{(i)}_{\alpha, \mu_i}. \label{Eq: tbt CP3}
\end{align}
Substituting this back into the eigendecomposition leads to the THC initialization:
\[
\tilde{g}_{i,p,q,j,r,s}^{\mathrm{THC}} = \sum_{\mu_i = 0}^{R_i-1} \sum_{\nu_j = 0}^{R_j-1}
\left(
\sum_\alpha \sigma_\alpha  \bar D^{(i)}_{\alpha, \mu_i} \bar D^{(j)}_{\alpha, \nu_j}\right) \xi^{(i)}_{p,\mu_i}\xi^{(i)}_{q,\mu_i} \xi^{(j)}_{r,\nu_j}\xi^{(j)}_{s,\nu_j},
\]
where we identify $\zeta^{(i,j)}_{\mu_i,\nu_j} = \sum_\alpha \sigma_\alpha  \bar D^{(i)}_{\alpha, \mu_i} \bar D^{(j)}_{\alpha, \nu_j}$.
When the number of modes is large and the THC optimization becomes computationally expensive, it can be more efficient to invest additional effort in obtaining a high-quality CP3 initialization and then perform only a few THC optimization iterations. 
For all systems analyzed in this work, the THC ranks were determined by performing multiple sweeps over increasing THC ranks in the CP3 decompositions of Eqs.~\eqref{Eq: 2nd trbt CP3} and \eqref{Eq: tbt CP3}.
Because the number of modals in the three-mode-coupling systems considered here was small, no additional THC optimization was necessary to achieve the target tensor accuracy of $1\,\mathrm{cm}^{-1}$.

\subsection{Complete cost of block-encoding THC LCU} \label{App: block encoding THC LCU}
\begin{figure}[!t]\centering    
    \includegraphics[width=0.7\linewidth]{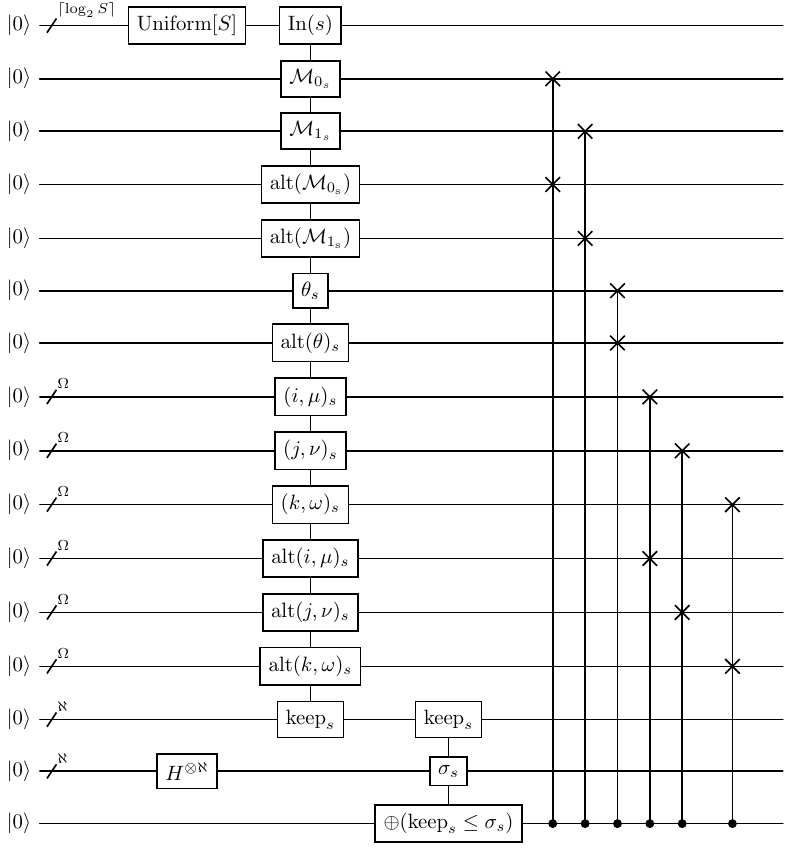}
    \caption{Quantum circuit detailing the $\mathrm{PREP}$ operation for block-encoding a second-quantized vibrational Hamiltonian expressed as a THC LCU. 
    As in the Pauli LCU, the first register from the top is used to create a uniform superposition over basis states that act as addresses to load all the necessary indices. 
    The next four qubits are used to load mode-coupling level flags and their alternate values. 
    The $\theta$ and $\mathrm{alt}(\theta)$ qubits store the primary and alternate coefficient-sign bits used in alias sampling.
    The six registers of size $\Omega$ are used to load THC factor indices and their alternate values. 
    Here, $\Omega = \lceil\log_2 M\rceil + \lceil\log_2 R_{\max}\rceil$ accommodates a single mode index and a single THC factor index, where $R_{\max}=\max\{N, \max_iR_i\}$.
    The parameter $\aleph$ represents the bit precision employed for the binary approximation of the coefficients.}\label{Fig: THC PREP}
\end{figure}

Consider the three-mode-coupling Hamiltonian of Eq.~\eqref{Eq: THC LCU} expressed as a THC LCU, which we rewrite below for convenience. 
\begin{align*}
    \hat H_{\mathrm{THC}} &=  \bm{\mathrm{C}} + \dfrac{1}{2}\sum_{i=0}^{M-1}\sum_{\mu_i=0}^{N-1} \bm \eta^{(i)}_{\mu_i} \hat W^{(i)}_{\mu_i} \hat Z_{0_i} \hat W^{(i)^\dagger}_{\mu_i}  + \dfrac{1}{4}\sum_{i<j=0}^{M-1}\sum_{(\mu_i, \nu_j)= 0}^{(R_i-1, R_j-1)} \bm{\zeta}_{\mu_i,\nu_j}^{(i,j)}\ \hat U^{(i)}_{\mu_i} \hat Z_{0_i}\hat U^{(i)^\dagger}_{\mu_i} \hat U^{(j)}_{\nu_j} \hat Z_{0_j}\hat U^{(j)^\dagger}_{\nu_j}\\
    &~+ \dfrac{1}{8}\sum_{i<j<k=0}^{M-1} \sum_{(\mu_i, \nu_j , \omega_k)= 0}^{(R_i-1, R_j-1, R_k-1)}  \bm{\Gamma}_{\mu_i,\nu_j,\omega_k}^{(i,j, k)}\ \hat V^{(i)}_{\mu_i} \hat Z_{0_i}\hat V^{(i)^\dagger}_{\mu_i} \hat V^{(j)}_{\nu_j} \hat Z_{0_j}\hat V^{(j)^\dagger}_{\nu_j} \hat V^{(k)}_{\omega_k} \hat Z_{0_k} \hat V^{(k)^\dagger}_{\omega_k}.
\end{align*}
The constant term can be ignored when block-encoding.
Unlike the Pauli LCU, the THC LCU depends on basis rotations in the modal subspace.
This requires Givens rotations in the SELECT circuit.
However, the idea of the PREP circuit remains the same as in the Pauli LCU.
Assuming that the average THC rank scales as $\mathcal O(N)$, we first prepare a uniform superposition over all possible unique coefficients $S = \mathcal{O}\left((MN)^{n}\right)$, use QROM to load the indices and corresponding coefficient values, and perform alias sampling to extract the coefficients.
Here as well, QROM refers to the ancilla-assisted constructions of Refs.~\cite{sparse_LCU,halving_qrom}.
We present the THC PREP circuit in Fig.~\ref{Fig: THC PREP}.

The state prepared by the PREP circuit can be given as
\begin{align}
\begin{split}
    \mathrm{PREP}\ket{0} &= \sum_{i=0}^{M-1}\ \sum_{\mu_i=0}^{N-1}\sqrt{\dfrac{|\bm \eta_{\mu_i}^{(i)}|}{2\lambda_{\mathrm{THC}}}}\ket{00}\ket{i\mu_i}\ket{00}\ket{00}\ket{\theta}\ket{\mathrm{garb}} \\
    &\quad+ \sum_{i<j=0}^{M-1}\ \sum_{(\mu_i, \nu_j)=0}^{(R_i-1, R_j-1)}\sqrt{\dfrac{|\bm \zeta_{\mu_i, \nu_j}^{(i,j)}|}{4\lambda_{\mathrm{THC}}}}\ket{01}\ket{i\mu_i}\ket{j\nu_j}\ket{00}\ket{\theta}\ket{\mathrm{garb}} \\
    &\quad+ \sum_{i<j<k=0}^{M-1}\ \sum_{(\mu_i, \nu_j, \omega_k)=0}^{(R_i-1, R_j-1, R_k-1)}\sqrt{\dfrac{|\bm \Gamma_{\mu_i, \nu_j, \omega_k}^{(i,j,k)}|}{8\lambda_{\mathrm{THC}}}}\ket{10}\ket{i\mu_i}\ket{j\nu_j}\ket{k\omega_k}\ket{\theta}\ket{\mathrm{garb}}.
\end{split} \label{Eq: THC PREP state}
\end{align}
The first two qubits flag the mode-coupling level of each term in the LCU.
The next three registers encode the mode and THC factor indices.
Although not shown for brevity, the $\theta$ register depends on the mode and factor indices and encodes the sign of the target THC coefficient.
Similarly, the register $\ket{\mathrm{garb}}$ is also entangled with the index qubits and represents all other qubits used in achieving this superposition state.
Since the SELECT operation does not depend on the garbage register, it does not interfere with the block-encoded Hamiltonian.
The Toffoli cost of implementing the PREP circuit can be obtained as follows.

\begin{itemize}
\item The circuit begins by preparing a uniform superposition over the $S$ basis states.
This operation incurs a cost of $3\ceil{\log_2 S} - 3\eta + 2b_r - 9$ \cite{thc}, where $\eta$ represents the largest integer such that $2^{\eta}$ divides $S$, and $b_r$ denotes the number of precision bits used to specify the required rotation angles for the state preparation.
\item All the indices and flags of the LCU are then loaded using QROM.
The cost of QROM with clean ancillas is given by $\mathrm{QRM}^{\mathrm T}_{\mathrm c}(S,m)$ defined in Eq.~\eqref{Eq: QROM clean Toffoli cost}, and QROM with dirty ancillas is given by $\mathrm{QRM}^{\mathrm T}_{\mathrm d}(S,m)$ defined in Eq.~\eqref{Eq: QROM dirty Toffoli cost}.
Here $m$ is defined as
\begin{align}
m = 2n(\ceil{\log_2 M} + \ceil{\log_2 R_{\max}}) + 2\lceil \log_2 n \rceil + 2 + \aleph, \label{Eq: THC QROM volume}
\end{align}
where $R_{\max}=\max\{N,\max_i R_i\}$. 
In practice, the maximum THC rank is at least $N$, so $R_{\max}=\max_i R_i$.
\item The target LCU coefficients are extracted from the loaded keep probabilities via coherent alias sampling \cite{linearT}.
This requires an inequality test between registers of size $\aleph$, leading to $\aleph$ Toffoli gates.
\item The result of the inequality test is then used to control the swaps between the primary index registers and their alternate values.
The Toffoli cost of performing these swaps equals $n(\ceil{\log_2 M} + \ceil{\log_2 R_{\max}}) + \lceil \log_2 n \rceil + 1$. 
\end{itemize}
In total, the Toffoli cost of the THC PREP circuit, parameterized by the choice of clean or dirty (c/d) ancillas, is given by
\begin{align}
\begin{split}
C^{\text{PREP}}_{\mathrm{THC}} &= 3\ceil{\log_2 S} - 3\eta + 2b_r - 9 + \mathrm{QRM}^{\mathrm T}_{\mathrm c/\mathrm d}(S,m) + \aleph\\ 
&\quad + n(\ceil{\log_2 M} + \ceil{\log_2 R_{\max}}) + \lceil \log_2 n \rceil + 1.
\end{split}\label{Eq: THC PREP Toffoli cost}  
\end{align}

Using the inverse QROM costs of Eqs.~\eqref{Eq: Inv QROM clean Toffoli cost} and \eqref{Eq: Inv QROM dirty Toffoli cost}, the cost of PREP$^\dagger$ can be given as
\begin{align}
\begin{split}
C^{\text{PREP}^\dagger}_{\mathrm{THC}} &= 3\ceil{\log_2 S} - 3\eta + 2b_r - 9 + \mathrm{QRM}^{\dagger \mathrm T}_{\mathrm c/\mathrm d}(S) + \aleph\\ 
&\quad + n(\ceil{\log_2 M} + \ceil{\log_2 R_{\max}}) + \lceil \log_2 n \rceil + 1. 
\end{split}\label{Eq: THC PREPdag Toffoli cost}  
\end{align}

Dominant contributions to the Toffoli costs of PREP and PREP$^\dagger$ come from QROM and inverse QROM, respectively.
With sufficient ancillas, the forward-QROM cost scales as $\mathcal{O}(\sqrt{Sm})$, whereas the inverse-QROM cost scales as $\mathcal{O}(\sqrt{S})$.
Thus, for fixed $n$ and precision parameters, $C^{\mathrm{PREP}}_{\mathrm{THC}}=\mathcal{O}\left((MN)^{n/2}\sqrt{m}\right)=\widetilde{\mathcal O}\left((MN)^{n/2}\right)$, while $C^{\mathrm{PREP}^\dagger}_{\mathrm{THC}} =\mathcal{O}\left((MN)^{n/2}\right)$.

The SELECT implementation for THC incurs a higher cost than the corresponding implementation for the Pauli LCU.
This increased cost arises from the requirement to implement modal rotations.
Specifically, the LCU unitaries that are implemented take the form $\hat {\mathcal{U}}^{(i),\mathcal{M}}_{\mu_i}
\hat Z_{0_i}
\left(\hat {\mathcal{U}}^{(i),\mathcal{M}}_{\mu_i}\right)^\dagger$,
where $\hat{\mathcal U}^{(i),\mathcal M}_{\mu_i}$ denotes a modal rotation operator.
The superscript $\mathcal{M}$ stands for the mode-coupling level of a given LCU term.
$\mathcal{M}$ can take values from 0 to $n-1$, where $n$ is the mode-coupling level of the Hamiltonian.
Thus, for one-, two-, and three-mode-coupling terms, $\mathcal{M}$ takes the values 0, 1, and 2, respectively.
We can relate $\hat {\mathcal U}^{(i),\mathcal{M}}_{\mu_i}$ to the modal rotation operators of Eq.~\eqref{Eq: THC LCU} as follows:
\begin{align}
    \hat {\mathcal U}^{(i),\mathcal{M}}_{\mu_i} =
    \begin{cases}
        \hat W^{(i)}_{\mu_i} & \mathcal{M} = 0 \\
        \hat U^{(i)}_{\mu_i} & \mathcal{M} = 1 \\
        \hat V^{(i)}_{\mu_i} & \mathcal{M} = 2
    \end{cases}.
\end{align}
Modal rotation is implemented by decomposing the operator into $N-1$ Givens rotations (see Appendix C of Ref.~\cite{majland2025fault}).
These rotations are realized by adding the corresponding rotation angles into a phase gradient state, thereby inducing the desired phase kickback.

\begin{figure}[!t]\centering    \includegraphics[width=1\linewidth]{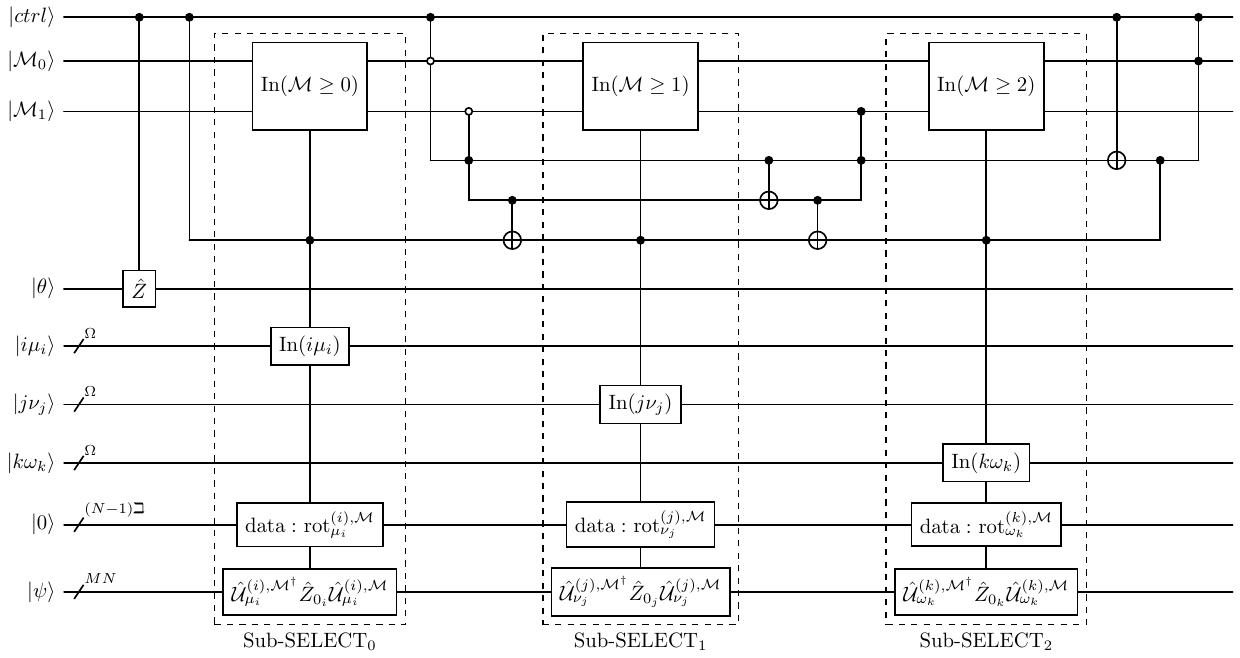}
    \caption{Quantum circuit of the $\mathrm{SELECT}$ operation for block-encoding a second-quantized vibrational Hamiltonian expressed as a THC LCU. 
    The controlled-$Z$ gate conditioned on the external control applies the phase associated with the coefficient sign. 
    The two-bit register $\mathcal{M}_0\mathcal{M}_1$ encodes the mode-coupling label $\mathcal M\in\{0,1,2\}$, corresponding to one-, two-, and three-mode-coupling terms.
    The accumulator qubit $\ket{\mathrm{acc}}$ is initialized conditionally on $\ket{\mathrm{ctrl}}$ and activates Sub-SELECT$_l$ when $l\leq \mathcal{M}$. 
    Each active Sub-SELECT applies the modal-rotation-conjugated $Z$ operator associated with the corresponding loaded mode and factor index.
    $\mathrm{rot}^{(i),\mathcal{M}}_{\mu_i}$ denotes the rotation angles corresponding to the modal rotation operator $\hat {\mathcal U}^{(i),\mathcal{M}}_{\mu_i}$. 
    The register of size $(N-1)\beth$ is used to store the rotation angles, where $\beth$ is the number of bits used to approximate the angles.}\label{Fig: THC SELECT}
\end{figure}

The overall SELECT circuit is shown in Fig.~\ref{Fig: THC SELECT}.
The controlled-$Z$ operation acts on the sign qubit to ensure that the correct sign is recovered in front of each LCU term.
To implement a ranged operation similar to the Pauli LCU, an ancilla is turned on, controlled on the external control qubit.
We label this ancilla as $\ket{acc}$, which stands for ``accumulator'' qubit.
This is followed by repeated application of a primitive operation referred to as Sub-SELECT.
Although each instance of Sub-SELECT differs slightly, the distinction arises mainly from the procedure used to perform unary iteration over the mode-coupling index and the mode index.
In successive applications of Sub-SELECT, the unary iteration over $\mathcal{M}$ and mode index $i$ begins from an integer value that is incremented by one relative to the previous instance.
We distinguish each Sub-SELECT by adding a subscript to it.
For example, Sub-SELECT$_{l}$ includes a unary iteration over the mode-coupling index that iterates over values $\mathcal{M} \geq l$ and a mode index $i \geq l$.
The iteration over mode-coupling qubits is used to implement distinct rotations corresponding to each mode-coupling level.
Sub-SELECT$_l$ is active iff $l\le\mathcal M$; equivalently, the accumulator remains active for the first $\mathcal M+1$ Sub-SELECT primitives.
Since an LCU term can have at most $\mathcal{M} = n-1$ and there are $n$ Sub-SELECT operations, it can be ensured that the $\ket{acc}$ qubit returns to zero after the action of the last Sub-SELECT primitive.

\begin{figure}[!t]\centering    \includegraphics[width=1\linewidth]{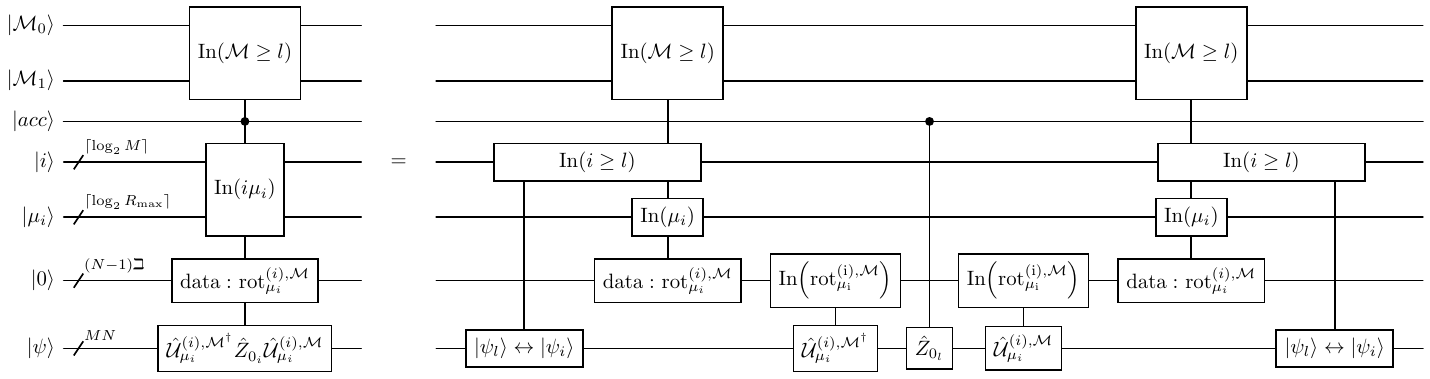}
    \caption{Quantum circuit of the Sub-SELECT$_l$ operation for block-encoding a second-quantized vibrational Hamiltonian expressed as a THC LCU. 
    The first two qubits encode the mode-coupling level of a given LCU term. 
    $\ket{acc}$ stands for the accumulator qubit. 
    The Sub-SELECT is performed by conjugating the action of Givens rotations by data loaders and swaps. 
    The iteration over index $i$ is used for swapping the system qubits while loading appropriate rotation angles in parallel. 
    The $\hat Z_{0_l}$ operator is controlled on the accumulator qubit to ensure that the whole Sub-SELECT primitive has a non-trivial action only if the accumulator qubit is switched on.
}\label{Fig: THC Sub-SELECT}
\end{figure}

The Toffoli count of the SELECT circuit is obtained as the sum of the Toffoli counts of each Sub-SELECT primitive and the additional overhead from ensuring a ranged operation.
This can be denoted as
\[ 
C^{\text{SEL}}_{\text{THC}} = \sum_{l=0}^{n-1} C^{\text{Sub-SEL}_l}_{\text{THC}} + n - 1,
\]
where $C^{\text{Sub-SEL}_l}_{\text{THC}}$ stands for the Toffoli cost of Sub-SELECT$_l$.
In Fig.~\ref{Fig: THC Sub-SELECT}, we provide the implementation for the Sub-SELECT$_{l}$ circuit.
It involves loading rotation angles as a function of $i, \mu_i$ and $\mathcal{M}$ values as well as swapping the system qubits for mode $l$ with mode $i$.
This can be efficiently performed by using QROM with clean ancillas to load the angles while keeping the iteration over the $i$ values at the top of the unary tree.
This allows us to use the same unary tree to perform the system qubit swaps in parallel to loading the rotation angles.
The gate $\mathrm{In}(i \geq l)$ in Fig.~\ref{Fig: THC Sub-SELECT} reflects this by having parallel branches for swapping and loading.
The reason for the restricted unary iteration over $i$ and $\mathcal{M}$ values directly follows from the restricted sum over mode indices in the LCU expansion of a vibrational Hamiltonian.
The rotations are performed by adding the angles to the phase gradient state.
A $\hat Z$ gate is applied to modal 0 of mode $l$, followed by the application of inverse rotations, inverse QROM, and CSWAPs.
The Toffoli cost of the Sub-SELECT$_l$ operation can be obtained as follows.

\begin{itemize}
    \item The cost of loading the rotation angles using QROM with clean ancillas is (see Eq.~\eqref{Eq: QROM clean Toffoli cost}) $\mathrm{QRM}^{\mathrm T}_{\mathrm c}(\tilde S_l, \tilde m)$, where $\tilde S_l$ is the number of values that are loaded and $\tilde m = (N-1)\beth$ is the size of the data that is loaded.
    $\beth$ is the number of bits used in the binary approximation of the rotation angles.
    The value of $\tilde S_l$ can be written as
    \begin{align}
        \tilde S_l &= (n-l)\sum_{i=l}^{M-1}R_i + \sum_{i=0}^{M-1}(N-R_i)\delta_{l,0} \nonumber \\
        &= (n - l)(M-l)\bar R_{\geq l} + M(N-\bar R)\delta_{l,0}, \label{Eq: THC SEL QROM S}
    \end{align}
    where $\bar R$ and $\bar R_{\geq l}$ stand for the average THC rank and average THC rank of the last $M-l$ modes, respectively.
    Thus, $\bar R = \bar R_{\geq 0}$.
    The presence of $\delta_{l,0}$ signifies that for the first Sub-SELECT, for $\mathcal{M} = 0$, which encodes the one-mode operator, the number of terms per mode in the diagonal representation is $N$ and not $R_i$. 
    \item While loading the rotation angles, by keeping the iteration over $i$ values at the top of the unary tree, we can also perform controlled swaps on the system qubits to bring the $i$-th subsystem qubits to the $l$-th position.
    This has a cost of $(M-l-1)N$. 
    \item The next step is to perform Givens rotations.
    This has a cost of $2(N-1)(\beth-2)$.
    \item The controlled $\hat Z$ operation does not require Toffoli gates.
    \item Inverting the rotations again costs $2(N-1)(\beth-2)$ Toffoli gates.
    \item The final step is to undo the data loading and controlled swaps from the first step.
    Using measurement-based uncomputation, the cost of unloading is $\mathrm{QRM}^{\dagger \mathrm T}_{\mathrm c}(\tilde S_l)$, while the cost of swaps is $(M-l-1)N$.

\end{itemize}
Thus, the cost of Sub-SELECT$_l$ can be given by
\begin{align}
    C^{\text{Sub-SEL}_l}_{\text{THC}} = 2(M-l-1)N + \mathrm{QRM}^{\mathrm T}_{\mathrm c}(\tilde S_l, \tilde m) +\mathrm{QRM}^{\dagger \mathrm T}_{\mathrm c}(\tilde S_l) + 4(N-1)(\beth-2),
\end{align}
where $\tilde S_l$ is given in Eq.~\eqref{Eq: THC SEL QROM S}.
Using this, we can write the cost of the SELECT circuit as
\begin{align}
    C^{\text{SEL}}_{\text{THC}} &= \sum_{l=0}^{n-1}\left(2(M-l-1)N + \mathrm{QRM}^{\mathrm T}_{\mathrm c}(\tilde S_l, \tilde m) +\mathrm{QRM}^{\dagger \mathrm T}_{\mathrm c}(\tilde S_l) + 4(N-1)(\beth-2) \right) + n - 1 \nonumber \\
    &= 2MNn - Nn(n+1) + n - 1 + 4n(N-1)(\beth -2 ) + \sum_{l=0}^{n-1}\left( \mathrm{QRM}^{\mathrm T}_{\mathrm c}(\tilde S_l, \tilde m) +\mathrm{QRM}^{\dagger \mathrm T}_{\mathrm c}(\tilde S_l) \right).
    \label{Eq: THC SELECT Toffoli cost}
\end{align}

For fixed $n$, constant $\beth$, and $\bar R=\mathcal O(N)$, the cost of the THC SELECT circuit scales as $\mathcal{O}(MNn)$.
The THC LCU block-encoding cost is defined as
\begin{align}
C^{\mathrm{BE}}_{\mathrm{THC}} = C^{\mathrm{PREP}}_{\mathrm{THC}} + C^{\mathrm{SEL}}_{\mathrm{THC}} + C^{\mathrm{PREP}^\dagger}_{\mathrm{THC}}. \label{Eq: THC BE Toffoli cost}
\end{align}
The qubits that need to be reflected on for construction of the qubitization walk operator are as follows: The $\lceil \log_2S \rceil$ qubits used in the uniform superposition state, 1 qubit used to flag the success of uniform superposition state preparation, and $\aleph$ qubits used in coherent alias sampling.
All other ancillas become unentangled with the system qubits after the application of PREP$^\dagger$.
These reflections have a cost of $\lceil \log_2S \rceil + \aleph - 1$ Toffolis.
We also need one Toffoli to control the SELECT operation on the success of preparing a uniform superposition state in PREP and one more for coupling the reflection with the right branch of the unary iterator of QPE.
Thus, the total Toffoli cost of estimating eigenvalues of a second-quantized vibrational Hamiltonian to accuracy $\epsilon$ using QPE is given by
\begin{align}
    C_{\mathrm{THC}} = \left\lceil\dfrac{\pi \lambda_{\mathrm{THC}}}{2\epsilon} \right\rceil (C^{\mathrm{BE}}_{\mathrm{THC}} + \lceil \log_2S \rceil + \aleph + 1). \label{Eq: THC LCU total Toffoli cost}
\end{align}
For optimal data-loading ancillas and fixed $n\ge 2$ and precision parameters, the asymptotic Toffoli cost based on Eq.~\eqref{Eq: THC LCU total Toffoli cost} is $\widetilde{\mathcal{O}}\left(\lambda_{\mathrm{THC}}(MN)^{n/2}/\epsilon\right)$.
Note that the asymptotic costs in both Pauli and THC LCUs are still dominated by the PREP circuit.
However, the requirement of modal rotations can provide significant overhead in the case of the THC LCU.

We now consider the qubit cost of implementing QPE for the THC LCU.
Some of the QPE-related costs remain identical to the Pauli LCU.
\begin{enumerate}
    \item The phase estimation requires a register of size $\lceil\log_2(\mathcal{I} + 1)\rceil$ where $\mathcal{I} = \left\lceil \dfrac{\pi\lambda_{\text{THC}}}{2\epsilon} \right \rceil$.
    \item A unary iterator on the phase estimation register requires $\lceil\log_2(\mathcal{I} + 1)\rceil - 1$ ancillas.
    \item Preparing the uniform superposition state requires $\ceil{\log_2(S)}$ qubits to store the uniform index, $1$ qubit for storing the success, $b_r$ qubits for the phase gradient state, and $1$ ancilla to perform the rotation on (see Appendix A of Ref.~\cite{thc}).
    Here $S$ is the number of terms in the LCU.
    \item A register of size $m$ (defined in Eq.~\eqref{Eq: THC QROM volume}) stores the bit strings loaded by QROM.
    The number of clean ancillas required when using clean or dirty (c/d) ancilla-based QROM is given by $\mathrm{QRM}^{\mathrm N}_{\mathrm{c/d}}(S,m)$, as defined in Eqs.~\eqref{Eq: QROM clean qubit cost} and \eqref{Eq: QROM dirty qubit cost}.
    Recall that we have defined $\mathrm{QRM}^{\mathrm N}_{\mathrm d}(S,m)\equiv \mathrm{QRM}^{\mathrm N}_{\mathrm d}(S)$ for notational convenience.
    For inverse QROM with measurement-based uncomputation, the corresponding clean ancilla cost is $\mathrm{QRM}^{\dagger \mathrm N}_{\mathrm c/ \mathrm d}(S)$; this contribution need not be counted separately because the inverse QROM can reuse the QROM ancillas together with any idle clean qubits available elsewhere in the circuit.
    \item $\aleph$ qubits are required for storing the uniform superposition state $\sigma$ used in coherent alias sampling.
    Another qubit is needed to store the result of the inequality test in coherent alias sampling.
    \item In SELECT, $\mathrm{QRM}^{\mathrm N}_{\mathrm c}(\tilde S_0, \tilde m)$ clean ancillas are required when utilizing QROM to load the rotation angles.
    $\tilde S_0$ is defined in Eq.~\eqref{Eq: THC SEL QROM S} and $\tilde m = (N-1)\beth$.
    Here, one can reuse the QROM ancillas from the PREP circuit.
    $\tilde m$ qubits are required to store the rotation angles.
    The qubit cost of unloading the angles can again be ignored since it will be smaller than the cost for loading them.
    \item While $b_r$ qubits are already counted for a phase gradient state, if $\beth > b_r$, then we will replace $b_r$ with $\beth$.
    \item $\beth - 2$ ancillas, which can again be borrowed from QROM ancillas if available, are required while adding to a phase gradient state.
    \item $MN$ qubits are required for encoding the system wavefunction.
\end{enumerate}
In total, the logical qubit cost for the THC LCU can be given as
\begin{align}
\begin{split}
    N_{\mathrm{THC}} &= 2\lceil\log_2(\mathcal{I} + 1)\rceil - 1 + \ceil{\log_2(S)} + \max(b_r, \beth) + 2 + m + \aleph + 1 + MN \\
    &\quad + \max\left( \mathrm{QRM}^{\mathrm N}_{\mathrm c / \mathrm d}(S,m),\ \max\big(\mathrm{QRM}^{\mathrm N}_{\mathrm c}(\tilde S_0, \tilde m),\ \beth-2\big)+\tilde m\right),
    \end{split}
    \label{Eq: THC LCU total qubit cost}
\end{align}
For fixed $n\ge 2$ and precision parameters, if an optimal number of clean or dirty ancillas is available to minimize the Toffoli count, then the qubit cost for the THC LCU scales as $\widetilde{\mathcal{O}}\left((MN)^{n/2}\right)$, which mainly comes from QROM.
Equations \eqref{Eq: THC LCU total Toffoli cost} and \eqref{Eq: THC LCU total qubit cost} together determine the quantum resources required to estimate eigenvalues of a vibrational Hamiltonian expressed as a THC LCU using QPE.
Utilization of BLISS only changes the value of the LCU 1-norm.

\section{Details of BLISS} \label{app: BLISS details}
This section provides a complete mathematical description of the BLISS technique, alongside the specific algorithmic choices used to construct the BLISS-Pauli and BLISS-THC LCUs.
We define the BLISS-modified Hamiltonian as
$$
\hat {H}^B = \hat H - \hat K_1 - \hat K_2 - \hat K_3,
$$
where $\hat H$ is assumed to be in the symmetrized form of Eq.~\eqref{Eq: second-quantized symmetrized vibrational Hamiltonian} and $\hat K_{n}$ is a killer operator at the mode-coupling level $n$.
We expand these killer operators using the number-operator symmetry as follows:
\begin{align}
\hat{K_1} &= \sum_{i=0}^{M-1} \mathcal{A}^{(i)} (\hat{\mathrm N}_i - 1), \label{Eq: obt killer}\\
\hat K_2 &= \sum_{i, j=0}^{M-1} \mathcal{B}^{(i,j)} (\hat{\mathrm N}_i \hat{\mathrm N}_j - 1) + \sum_{i,j=0}^{M-1} \sum_{r,s=0}^{N-1} \mathcal{C}_{r,s}^{(i,j)} (\hat{\mathrm N}_i - 1) \hat{E}^{r_j}_{s_j}, \label{Eq: tbt killer}\\
\begin{split}
\hat K_3&= \sum_{i,j,k=0}^{M-1} \mathcal{D}^{(i,j,k)} (\hat{\mathrm N}_i \hat{\mathrm N}_j \hat{\mathrm N}_k - 1) + \sum_{i,j,k=0}^{M-1} \sum_{t,u=0}^{N-1} \mathcal{E}_{t,u}^{(i,j,k)} (\hat{\mathrm N}_i \hat{\mathrm N}_j - 1) \hat{E}^{t_k}_{u_k} \\
&\quad + \sum_{i,j,k=0}^{M-1} \sum_{r,s,t,u=0}^{N-1} \mathcal{F}_{r,s,t,u}^{(i,j,k)} (\hat{\mathrm N}_i - 1) \hat{E}^{r_j}_{s_j} \hat{E}^{t_k}_{u_k},
\end{split} \label{Eq: trbt killer}
\end{align}
where $\mathcal{A}, \mathcal{B}, \mathcal{C}, \mathcal{D}, \mathcal{E}$, and $\mathcal{F}$ are parameterized tensors that can be optimized to minimize the overall 1-norm.
Because the excitation operators natively conserve excitation count within their mode, each of these killers annihilates states within the one-excitation-per-mode physical subspace, regardless of the tensor coefficient values.
To ensure that $\hat{H}^B$ remains Hermitian and that the modified tensors retain the permutation symmetries of the original $\hat H$, we impose the following restrictions on the coefficients:
\begin{gather*}
\mathcal{B}^{(i,j)} = \mathcal{B}^{(j,i)}, \\
\mathcal{C}_{r,s}^{(i,j)} = \mathcal{C}_{s,r}^{(i,j)}, \\
\mathcal{D}^{(i,j,k)} = \mathcal{D}^{(k,i,j)} = \mathcal{D}^{(j,k,i)} = \mathcal{D}^{(j,i,k)} = \mathcal{D}^{(k,j,i)} = \mathcal{D}^{(i,k,j)}, \\
\mathcal{E}_{u,t}^{(i,j,k)} = \mathcal{E}_{t,u}^{(i,j,k)} = \mathcal{E}_{t,u}^{(j,i,k)} = \mathcal{E}_{u,t}^{(j,i,k)}, \\
\mathcal{F}_{s,r,u,t}^{(i,j,k)} = \mathcal{F}_{r,s,u,t}^{(i,j,k)} = \mathcal{F}_{s,r,t,u}^{(i,j,k)} = \mathcal{F}_{r,s,t,u}^{(i,j,k)} = \mathcal{F}_{t,u,r,s}^{(i,k,j)} = 
\mathcal{F}_{u,t,r,s}^{(i,k,j)} =
\mathcal{F}_{t,u,s,r}^{(i,k,j)} =
\mathcal{F}_{u,t,s,r}^{(i,k,j)}.
\end{gather*}
Additionally, these coefficients are strictly zero whenever any two of the mode indices ($i, j, k$) are identical.
Applying these killers to the Hamiltonian in Eq.~\eqref{Eq: second-quantized symmetrized vibrational Hamiltonian} yields the BLISS-modified tensors
\begin{align}
\begin{split}
h_{i,p,q}^B &= h_{i,p,q} - \mathcal{A}^{(i)}\delta_{p,q} + \sum_{j=0}^{M-1}\mathcal{C}_{p,q}^{(j,i)} + \sum_{j,k=0}^{M-1}\mathcal{E}_{p,q}^{(j,k,i)}, \\
g_{i,p,q,j,r,s}^B &= g_{i,p,q,j,r,s} 
-  \mathcal{B}^{(i,j)} \delta_{p,q} \delta_{r,s} - \frac{1}{2} \Bigl( \mathcal{C}_{r,s}^{(i,j)} \delta_{p,q} + \mathcal{C}_{p,q}^{(j,i)} \delta_{r,s} \Bigr) 
+ \sum_{k=0}^{M-1} \mathcal{F}_{p,q,r,s}^{(k,i,j)}, \\
v_{i,p,q,j,r,s,k,t,u}^B &= v_{i,p,q,j,r,s,k,t,u} -  \mathcal{D}^{(i,j,k)} \delta_{p,q} \delta_{r,s} \delta_{t,u}  - \frac{1}{3} \Bigl( \mathcal{E}_{t,u}^{(i,j,k)} \delta_{p,q} \delta_{r,s} + \mathcal{E}_{r,s}^{(i,k,j)} \delta_{p,q} \delta_{t,u} + \mathcal{E}_{p,q}^{(k,j,i)} \delta_{r,s} \delta_{t,u} \Bigr) \\
&\quad - \frac{1}{3} \Bigl( \mathcal{F}_{r,s,t,u}^{(i,j,k)} \delta_{p,q} + \mathcal{F}_{p,q,t,u}^{(j,i,k)} \delta_{r,s} + \mathcal{F}_{r,s,p,q}^{(k,j,i)} \delta_{t,u} \Bigr),
\end{split} \label{Eq: BLISS modified tensors}
\end{align}
along with a constant global shift to the Hamiltonian given by
\begin{align*}
C^B = \sum_i \mathcal{A}^{(i)} +  \sum_{i, j} \mathcal{B}^{(i,j)} + \sum_{i, j, k} \mathcal{D}^{(i,j,k)}.
\end{align*}
Optimizing the Pauli LCU via BLISS proceeds directly from these modified tensors.
Once the Pauli LCU for $\hat H^B$ is derived, the resulting 1-norm becomes a cost function of the BLISS parameters, which are optimized classically.
For the three-mode-coupling Hamiltonians studied here, we enforce $\mathcal{F} = 0$.
We observed that including the $\mathcal{F}$ tensor decreased the sparsity of the Hamiltonian while offering only marginal reductions in the 1-norm.
Similarly, for Hamiltonians truncated at the two-mode-coupling level, we set $\hat K_3 =0$.

For the BLISS-THC approach, we observed that applying the higher-order killers $\hat K_2$ or $\hat K_3$ from Eqs.~\eqref{Eq: tbt killer} and \eqref{Eq: trbt killer} requires substantially larger THC ranks to maintain the same accuracy.
Consequently, we construct the BLISS-THC Hamiltonian using only the one-mode killer, $\hat K_1$.
This modifies the parameters $\bm{\eta}^{(i)}_{\mu_i}$ of Eq.~\eqref{Eq: THC LCU} to $\bm{\eta}^{(i)B}_{\mu_i} = \bm{\eta}^{(i)}_{\mu_i} + \mathcal{A}^{(i)}$.
Because the $\bm{\eta}^{(i)}_{\mu_i}$ parameters correspond to the negative of the eigenvalues of the THC-modified one-mode-coupling tensor, the 1-norm is optimally minimized by setting $\mathcal{A}^{(i)} = -\operatorname{median}_{\mu_i}\big(\bm{\eta}^{(i)}_{\mu_i}\big)$.
Thus, the final 1-norm for the BLISS-THC LCU retains the exact mathematical form of Eq.~\eqref{Eq: THC 1-norm}, with $\bm{\eta}^{(i)}_{\mu_i}$ replaced by $\bm{\eta}^{(i)B}_{\mu_i}$.

\section{Comparing Qubitization with Trotter Approximation} \label{App: Trotter}
Qubitization of the block-encoded Hamiltonian enables efficient implementation of QPE.
Alternatively, one can approximate the time-evolution operator using a Trotter decomposition and perform phase estimation directly on the resulting operator.
This approach can be highly favorable because Trotterization does not require additional ancilla qubits.
However, previous studies on the electronic Hamiltonian have shown that qubitization often leads to a smaller Toffoli cost than Trotter-based methods \cite{thc}, though this performance gap is narrowing \cite{SPRINT}.
In Ref.~\cite{malpathak2025trotter}, different Hamiltonian fragmentations and their associated costs were studied for implementing a Trotterized time propagator.
In this section, we consider these costs and compare them with the cost of qubitization-based phase estimation using the LCUs developed here.

The second-order Trotter approximation evaluates the time evolution of a Hamiltonian, $\hat{H} = \sum_{i=0}^{L-1} \hat{H}_i$, by expressing the exact time propagator for a time $t$ as a product of the evolutions of its individual fragments for a time step $\Delta t = t/r$, repeated $r$ times in the following form:
\[e^{-\mathrm{i}\hat{H}t} \approx \left(\prod_{i=0}^{L-1} e^{-\mathrm{i}\hat{H}_i \Delta t/2} \prod_{i=L-1}^0 e^{-\mathrm{i}\hat{H}_i \Delta t/2} \right)^r.\]
The resulting approximate time propagator can then be viewed as being generated exactly by a different, effective Hamiltonian, $\hat{H}_{\mathrm{eff}}(\Delta t)$, such that
\[\prod_{i=0}^{L-1} e^{-\mathrm{i}\hat{H}_i \Delta t/2} \prod_{i=L-1}^0 e^{-\mathrm{i}\hat{H}_i \Delta t/2} = e^{-\mathrm{i}\hat{H}_{\mathrm{eff}}(\Delta t)\Delta t}.\]
By increasing $r$, the difference between the eigenvalues of $\hat{H}_{\mathrm{eff}}(\Delta t)$ and $\hat{H}$ can be kept below a target error threshold, $\epsilon_{\mathrm{Trot}}$.
Reference~\cite{malpathak2025trotter} assumes the state of the system undergoing QPE has a spectral support of width $\alpha$.
To~prevent the aliasing of Hamiltonian energies, QPE can be performed on the propagator $\exp(-\mathrm{i}\hat{H}_{\mathrm{eff}}2\pi/\alpha)$ rather than $\exp(-\mathrm{i}\hat{H}_{\mathrm{eff}}2\pi/\Vert{}\hat{H}\Vert{})$.
Because $\Vert{}\hat{H}\Vert{} > \alpha$, using the latter propagator in QPE would require more queries, thereby inflating the final resource estimates.

The qubitization costs evaluated in this work employ optimizations established in Refs.~\cite{thc} and \cite{linearT}.
Specifically: (1) instead of initializing a uniform superposition state on the ancilla qubits, QPE is performed on a state that improves precision by either minimizing the spectral leakage \cite{dpss_prep} or the Holevo variance \cite{optimal_phase_shift}; (2) rather than applying a controlled time propagator conditioned solely on the QPE ancilla being in the $\ket{1}$ state, a directionally controlled forward or backward evolution is conditioned on the ancilla being in either the $\ket{1}$ or $\ket{0}$ state; and (3) the number of queries to the walk operator is not restricted to a power of 2.
Under these assumptions, the required number of walk operator queries is $\lceil \pi \lambda_{\mathrm{LCU}}/(2\epsilon_{\mathrm{QPE}}) \rceil$, where $\lambda_{\mathrm{LCU}}$ is the LCU 1-norm and $\epsilon_{\mathrm{QPE}}$ is the target energy resolution for QPE.

To ensure a fair comparison, we apply the same optimizations to Trotterization.
The directionally controlled Trotter evolution can be implemented at the same cost as an uncontrolled time propagator \cite{directional_evoln_1, directional_evoln_2}. 
If QPE is performed on the operator $\exp(-\mathrm{i}\hat{H}_{\mathrm{eff}}2\pi/\alpha)$, the algorithm estimates the dimensionless parameter $\phi = E/\alpha$, where $E$ is the target eigenvalue.
Therefore, the error in the energy, $\epsilon_{\mathrm{QPE}}$, relates to the error in the measured value, $\delta\phi$, as
\begin{align*}
    \epsilon_{\mathrm{QPE}} = \alpha \delta \phi.
\end{align*}
Assuming the phase error is dominated by the Holevo variance \cite{linearT}, we can relate it to $N_{\mathrm{calls}}$, the number of QPE queries to the propagator, as
\begin{align*}
    \delta \phi = \dfrac{1}{2\pi}\tan\left(\dfrac{\pi}{2N_{\mathrm{calls}} + 1}\right) \approx \dfrac{1}{4N_{\mathrm{calls}}}.
\end{align*}
The prefactor of $1/(2\pi)$ arises from explicitly including the $2\pi$ factor in the exponential definition of the propagator.
This factor is absent in Ref.~\cite{linearT}, which defines the time propagator as $\exp(-\mathrm{i}\hat{H}/\alpha)$.
Thus, $N_{\mathrm{calls}}$ can be expressed in terms of the target QPE energy accuracy as
\begin{align*}
    N_{\mathrm{calls}} = \left \lceil \frac{\alpha}{4\epsilon_{\mathrm{QPE}}} \right \rceil. 
\end{align*}
Because each Trotter oracle $\exp(-\mathrm{i}\hat{H}_{\mathrm{eff}}2\pi/\alpha)$ is implemented using $r$ discrete steps of $\exp(-\mathrm{i}\hat{H}_{\mathrm{eff}}2\pi/(\alpha r))$, the total cost of estimating the energy to an accuracy $\epsilon$ using QPE on the Trotterized time evolution is
\begin{align*}
    C_{\mathrm{Trot}} = \left \lceil \frac{\alpha}{4\epsilon_{\mathrm{QPE}}} \right \rceil r C_{\mathrm{step}},
\end{align*}
where $C_{\mathrm{step}}$ is the cost of implementing a single Trotter step, $r$ is chosen such that the energy error of $\hat{H}_{\mathrm{eff}}$ is less than $\epsilon_{\mathrm{Trot}}$, and the total target accuracy is $\epsilon = \epsilon_{\mathrm{QPE}} + \epsilon_{\mathrm{Trot}}$.
If $\delta t$ is the evolution time of a single Trotter step required to bound the energy error below $\epsilon_{\mathrm{Trot}}$, then $r = \max(1, \lceil 2\pi/(\alpha \delta t) \rceil)$.
For a second-order Trotter product formula, the step duration scales as $\delta t \propto \sqrt{\epsilon_{\mathrm{Trot}}}$.

Table 3 in Ref.~\cite{malpathak2025trotter} lists the values of $C_{\mathrm{step}}$ and $r$ for four molecules (H$_2$S, CO$_2$, CH$_2$O, and CH$_4$).
To enable a direct comparison with the reported Trotter resource estimates, we retain the parameters used there, including $\epsilon_{\mathrm{Trot}}=7\,\mathrm{cm}^{-1}$ and a spectral width of $\alpha=5516\,\mathrm{cm}^{-1}$.
Among the various Hamiltonian fragmentation schemes evaluated, Christiansen Greedy Fragments (CGFs) yield the lowest resource estimates.
The CGFs were obtained via a greedy factorization of two-mode-coupling tensors to a relative accuracy of approximately 5\%.
Accordingly, we generated THC and Pauli LCUs of the same Hamiltonians to an accuracy of $\sim5\% \pm 1\%$ for the data presented here.
Figure~\ref{Fig: qubitization_vs_trotter} in the main text illustrates the results for estimating the energy to a combined QPE-plus-Trotter error budget of $\epsilon=10\,\mathrm{cm}^{-1}$.

\section{Numerical Data} \label{App: data}

\begin{table}
\centering
\begin{adjustbox}{max width=\linewidth}
\setlength{\tabcolsep}{2pt}
\begin{tabular}{|c|c|c|c|c|c|c|c|c|}
\hline
\textbf{Metric} & \textbf{Method} & \textbf{PAH1} & \textbf{PAH2} & \textbf{PAH3} & \textbf{PAH4} & \textbf{PAH5} & \textbf{PAH6} & \textbf{PAH7} \\ \hline
\multirow{5}{*}{\begin{tabular}[c]{@{}c@{}}1-norms \\ ($\times 10^{6}\,\mathrm{cm}^{-1}$)\end{tabular}} & Triangular & 2.19 & 3.14 & 4.27 & 5.19 & 6.30 & 7.22 & 8.33 \\
 & Pauli & 1.05 & 1.48 & 2.05 & 2.47 & 3.00 & 3.43 & 3.97 \\
 & BLISS-Pauli & 0.53 & 0.75 & 1.06 & 1.40 & 1.72 & 1.92 & 2.10 \\
 & THC & 0.48 & 0.72 & 0.96 & 1.23 & 1.47 & 1.70 & 1.96 \\
 & BLISS-THC & \textbf{0.39} & \textbf{0.57} & \textbf{0.76} & \textbf{0.97} & \textbf{1.15} & \textbf{1.33} & \textbf{1.53} \\ 
 \hline
\multirow{3}{*}{Qubit cost} & Triangular & \textbf{301} & \textbf{395} & \textbf{487} & \textbf{579} & \textbf{671} & \textbf{763} & \textbf{853} \\
 & Pauli & 334$\rightarrow$2567 & 432$\rightarrow$2789 & 529$\rightarrow$5570 & 619$\rightarrow$5660 & 711$\rightarrow$5752 & 803$\rightarrow$10963 & 897$\rightarrow$11565 \\
 & THC & 401$\rightarrow$1154 & 498$\rightarrow$2271 & 593$\rightarrow$2491 & 684$\rightarrow$4630 & 776$\rightarrow$4722 & 867$\rightarrow$4814 & 961$\rightarrow$5160 \\ 
 \hline
\multirow{5}{*}{\begin{tabular}[c]{@{}c@{}}Toffoli cost \\ ($\times 10^{11}$)\end{tabular}} & Triangular & 6.63 & 22.40 & 62.90 & 104.00 & 184.00 & 293.00 & 446.00 \\
 & Pauli & 0.45$\rightarrow$0.09 & 1.03$\rightarrow$0.19 & 2.62$\rightarrow$0.39 & 3.24$\rightarrow$0.55 & 4.79$\rightarrow$0.82 & 6.35$\rightarrow$1.12 & 9.54$\rightarrow$1.49 \\
 & BLISS-Pauli & 0.28$\rightarrow$0.05 & 0.70$\rightarrow$0.11 & 1.71$\rightarrow$0.22 & 2.59$\rightarrow$0.37 & 3.91$\rightarrow$0.56 & 4.94$\rightarrow$0.72 & 7.14$\rightarrow$0.92 \\
 & THC & 0.07$\rightarrow$0.03 & 0.20$\rightarrow$0.07 & 0.40$\rightarrow$0.13 & 0.66$\rightarrow$0.21 & 1.10$\rightarrow$0.29 & 1.39$\rightarrow$0.39 & 2.03$\rightarrow$0.54 \\
 & BLISS-THC & \textbf{0.06$\rightarrow$0.03} & \textbf{0.16$\rightarrow$0.06} & \textbf{0.32$\rightarrow$0.10} & \textbf{0.52$\rightarrow$0.16} & \textbf{0.86$\rightarrow$0.23} & \textbf{1.09$\rightarrow$0.31} & \textbf{1.59$\rightarrow$0.42} \\
 \hline
\end{tabular}
\end{adjustbox}
\caption{LCU 1-norms and QPE quantum resource estimates underlying Figs.~\ref{Fig: 1-norms} and \ref{Fig: Toffoli and qubit costs for two-mode-coupling PAHs} for seven two-mode-coupling PAH Hamiltonians represented using five VSCF modals per mode, with a target QPE accuracy of $1\,\mathrm{cm}^{-1}$.
Arrows show the resource tradeoff from dirty-ancilla-assisted QROM to the maximum retained clean-ancilla QROM configuration.
BLISS variants are omitted from the qubit rows, consistent with the plotted qubit costs.}
\label{tab: two-mode coupling quantum resources table}
\end{table}

The PAH molecules considered here are linear chains of fused benzene rings called polyacenes. 
We consider up to seven benzene rings. 
Hamiltonians for these systems were obtained directly from the authors of Ref.~\cite{majland2025fault}. 
As such, the electronic structure calculations used to obtain these systems are given in Appendix F of the same paper. 
MidasCpp~\cite{midascpp} was then utilized to perform VSCF calculations and generate the explicit mode-coupling tensors. 
For the four small molecules used in obtaining Trotter-based resource estimates, the Hartree--Fock level of theory was used to generate the electronic PES.
In Table~\ref{tab: two-mode coupling quantum resources table}, we provide the numerical values used in plotting Figs.~\ref{Fig: 1-norms} and \ref{Fig: Toffoli and qubit costs for two-mode-coupling PAHs}.
Table~\ref{tab: trotter qubitization resources table} lists the numerical values used in plotting Fig.~\ref{Fig: qubitization_vs_trotter}.

\begin{table}[!t]
\centering
\begin{tabular}{|c|c|c|c|c|c|}
\hline
\textbf{Metric} & \textbf{Method} & \textbf{H$_2$S} & \textbf{CO$_2$} & \textbf{CH$_2$O} & \textbf{CH$_4$} \\ \hline \hline
\multirow{3}{*}{Qubits} & Trotter & \textbf{30} & \textbf{34} & \textbf{42} & \textbf{54} \\
 & BLISS-Pauli & 148 & 152 & 170 & 190 \\
 & BLISS-THC & 202 & 207 & 223 & 242 \\ \hline
\multirow{3}{*}{\shortstack{T gates \\ $(\times 10^{7})$}} & Trotter & 1.57 & 3.56 & 10.21 & 52.44 \\
 & BLISS-Pauli & 0.76 & \textbf{0.63} & 3.32 & 11.92 \\
 & BLISS-THC & \textbf{0.73} & 0.98 & \textbf{2.65} & \textbf{6.10} \\ \hline
\end{tabular}
\caption{QPE resource estimates underlying Fig.~\ref{Fig: qubitization_vs_trotter} for $\mathrm{H_2S}$, $\mathrm{CO_2}$, $\mathrm{CH_2O}$, and $\mathrm{CH_4}$, represented in a local-mode harmonic-oscillator basis with four modals per mode.
The combined error budget from QPE and the Trotter approximation is $10\,\mathrm{cm}^{-1}$, with $\epsilon_{\rm QPE}=3\,\mathrm{cm}^{-1}$ and $\epsilon_{\rm Trot}=7\,\mathrm{cm}^{-1}$.
The Trotter estimates correspond to the Christiansen Greedy Fragment construction of Ref.~\cite{malpathak2025trotter}, while the BLISS-Pauli and BLISS-THC estimates use QROM without the Toffoli--qubit tradeoff.}
\label{tab: trotter qubitization resources table}
\end{table}

\FloatBarrier

\bibliographystyle{unsrt}

\bibliography{ref}

\end{document}